\documentclass[reprint,floatfix]{revtex4-2}
\usepackage{amsmath}
\usepackage{graphicx}
\usepackage{dcolumn}
\usepackage{physics}
\usepackage{bm}
\usepackage{hyperref}
\usepackage{epsfig,colordvi}
\usepackage{color}
\usepackage{natbib}
\usepackage{subfiles}
\usepackage{float}
\usepackage{booktabs}
\usepackage{array}
\usepackage{soul}
\usepackage[normalem]{ulem}

\usepackage{lmodern}

 \def\be{\begin{equation}}
 \def\ee{\end{equation}}
 \def\bea{\begin{eqnarray}}
 \def\eea{\end{eqnarray}}
 \def\bean{\begin{eqnarray*}}
 \def\eean{\end{eqnarray*}}
 \def\gsim{\mathrel{\rlap{\lower0.2em\hbox{$\sim$}}\raise0.2em\hbox{$>$}}}
 \def\ksim{\mathrel{\rlap{\lower0.2em\hbox{$\sim$}}\raise0.2em\hbox{$<$}}}
 \def\kg{\mathrel{\rlap{\lower0.25em\hbox{$>$}}\raise0.25em\hbox{$<$}}}

\begin{document}
\title{
Systematic study of flow of protons and light clusters in intermediate-energy heavy-ion collisions with momentum-dependent potentials.}

\date{\today}

\author{Viktar Kireyeu$^{1}$, Vadim Voronyuk$^{1}$, Michael Winn$^{2}$,
 Susanne Gl{\"a}{\ss}el$^{3}$, J\"org Aichelin$^{2,4}$, Christoph Blume$^{3,5,6}$, Elena Bratkovskaya$^{5,6,7}$, Gabriele Coci$^{8}$, Jiaxing Zhao$^{6,7}$
   }

\affiliation{$^{1}$ Joint Institute for Nuclear Research, Joliot-Curie 6, 141980 Dubna, Moscow region, Russia}
\affiliation{$^{2}$ SUBATECH, Nantes University, IMT Atlantique, IN2P3/CNRS
4 rue Alfred Kastler, 44307 Nantes cedex 3, France}
\affiliation{$^{3}$ Institut f\"ur Kernphysik, Max-von-Laue-Str. 1, 60438 Frankfurt, Germany}
\affiliation{$^{4}$ Frankfurt Institute for Advanced Studies, Ruth Moufang Str. 1, 60438 Frankfurt, Germany} 
\affiliation{$^{5}$ GSI Helmholtzzentrum f\"ur Schwerionenforschung GmbH,
  Planckstr. 1, 64291 Darmstadt, Germany}
\affiliation{$^{6}$ Helmholtz Research Academy Hessen for FAIR (HFHF),GSI Helmholtz Center for Heavy Ion Physics. Campus Frankfurt, 60438 Frankfurt, Germany}  
\affiliation{$^{7}$ Institut f\"ur Theoretische Physik, Johann Wolfgang Goethe University,
Max-von-Laue-Str. 1, 60438 Frankfurt, Germany}
\affiliation{$^{8}$ Dipartimento di Fisica e Astronomia ``E. Majorana'', Università degli Studi di Catania, Via S. Sofia, 64, I-95125 Catania, Italy}

\begin{abstract} \noindent
We study the influence of the nuclear equation-of-state (EoS) on collective observables — the directed ($v_1$) and elliptic flow ($v_2$) of nucleons and light clusters — in heavy-ion collisions at GeV energies using the Parton-Hadron-Quantum-Molecular Dynamics (PHQMD) approach.  
A novel development in this work is the inclusion of a momentum-dependent nucleon potential in the PHQMD in addition to the static, density-dependent Skyrme interaction. This enables three distinct EoS scenarios: two static (“soft” and “hard,” differing in compressibility) and a soft, momentum-dependent EoS calibrated to $pA$ elastic scattering data.  
In PHQMD, clusters form during the entire heavy-ion collision via nucleon interactions and are identified using the Minimum Spanning Tree (MST) algorithm, including additional deuteron production from hadronic kinetic reactions.  
We find a strong EoS sensitivity in proton and cluster rapidity and $p_T$ distributions: soft and soft momentum-dependent EoS yield similar results, markedly different from the hard EoS. Softening the EoS reduces proton yields at midrapidity while enhancing light-cluster production.  
The EoS also affects flow observables differently for nucleons and clusters. For protons, a soft momentum-dependent potential increases slightly the magnitude of $v_1$ and $v_2$ relative to the hard EoS, whereas cluster flows are nearly similar. The soft momentum-dependent EoS provides an overall good agreement with experimental data from HADES and FOPI Collaborations while the soft EOS is not in line with the data. A scaling of $v_2$ with cluster mass number $A$ is observed at midrapidity for low $p_T$, which breaks at higher $p_T$. 
Finally, we examine the sensitivity of flow observables to deuteron production mechanisms. Deuterons formed via MST clustering exhibit different flow patterns from those produced by coalescence at freeze-out, indicating that flow harmonics may help discriminate between cluster formation scenarios.
\end{abstract}

\maketitle

\section{Introduction}

The exploration of the nuclear equation-of-state (EoS) is one of the major objectives in nuclear physics. Experimentally its study started with the first heavy-ion experiments at the Bevalac accelerator in Berkeley \cite{Gustafsson:1984ka} , where for the first time nuclear densities well above  normal nuclear matter density could be obtained in the laboratory. Later also complimentary information has been obtained from the observed mass-radius relations of neutron stars \cite{Koehn:2024set,Ozel:2010bz,Lastowiecki:2011hh,Bogdanov:2021yip} and even more recently first attempts have been made to obtain information about the nuclear equation-of-state from gravitational waves \cite{Bauswein:2012ya,Most:2022wgo}.  

At densities well above normal nuclear matter density, $\rho_0$, the EoS cannot be determined by nuclear matter calculations because there the expansion schemes of the Brückner G-matrix \cite{Botermans:1990qi}, as well as of the chiral perturbation approach, break down  \cite{Fuchs:1998zz}.
Therefore, the theoretical interpretation of 
heavy-ion experiments is the only way to study the EoS systematically.  In these experiments the EoS is, however, not directly measured and consequently it is a challenge for theory
to identify those experimental observables, which are sensitive to the EoS and to predict them with sufficient precision that a comparison with experimental results is meaningful.  It turned out that at  beam energies around 1 GeV/nucleon, where, on the one side, densities up to three times nuclear matter density can be reached and, on the other side, meson production is not very frequent, the nucleon dynamics is very sensitive to the potential interaction, which is directly related to the EoS of nuclear matter 
\cite{Gale:1987zz,Aichelin:1987ti,Ko:1987gp,Aichelin:1991xy,Weber:1993et,Fuchs:1998zz,Sahu:1998vz,Danielewicz:1998vz,Kuhrts:2000zs,Sahu:1999mq,Cassing:2000bj,Sahu:2002ku,Nara:2020ztb}.
These studies revealed  that - besides subthreshold kaon production - the directed and elliptic flow are among the the most promising experimental signals, which can be used for the determination of the EoS. 

The theoretical prediction within microscopic transport approaches is challenging since flow coefficients are very sensitive not only to the employed potentials, which reflect the equation-of-state, but also to the properties of hadrons in the medium and their collisions. They depend also strongly on the centrality of the reaction what makes a detailed comparison with experiments complicated. In addition, the signals are that tiny that different numerical realizations of the transport approaches start to play a role. 

To determine the EoS by analyzing the experimental results with the help of transport approaches is therefore a complicated task.
In earlier transport calculations the EoS had been considered as static and from the first Plastic Ball data \cite{Gustafsson:1984ka} it has been concluded that the nuclear EoS is rather hard \cite{Molitoris:1986pp}. Later one realized that statements about the nuclear EoS cannot be made without including the strong momentum dependence of the nucleon-nucleon potential \cite{Cooper:2009zza}. This influence of the momentum dependence has been studied in Quantum Molecular Dynamics (QMD) models \cite{Aichelin:1987ti,LeFevre:2015paj,Hillmann:2019wlt}, where the implementation is straight forward, as well as in BUU type models \cite{Welke:1988zz,Gale:1987zz,Gale:1989dm, Zhang:1994hpa,Sahu:1998vz,Tarasovicova:2024isp,Mohs:2024gyc}, where the implementation is more demanding. Two common conclusions emerged \cite{Aichelin:1987ti,Welke:1988zz,Gale:1987zz,Gale:1989dm,Zhang:1994hpa,Tarasovicova:2024isp,Mohs:2024gyc} : i) the flow of nucleons depends strongly on this momentum dependence: a static hard and a soft momentum-dependent interaction give roughly the same directed flow of nucleons whereas the flow for a static soft EoS gives considerably lower flow values  and ii) the experimentally measured flow of nucleons is approximately described by a soft momentum dependent interaction. For reviews we refer to  \cite{Bleicher:2022kcu,TMEP:2022xjg,Sorensen:2023zkk}.

In the meantime the transport approaches have been advanced and, even more important, several data sets for Au+Au collisions at very similar energies from the HADES and FOPI Collaborations became available.  In these experiments also the directed $v_1$ and elliptic $v_2$ flow of light clusters has been measured, which
allows for the first time  to extend the EoS studies to light clusters.  This is possible despite of the fact that the origin of cluster production at midrapidity is still debated.  Even more, the study of the flow of light clusters may help to identify the way they are produced, as we will show in this article. 

Light clusters at midrapidity have been found in heavy-ion experiments from beam energies of a couple of hundred MeV per nucleon up to the highest presently available energies of $\sqrt{s}$=5.02 TeV \cite{Reisdorf:2010aa,NA49:2004mrq,NA49:2016qvu,STAR:2019sjh,ALICE:2015wav}.  The slope of the transverse momentum spectra of all particles at midrapidity is - from LHC energies down to a beam energy of 1 A GeV - of the order of 100 MeV \cite{Adam:2019wnb, Melo:2020gbb, HADES:2020ver}. The excitation function of the multiplicity is rather smooth \cite{STAR:2019sjh,STAR:2022hbp}. Both observations point to an energy independent  formation mechanism.  One of the challenges for theory is how such clusters, weakly bound objects with a binding energy of the order of a couple of MeV (and in the case of hypertriton as low as about 130 keV), can survive in such an environment. 

Several propositions have been advanced to  understand the production of light clusters at midrapidity:

i)
    Clusters are formed according to phase space at a given temperature and a given chemical potential \cite{Andronic:2010qu}. These statistical models assume implicitly that the clusters are
    formed in a thermal environment and that after creation they do not interact anymore.  Statistical models are not able to make predictions on collective flow without additional assumptions.

ii)
    Clusters are formed by coalescence \cite{Butler:1963pp,Zhu:2015voa,Zhao:2020irc,Sun:2021dlz,Sombun:2018yqh,Kireyeu:2022qmv} assuming that after their last collision nucleons form clusters if the relative distance between the entrained nucleons in momentum and coordinate space is smaller than a given $\Delta r_0$ and $\Delta p_0$.
    
iii)
    Clusters are formed by the same potential interaction \cite{Aichelin:1991xy,Aichelin:2019tnk,Glassel:2021rod,Coci:2023daq}, which also determines the time evolution of the baryons in the semi-classical heavy-ion transport approaches.  Such clusters are recognized using the Minimum Spanning Tree (MST) algorithm \cite{Aichelin:1991xy}.
    
iv)
    Light clusters - as deuterons and  $A=3$ clusters - are formed by three-body collisions (like $NNN\to dN$ and $NN\pi \to d\pi$ for deuterons) and  destroyed by the inverse reaction  \cite{Danielewicz:1991dh,Danielewicz:1992mi,Oliinychenko:2020znl,Staudenmaier:2021lrg,Coci:2023daq,Sun:2021dlz,Ege:2024vls}. The cross sections for $\pi d $ and $Nd$ are known but the rate for the inverse reaction can only be obtained with additional approximations, which differ in the various approaches.

Although there are observables, which are sensitive to these cluster production mechanisms, the presently available data do not allow for an experimental distinction on the basis of the measured rapidity and $p_T$ distribution of the clusters \cite{Kireyeu:2024woo}. 
 
The goal of our study here is therefore twofold: \\
- to investigate the equation-of-state (EoS) of strongly interacting hadronic matter, created in heavy-ion collisions, by analyzing the directed flow $v_1$ and the elliptic flow $v_2$  of protons and light clusters. \\
- to investigate whether the collective flow observables $v_1$ and $v_2$ can distinguish between different theoretical models for cluster production in heavy-ion collisions.\\

For this purpose we calculate  $v_1$ and $v_2$ of protons and different clusters as a function of rapidity $y$ and transverse momentum $p_T$ for different EoS and for different cluster production mechanisms employing  Parton-Hadron-Quantum-Molecular Dynamics (PHQMD), a  microscopic transport approach \cite{Aichelin:2019tnk,Glassel:2021rod,Kireyeu:2022qmv,Coci:2023daq,Kireyeu:2024woo}.  
In the PHQMD approach we have the possibility to compare the  MST + kinetic  and  the coalescence  production mechanisms directly, because both are applied to the same physical events and therefore the environment is identical. This allows to study directly the consequences of the different production mechanisms. 
We confronted our calculations with the experimental data from the HADES  \cite{HADES:2020lob,HADES:2022osk} and  FOPI Collaborations \cite{FOPI:2011aa} at SIS energies. 
Also we relate our observations to recent results of UrQMD and SMASH. In UrQMD \cite{Hillmann:2019wlt,Steinheimer:2024eha} a static hard equation-of-state is used, which gives for flow observables quite similar results as a momentum dependent soft EoS \cite{Aichelin:1987ti}. SMASH \cite{Tarasovicova:2024isp} is a BUU type approach and uses a momentum dependent mean field, however with an implementation \cite{Mohs:2024gyc,Mohs:2025wrp}, which differs from that in PHQMD.
At lower beam energies the determination of the equation-of-state has been also addressed by Cozma in \cite{Cozma:2024cwc} in a different QMD model as well as by Danielewicz et al in \cite{Danielewicz:1999zn, Danielewicz:2002pu}. 

Our paper is organized as follows: we start with the model description in Section II.  In Section III we discuss the consequences of the different EoS for heavy-ion dynamics at SIS energies and show our results on flow coefficients in comparison to the HADES and FOPI data. The comparison of the flow of deuterons for different deuteron production mechanisms is discussed in Section IV  while our findings are summarized in Section V.  \\
We note that we  use the $\hbar = c = 1$  convention.
 
\section{Model description: PHQMD}

The Parton-Hadron-Quantum-Molecular Dynamics (PHQMD)   \cite{Aichelin:2019tnk,Glassel:2021rod,Kireyeu:2022qmv,Coci:2023daq,Kireyeu:2024woo} is  a microscopic n-body transport model, which combines the characteristics of baryon propagation from the Quantum Molecular Dynamics (QMD) model~\cite{Aichelin:1991xy,Aichelin:1987ti,Aichelin:1988me,Hartnack:1997ez} and the dynamical properties and interactions in- and out-of-equilibrium of hadronic and partonic degrees-of-freedom of the Parton-Hadron-String-Dynamics (PHSD) approach~\cite{Cassing:2008sv,Cassing:2008nn,Cassing:2009vt,Bratkovskaya:2011wp,Linnyk:2015rco,Moreau:2019vhw}. We refer to these papers for the details and discuss here only the novelties and their consequences.

\subsection{Momentum-dependent potential}

\begin{figure}[h!]
    \centering
    \resizebox{0.45\textwidth}{!}{\includegraphics{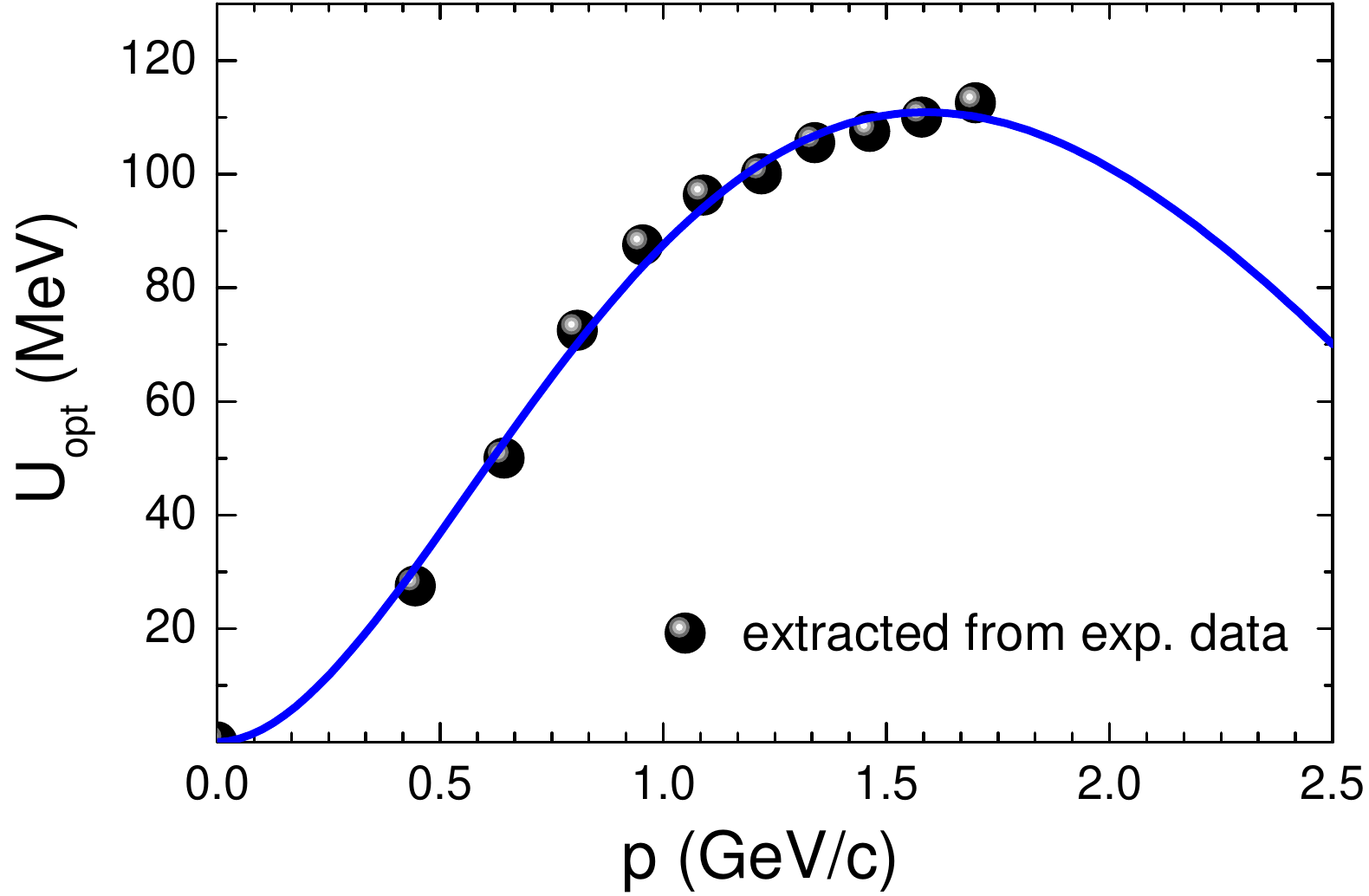}}
    \caption{Schrödinger equivalent optical potential $U_{opt}$, extracted from pA collisions, 
    versus total momentum $p$ of the proton  \cite{Clark:2006rj,Cooper:1993nx}.}
    \label{uopt}
\end{figure}

In the previous PHQMD calculations we employed a static nucleon-nucleon interaction. In reality this interaction is momentum-dependent as can be seen by analysing the beam energy dependence of elastic pA scattering data \cite{Clark:2006rj,Cooper:1993nx}.  Usually these data are analyzed by comparing the data with solutions of the Dirac equation with scalar $U_s$ and vector potentials with the zero component $U_0$. 
These are displayed in ref.\cite{Clark:2006rj,Cooper:1993nx}. To obtain a nucleon-nucleon potential, which we can employ in QMD type calculations, we have to calculate first the Schr\"odinger equivalent potential $U_{SEP}$ \cite{Jaminon:1980vk}
\begin{equation}
    U_{SEQ}(p) =U_s+U_0+\frac{2}{m_N}(U_s^2-U_0^2)+\frac{U_0}{m}(\sqrt{p^2+m^2}-m), 
\end{equation}
where $p$ is the momentum of the incoming proton in the target rest system. Our fit of this potential is displayed in Fig. \ref{uopt}  together with the points extracted from the analysis of the pA scattering data in the framework of the Dirac equation. 
There are no data available beyond a proton momentum of 1 GeV/c.
We checked that different extrapolations of the form of $U_{opt}(p)$ to higher proton momentum have very little influence on flow observables for beam energies $E_{kin} \le$ 1.5 AGeV.

In a second step we have to get the two body interaction between two nucleons. The Schr\"odinger equivalent potential  is obtained by averaging the two body potential $V({\bf p},{\bf p}_1) $  over the Fermi distribution of the cold target nucleons
\begin{equation}
    U_{SEQ}(p) =\frac{\int^{p_F} V({\bf p},{\bf p}_1) dp_1^3}{\frac{4}{3}\pi p_F^3}
\end{equation}
With  $ \Delta p = \sqrt{({\bf p}_{01}-{\bf p}_{02})^2}$ we find that a functional form of two-body potential is
\begin{equation}
\label{Umom12}
  V(\Delta p)=   V({\bf p}_{01},{\bf p}_{02})  =(a \Delta p + b\Delta p^2)\  exp [-c \sqrt{\Delta p}  ]
\end{equation}

In the Dirac analysis the vector potentials depend linearly on the baryon density in the nucleus . Therefore we assume that the momentum-dependent part of the two body potential has also a linear dependence on the baryon density, and obtain finally:
 \begin{equation}
    V({\bf r}_1,{\bf r}_2,{\bf p}_{01},{\bf p}_{02})  = (a \Delta p + b\Delta p^2)\  exp [-c \sqrt{\Delta p}]\ \delta({\bf r}_1-{\bf r}_2 ).
\end{equation}
The $\delta$ function creates a linear density dependence. The  energy of the system is
\begin{eqnarray}
   E &=& \langle \psi(t) |(T+V) |\psi(t) \rangle\nonumber \\
&=&\sum_i[\langle i | \frac{p^2}{2m} |i\rangle + \sum_{i\neq j} \langle ij| V_{ij} |ij \rangle] \nonumber \\
&=& \int H(r) d^3r.
\label{eq:energy}
\end{eqnarray}
$|\psi(t)\rangle =\prod_i |\psi_i\rangle$ is the n-body wave function, which is taken as the direct product of the single particle wave functions of the nucleons. The momentum-dependent potential has been introduced in QMD type transport approaches in Ref. \cite{Aichelin:1987ti} and explored later in Refs. \cite{Aichelin:1991xy,LeFevre:2016vpp,Hillmann:2019wlt,Nara:2020ztb}
and for BUU type approaches in Ref. \cite{Gale:1987zz} and has been widely applied later in different forms \cite{Ko:1987gp,Weber:1993et,Sahu:1998vz,Danielewicz:1998vz,Sahu:1999mq,Sahu:2002ku,Cassing:2000bj,Buss:2011mx,Mohs:2020awg}.

To include a momentum-dependent interaction in  mean field approaches, like it is done in  \cite{Welke:1988zz,Gale:1989dm,Danielewicz:1991dh,Sahu:2002ku,Tarasovicova:2024isp} is a challenging task and yield. Depending  on the implementation,  the forces acting on nucleons are different in  the QMD and BUU approaches, even for the same nucleon-nucleon potential. This one should keep in mind when comparing mean field (BUU) and QMD results. The dependence on the implementation has already been evoked in \cite{Welke:1988zz,Mohs} .

The total potential energy of nucleons in PHQMD has three parts, a local static Skyrme type interaction, a local momentum-dependent interaction and a Coulomb interaction
\begin{eqnarray}
V_{ij}&=& V({\bf r}_i, {\bf r}_j,{\bf r}_{i0},{\bf r}_{j0},{\bf p}_{i0},{\bf p}_{j0},t) \\
  &=& V_{\rm Skyrme\ loc}+ V_{\rm mom}+ V_{\rm Coul} \nonumber \\  
 &=&\frac{1}{2} t_1 \delta ({\bf r}_i - {\bf r}_j)  +  \frac{1}{\gamma+1}t_2 \delta ({\bf r}_i - {\bf r}_j)  \,     
    \rho_{int}^{\gamma-1}({\bf r}_{i0},{\bf r}_{j0},t) \nonumber \\  
&& + V({\bf r}_{i},{\bf r}_{j},{\bf p}_{i0},{\bf p}_{j0})+ \frac{1}{2}  \frac{Z_i Z_j e^2}{|{\bf r}_i-{\bf r}_j|} . \nonumber
\label{eq:ep} 
\end{eqnarray}
In QMD we use a single-particle Wigner density of the nucleon wave function $\psi_i$, which is given by  
\begin{equation} 
 f ({\bf r_i, p_i,r_{i0},p_{i0}},t) = \frac{1}{\pi^3 \hbar^3 }
 {\rm e}^{-\frac{2}{L} ({\bf r_i} - {\bf r_{i0}} (t) )^2   }
 {\rm e}^{-\frac{L}{2\hbar^2} ({\bf p_i - p_{i0}} (t) )^2},
 \label{Wignerdens}
\end{equation}
where the Gaussian width $L$ is taken as  $L=2.16$ fm$^2$. The total one-body Wigner density is the sum over the densities of all nucleons, whereas the $\psi(t)$ in eq. \ref{eq:energy} is the n-body wave function, which is chosen in PHQMD as the direct product of the single particle wave functions.  The corresponding single particle density at ${\bf r}$ is obtained by integrating the single-particle Wigner density over 
momentum and summing up the contribution of all nucleons:
\begin{eqnarray}
\rho_{sp}({\bf r},t)= \sum_i
\int d{\bf p_i}  f ({\bf r, p_i,r_{i0},p_{i0}},t) \nonumber\\
= \sum_i\Big(\frac{2}{\pi L}  \Big)^{3/2}{\rm e}^{-\frac{2}{L} ({\bf r} - {\bf r_{i0}} (t) )^2}.
\label{rhosp}
\end{eqnarray}

The expectation value of the potential energy $V_{ij}$,
between the nucleons i and j is given by
\begin{eqnarray}
&&\langle V_{ij}({\bf r_{i0},p_{i0},r_{j0},p_{j0}},t)\rangle  = \nonumber \\ 
&& =\int d^3r_id^3r_j
 V_{ij}({\bf r_i, r_j,p_{i0},p_{j0}}) \nonumber \\ 
&&\times f ({\bf r_i,r_{i0}},\bf{p}_{i0},t) f ({\bf r_j, r_{j0},p_{j0}},t).
\label{Vpot}
\end{eqnarray}
and the interaction density is given by
\begin{eqnarray}
&& \rho_{\rm int}({\bf r}_{i0},t)=\sum_{j \neq i}\int d^3r_id^3r_jd^3p_id^3p_j \delta({\bf r}_i-{\bf r}_j )
\nonumber \\ 
&&\times f ({\bf r_i, p_i,r_{i0}},\bf{p}_{i0},t) f ({\bf r_j, p_j, r_{j0},p_{j0}},t).
\label{rhoint}
\end{eqnarray}

In order to extend  PHQMD to relativistic energies we take into account the Lorentz contraction of the initial nuclei. This is done in an approximate way, as explained in Ref. \cite{Aichelin:2019tnk}, by introducing a modified single-particle Wigner density for each nucleon $i$:
\bea
&& \tilde f (\mathbf{r}_i, \mathbf{p}_i,\mathbf{r}_{i0},\mathbf{p}_{i0},t) =  \label{fGam} \\
&& =\frac{1}{\pi^3} {\rm e}^{-\frac{2}{L} [ \mathbf{r}_{i}^T(t) - \mathbf{r}_{i0}^T (t) ]^2} 
  {\rm e}^{-\frac{2\gamma_{cm}^2}{L} [ \mathbf{r}_{i}^L(t) - \mathbf{r}_{i0}^L (t) ]^2}  \nonumber \\
&& \times {\rm e}^{-\frac{L}{2} [ \mathbf{p}_{i}^T(t) - \mathbf{p}_{i0}^T (t) ]^2} 
  {\rm e}^{-\frac{L}{2\gamma_{cm}^2} [ \mathbf{p}_{i}^L(t) - \mathbf{p}_{i0}^L (t) ]^2},  \nonumber 
\eea
which accounts for the Lorentz contraction of the nucleus in the beam $z$-direction
in coordinate and momentum space by  including $\gamma_{cm} =1/\sqrt{1-v_{cm}^2}$, where $v_{cm}$ is the velocity of projectile and target in the computational frame, which is the center-of-mass system of the heavy-ion collision.  
Accordingly, the interaction density modifies as 
\bea
 \tilde \rho_{int} (\mathbf{r}_{i0},t) 
 &\to & C  \sum_j \Big(\frac{1}{\pi L}\Big)^{3/2} \gamma_{cm} 
 {\rm e}^{-\frac{1}{L} [ \mathbf{r}_{i0}^T(t) - \mathbf{r}_{j0}^T (t) ]^2} \nonumber \\ 
 &&\times {\rm e}^{-\frac{\gamma_{cm}^2}{L} [ \mathbf{r}_{i0}^L(t) - \mathbf{r}_{j0}^L (t) ]^2}.  
 \label{densGam}
\eea
For the energies considered here the relativistic correction are not important as follows from our numerical calculations with and without  $\gamma_{cm}$ in Eq. (\ref{densGam}).

\subsection{Relation of the potential to the EoS of nuclear matter }
In infinite nuclear matter, momentum and position are not correlated and one can calculate from the potential the equation-of-state of cold nuclear matter.  
In infinite matter the static part of the QMD potential is given, as in \cite{Aichelin:2019tnk}, by 
\begin{equation}
    V_{Skyrme\ stat} =\alpha\frac{\rho}{\rho_0}+\beta \Big(\frac{\rho}{\rho_0}\Big)^\gamma,
\end{equation}
to this the momentum-dependent part for cold nuclear matter is added, which  can be obtained by
\begin{equation}
    V_{mom}(p_F)=\frac{\int^{p_F}\int^{p_F} dp_1^3dp_2^3 V({\bf p_2}-{\bf p}_1) }{(\frac{4}{3}\pi p_F^3)^2}\frac{\rho}{\rho_0}.
\end{equation}

\begin{figure}[t]
    \centering
    \resizebox{0.45\textwidth}{!}{
        \includegraphics{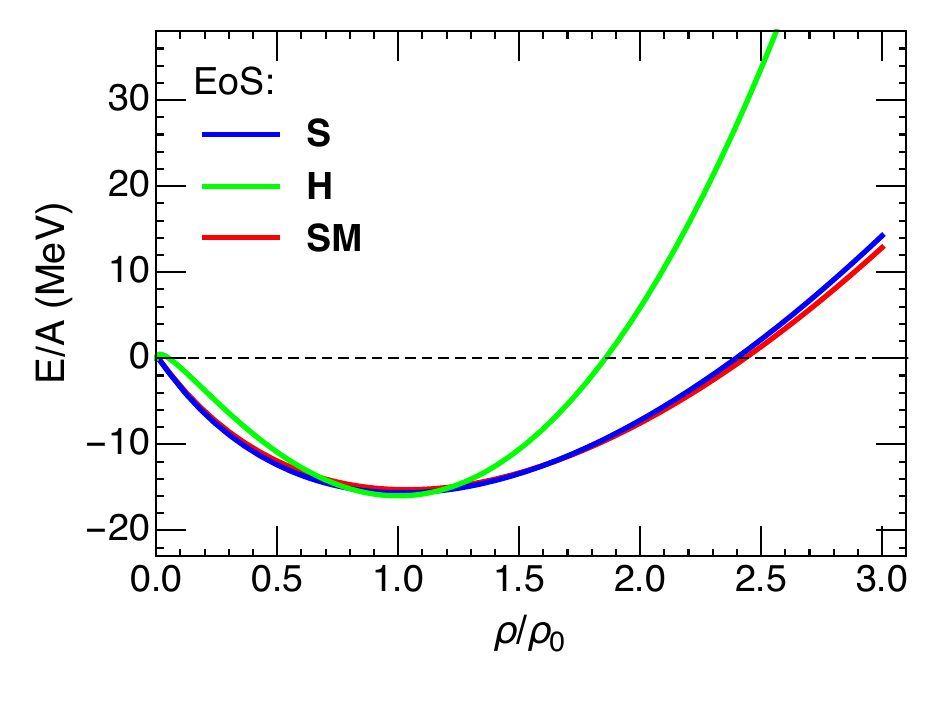}}
    \caption{Equation-of-state for $T=0$ for the hard (green line), soft (blue line) and the soft momentum-dependent potential (red line). } 
    \label{fig:eos}
\end{figure}

The Fermi momentum is a function of the density and therefore one obtains the total strong interaction potential
\begin{equation}
    V_{Skyrme}(\rho) = V_{Skyrme \ stat}(\rho)+V_{mom}(\rho).
\end{equation}
To calculate the energy per nucleon, we introduce  $U=\int V(\rho) d\rho $ . This allow to  to write
\begin{equation}
    \frac{E}{A}(\rho)=  \frac{3}{5}  E_{Fermi}(\rho) + \frac{U}{\rho}.
\end{equation}
This equation contains the 3 parameters $\alpha, \beta,\gamma$, which have to be determined.  Two of them can be obtained by the requirement that at normal nuclear density  $E/A = -16$ MeV. The third parameter is traditionally determined by fixing  the compression modulus $K$ of nuclear matter,  the inverse of the compressibility $\chi = \frac{1}{V}\frac{d V}{d P}$, which corresponds to the curvature of the energy at $\rho=\rho_0$ (for $T=0$):  
\begin{equation}
K  = 9\rho\frac{d P}{d\rho}\vert_{\rho=\rho_0}=\left. 9 \rho^2
\frac{{\rm \partial}^2(E/A(\rho))}{({\rm \partial}\rho)^2} \right\vert_{\rho=\rho_0} .\qquad
\end{equation}
Here $P$ is the pressure of the system ($P=\rho^2\frac{\partial E/A}{\partial \rho}$).  An EoS with a rather low value of the compression modulus $K$ yields a weak repulsion against the compression of nuclear matter and thus describes "soft" matter (denoted by "S"). A high value of $K$ causes a strong repulsion of nuclear matter under compression (called a hard EoS, "H").  The hard, the soft and the soft momentum-dependent equations-of-state used in this study are illustrated in Fig. \ref{fig:eos} and the parameters for the three equations-of-state are presented in Table \ref{table_eos}.
Soft and soft momentum-dependent EoS have for cold nuclear matter the same $\frac{E}{A}(\rho)$.

\begin{table}[h]
    \centering
      \begin{tabular}{c c c c c}
      \toprule
      EoS & $\alpha$ [MeV] & $\beta$ [MeV] & $\gamma$ & K [MeV]\\
      \midrule
      S  &  -383.5 &  329.5 & 1.15 & 200 \\
      H  &  -125.3 &   71.0 &  2.0 & 380 \\
      SM & -478.87 & 413.76 &  1.1 & 200 \\
      \cmidrule{2-4}
       ~ & $a$ [MeV$^{-1}$] & $b$ [MeV$^{-2}$] & $c$ [MeV$^{-1}$]  & ~\\
       ~ & 236.326 & -20.73 &  0.901 & ~ \\
      \bottomrule
      \end{tabular}
    \caption{Parameters of the potential used in PHQMD.} 
    \label{table_eos}
    \end{table}

\subsection{QMD Propagation}

For the time evolution of the wave function we use the Dirac-Frenkel-McLachlan approach \cite{raab:2000,BROECKHOVE1988547}, which is based on the variation
\begin{equation}
\delta \int_{t_1}^{t_2} dt <\psi(t)|i\frac{d}{dt}-H|\psi(t)> = 0
\end{equation}
and has been developed in chemical physics. It has also been applied in nuclear physics for QMD like models  \cite{Feldmeier:1989st,Aichelin:1991xy,Ono:1992uy,Hartnack:1997ez}. This approach
provides the time derivatives of the centers of the wave functions ${\bf r}_{i0},{\bf p}_{i0}$. Being a n-body approach it conserves the correlations in the system and does not suppress fluctuations as mean-field calculations do.  Since clusters are n-body correlations it is well suited to address their creation and time evolution.
With our assumption that the wave functions have a Gaussian form and that the width of the wave function is time independent, one obtains for the time evolution of the centroids of the Gaussian single particle Wigner density two equations, which resemble the equation-of-motion of a classical particle with the phase space coordinates ${\bf r_{i0},p_{i0}}$ \cite{Aichelin:1991xy}.
The difference is that here the expectation value of the quantal Hamiltonian is used and not a classical Hamiltonian: 
\begin{equation}
\dot{r_{i0}}=\frac{\partial\langle H \rangle}{\partial p_{i0}} \qquad
\dot{p_{i0}}=-\frac{\partial \langle H \rangle}{\partial r_{i0}} \quad .
\label{prop}
\end{equation}
The Hamiltonian of the nucleus is the sum of the Hamiltonians of the nucleons, composed of kinetic and two-body potential energy, which has a strong interaction and a Coulomb part
\begin{equation}
H = \sum_i H_i  = \sum_i  (T_i + \sum_{j\neq i}  V_{ij}).
\end{equation}

The expectation value  of the Coulomb interaction can also be calculated analytically.
The expectation value of the  Hamiltonian, which enters Eq. \ref{prop} is finally given by
\begin{eqnarray}
\langle H \rangle &=& \langle T \rangle + \langle V \rangle
\label{ham} \\
&=& \sum_i \big(\sqrt{p_{i0}^2+m^2}-m\big) \\ \nonumber &+&\sum_{i}  \langle V_{Skyrme}({\bf r}_{i0},t)+V_{mom}({\bf r}_{i0},{\bf p}_{i0}t)+V_{coul}({\bf r}_{i0},t)\rangle .
\nonumber 
\end{eqnarray}

\subsection{Cluster production in PHQMD}

In the PHQMD 3 mechanisms for the   clusters production are available:

1) {\bf MST clusters:} The attractive  potential between baryons with a small relative momentum keeps them close together and can lead to a group of bound nucleons. Such groups of co-moving nucleons can be identified as clusters during the dynamical evolution, using  the advanced Minimum Spanning Tree (aMST) method, as detailed in Ref. \cite{Coci:2023daq}. 

MST \cite{Aichelin:1991xy} collects nucleons, which are close in coordinate space. At a given time $t$ a snapshot of the positions and momenta of all nucleons is recorded and the MST clusterization algorithm is applied: two nucleons $i$ and $j$ are considered as ``bound" to a deuteron or to a larger cluster $A>2$ if they fulfill the condition
\begin{equation}\label{eq:MSTcond}
| \mathbf{r}_i^* - \mathbf{r}_j^* | < r_{clus} \, ,
\end{equation}
where on the left hand side the positions are boosted in the center-of-mass of the $ij$ pair. The maximal distance between cluster nucleons, $r_{clus}=4$ fm, corresponds roughly to the range of the attractive $NN$ potential. Additionally, in aMST the clusters have to be bound ($E_B>0)$. 

For the beam energies considered here the difference between aMST and MST is small, less the 10\%.  aMST corrects here only the fact that in semiclassical systems a nucleon can acquire more kinetic energy than in the corresponding
quantum system because all other nucleons can have negligible momenta what is not the case in the quantum system. This leads to an enhanced emission of nucleons from semiclassical clusters, which has to be corrected.
It is important to highlight that MST serves as a tool for cluster recognition  at each time step by determining the nucleons, which form a cluster under the above conditions. It is not a mechanism for ‘building’ clusters, which are then propagated,   since the QMD transport model propagates baryons and not  clusters. At the end of the heavy ion reaction the clusters as well as the nucleons, which are part of a cluster, do not change anymore. This allows to determine the asymptotic cluster observables.

2)  {\bf kinetic mechanism:}  Deuterons can be created in catalytic hadronic reactions as $\pi NN \leftrightarrow \pi d$ and $NNN \leftrightarrow N d$  in different isospin channels. The quantum nature of the deuteron is considered through an excluded volume, which forbids its production if another hadron is localized in this volume, i.e. a deuteron with a rms radius of about $\sqrt{<r^2_d>} \simeq 2.1$ fm cannot be formed if between the $p$ and the $n$ other hadrons are located.
We project furthermore the relative momentum of the incoming nucleons onto the deuteron wave function in momentum space.  These quantum corrections  lead to a significant reduction of deuteron production by kinetic mechanism, particularly at target/projectile  rapidities.
We note that the kinetic deuterons are propagated explicitly in the PHQMD as a degree-of-freedom (contrary to the nucleons in MST clusters, which are propagated as nucleons. We refer the reader to Ref. \cite{Coci:2023daq} for the details.

3) {\bf coalescence mechanism:} Additionally  to the  cluster production by potential interaction and kinetic reactions occurring during the whole time evolution of the system,  we have an option in PHQMD to apply a coalescence procedure at the freeze-out time of the nucleons.  We recall that the coalescence framework in PHQMD is adopted from  UrQMD  \cite{Sombun:2018yqh} and described in  Ref. \cite{Kireyeu:2022qmv}.  A proton and a neutron can form a deuteron if their distance at the time, when the last one of the two freezes out,  in phase space is less than (see Ref. \cite{Kireyeu:2022qmv})  $|r_1-r_2| \le 3.575$ fm and $|p_1-p_2|\le 285$ MeV$/c$. These radii have been fitted to data in order to reproduce the deuteron multiplicity, if a spin degeneracy factor of 3/8 \cite{Kireyeu:2022qmv} is applied as in UrQMD  \cite{Sombun:2018yqh}.

We stress that the aMST is a {\em cluster recognition algorithm}, which is applied at different times during the time evolution of the system. It has, as said, no influence on the propagation of baryons. The PHQMD propagates baryons and kinetic deuterons as degrees-of-freedom, i.e. the identification of individual nucleons as a cluster member, recognized by aMST, doesn't influence its dynamics. Technically the history of clusters, identified  by aMST, is written on a separate file, which can be analyzed later on. We can also analyze whether the clusters in subsequent time steps contain the same nucleons.

The PHQMD results for the light clusters in this study are based on different algorithms  of  cluster formation, including both  MST and kinetic mechanisms for deuterons. While kinetic deuterons are propagated as a degree-of-freedom in PHQMD, the MST clusters are considered in a perturbative way, i.e. PHQMD propagates nucleons which can be identified as a cluster members via MST algorithm and stored in perturbative vectors.
If not otherwise stated the calculations are performed in the standard setup of PHQMD, a combination of MST and kinetic clusters \cite{Coci:2023daq}.
In Section \ref{mst_coal}, we also present deuterons formed through the coalescence mechanism applied to the events generated without kinetic deuterons for a better comparison with other microscopic models.

\section{The PHQMD results in comparison to experimental data}

\subsection{Directed and elliptic flow}
In this section we proceed to the comparison of our PHQMD calculation with experimental observables characterizing the anisotropy of the system in momentum space. The later is related to the  the azimuthal distribution of particles in momentum space at the end of a heavy-ion collisions, which can be  be analyzed by a Fourier series with the flow coefficient  $v_n$, defined as
\begin{equation} 
\label{eq:azim}
\frac{dN}{d\phi}\propto 1 + 2 v_1 cos(\phi-\Psi_R) +2v_2 cos(2(\phi-\Psi_R)) + ...
\end{equation}
$\phi$ is the azimuthal angle of the particle measured with respect to the event plane (or a "reaction plane") $\Psi_R$. The flow coefficients $v_n, \ n=1,2,...$  are defined with respect to $\Psi_R$ as average over all particles in all events for a given centrality range \cite{Ollitrault:1997di,Poskanzer:1998yz}:
\begin{equation} 
\label{eq:vn}
 v_n = <\cos(n(\phi-\Psi_R))>.
\end{equation}

In this study we concentrate on the comparison of the PHQMD results with the fixed target experiments at SIS energies, HADES and FOPI, which have the advantage that they can define the reaction plane with high accuracy due to the HADES forward wall detector \cite{HADES:2022osk}, and  the FOPI $4\pi$ geometry \cite{Reisdorf:2010aa}.
We study the directed flow, $v_1$, and the elliptic flow, $v_2$ for Au+Au collisions, at three different but neighboring energies around $E_{kin} \simeq 1.2$ GeV. For the comparison of the PHQMD results with the HADES experimental data the HADES  pseudo-rapidity acceptance cuts, $-0.79 < \eta_{CM} < 0.96$, are applied \cite{HADES:2020lob}.

While in the theoretical model calculations the geometry of each collision is well defined because the reaction plane is linked to the coordinate system (i.e. $\Psi_{R}=0$), this is not the case for experiments, which measure only a subset of the final hadrons in a selected part of the phase-space. 
They are used to experimentally define the so-called event plane angle $\Psi_{EP}$, which has a dispersion around the ideal reaction plane angle $\Psi_{R}$. However, this dispersion can be estimated from data and is used to correct the measured raw flow coefficients. There are different methods employed to extract these corrections
\cite{Ollitrault:1997di,Poskanzer:1998yz}; they might affect 
the flow $v_n$ at large $p_T$ in different ways (cf. the study within the PHSD model in Ref. \cite{Towseef:2023ytc}). 

In this study we calculate the flow coefficients using the theoretically defined reaction plane ($\Psi_{R}=0$) and compute the directed and elliptic flow as  
\begin{equation}
 \begin{aligned}
  v_1 = <\frac{p_x}{p_T}> ,  \   v_2 = <\frac{p_x^2 - p_y^2}{p_T^2}>   
 \end{aligned}
\end{equation}
where $p_T$ is the transverse momentum  $p_T=(p_x^2 + p_y^2)^{1/2}$ of the hadron with 4-momentum $p=(E,p_x,p_y,p_z)$.


\subsection{Consequences of the EoS for heavy-ion dynamics at SIS energies}
\label{conseq}

Before we compare the PHQMD results to the experimental data we discuss how the bulk properties in heavy-ion collisions depend on the different EoS. 
To demonstrate this we use Au+Au collisions at  $E_{kin}=1.5$ A GeV. 
We start with the maximum density attained during the collision.  
\begin{figure}[h!]
\includegraphics[scale=0.45]{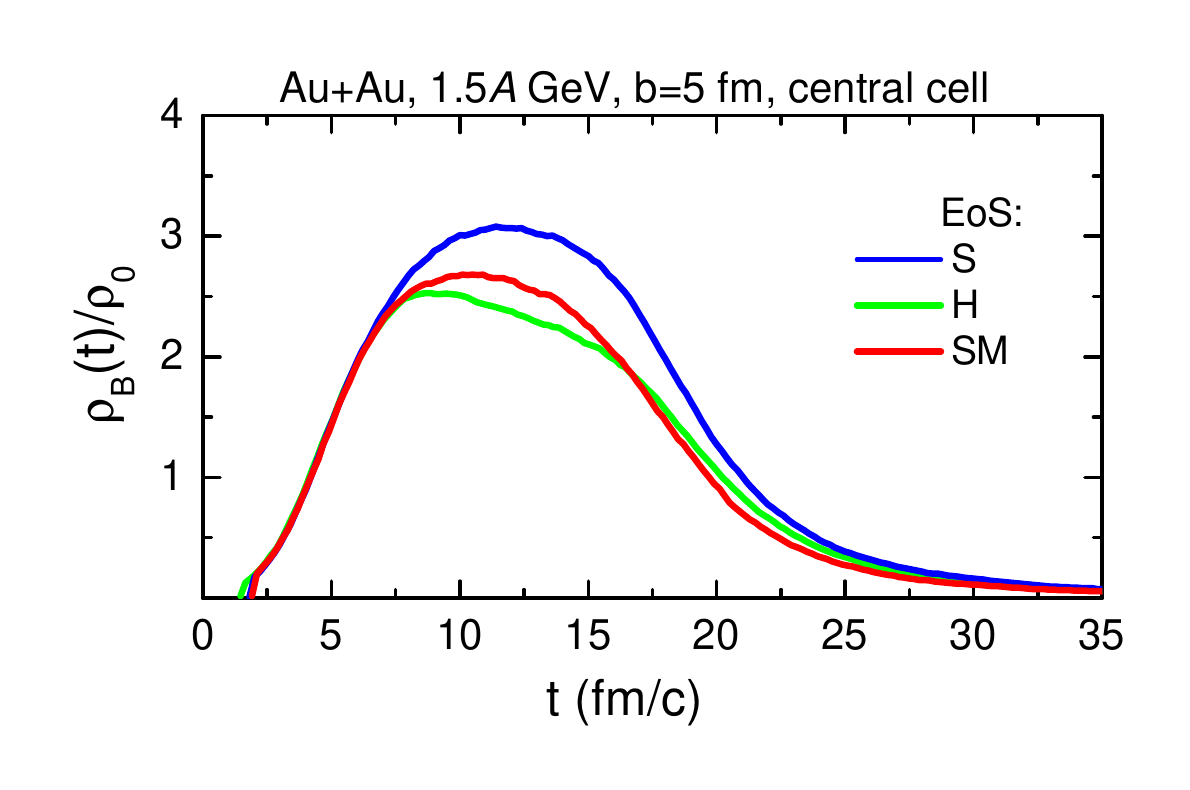}
   \caption{\label{fig:dens} Time evolution of the normalized baryon density distributions in the central cell for Au+Au collisions at $b=5$ fm  and  $E_{kin}=1.5$ A GeV.
 The blue lines "S" correspond to the PHQMD calculations with the "soft" EoS, the green lines "H" show the "hard" EoS, the red lines the "SM" represent the momentum-dependent "soft" EoS. }
\end{figure}

 In Fig. \ref{fig:dens} we show the time evolution of the normalized baryon density (to $\rho_0=0.168$ nucleons/fm$^3$) in the central cell in Au+Au collisions for $b=5$ fm at $E_{kin}=1.5$ A GeV.
The calculations are done for three different EoS: the blue lines "S" correspond to PHQMD calculations with the "soft" EoS, the green lines "H" show the "hard" EoS, the red lines "SM" represent the " momentum-dependent soft" EoS.
 One can see that the baryon density in the central cell does not only depend on the compression modulus - which determines whether an equation-of-state is soft or hard - but also on the momentum-dependence of the potential. Despite of being soft, the maximal density for the soft momentum-dependent EoS is very close to that obtained for a hard EoS and much lower than that for a soft EoS without momentum-dependence. This has of course consequences for the mean free path and quantities, which characterize the collisions as directed flow  $v_1$. 
 
\begin{figure}[h!]
\includegraphics[scale=0.4]{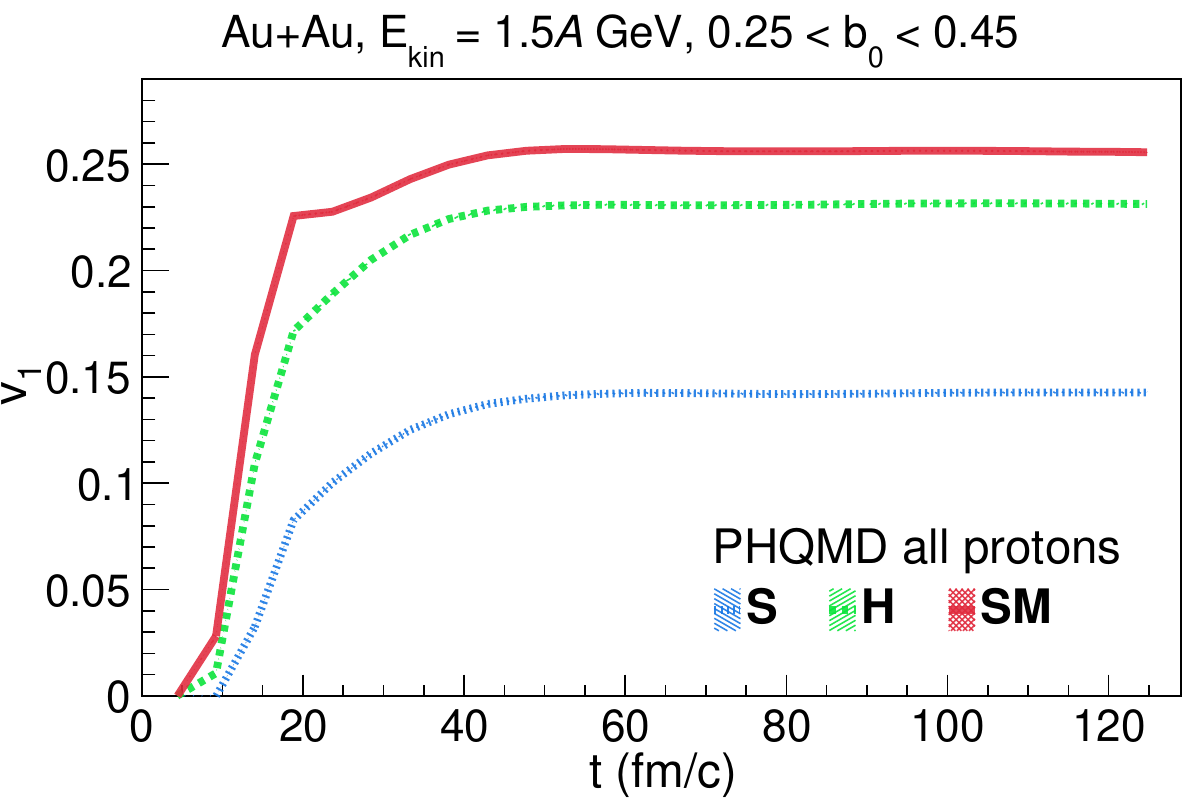}
    \caption{$v_1$ of all protons (free and bound in MST clusters) for Au+Au collisions at $E_{kin}=1.5$ A GeV  for centrality $0.25<b_0<0.45$  as a function of time. The colour code is the same as in Fig. \ref{fig:dens}. }
    \label{Fig:v1t}     
\end{figure}

\begin{figure}[h!]
\includegraphics[scale=0.4]{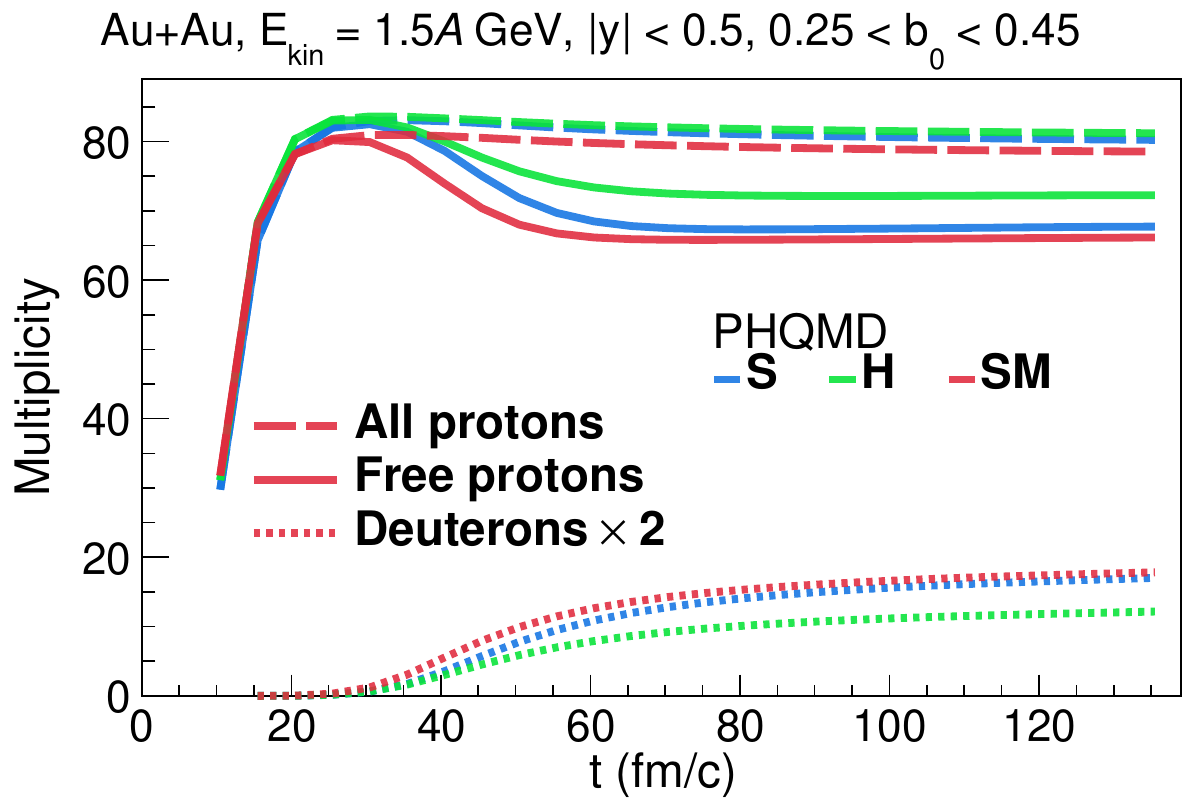}
    \caption{Time evolution of the midrapidity multiplicities of the protons and deuterons for Au+Au collisions  at $E_{kin}=1.5$ A GeV  for centrality $0.25<b_0<0.45$.
    The solid lines are for the free unbound protons, the long dashed lines show all (free + bound) protons, the short dashed lines present deuterons. The deuteron lines are multiplied by a factor of 2 for  better visibility. The colour code is the same as in Fig. \ref{fig:dens}. }
    \label{Fig:ntime}     
\end{figure}

In Fig. \ref{Fig:v1t} we present the time evolution of $v_1$ (for $\Psi_R = 0$) for all protons (free and bound in clusters) for the three EoS in the FOPI mid-centrality interval ($0.25<b_0=b/b_{max}<0.45$) integrated over $y>0$ and $p_T$. 
One can see that $v_1$ for the soft EoS is considerably smaller than that for the SM EoS, which is close to that for a hard EoS. For SM $v_1$ develops also earlier when the two nuclei touch each other and before the system has reached its maximum density.  $v_1$ for the hard EoS  starts slightly later when the system gets compressed. H and SM have almost the same time profile although due to very different origins: 
Whereas for the hard EoS it is the density gradient, which creates the directed flow, for the SM EoS  it is the momentum-dependence of the potential, which is at the origin for the increase of $v_1$ \cite{Aichelin:1991xy}. $v_1$ of the S EoS develops later and does not reach the same maximal value due to smoother density gradients. 

This can be further illustrated by showing  in Fig. \ref{Fig:ntime}  the time evolution of the midrapidity multiplicities of the protons and deuterons for Au+Au collisions  at $E_{kin}=1.5$ A GeV for the centrality $0.25<b_0<0.45$.  Here the solid lines are for the unbound protons, the long dashed lines show all (unbound + bound) protons, the short dashed lines present deuterons. 
We observe that the number of deuterons stabilizes after freeze-out, with a multiplicity being very similar for the S and SM EoS and noticeably different for the H EoS.

\begin{figure}[h!]
\includegraphics[scale=0.4]{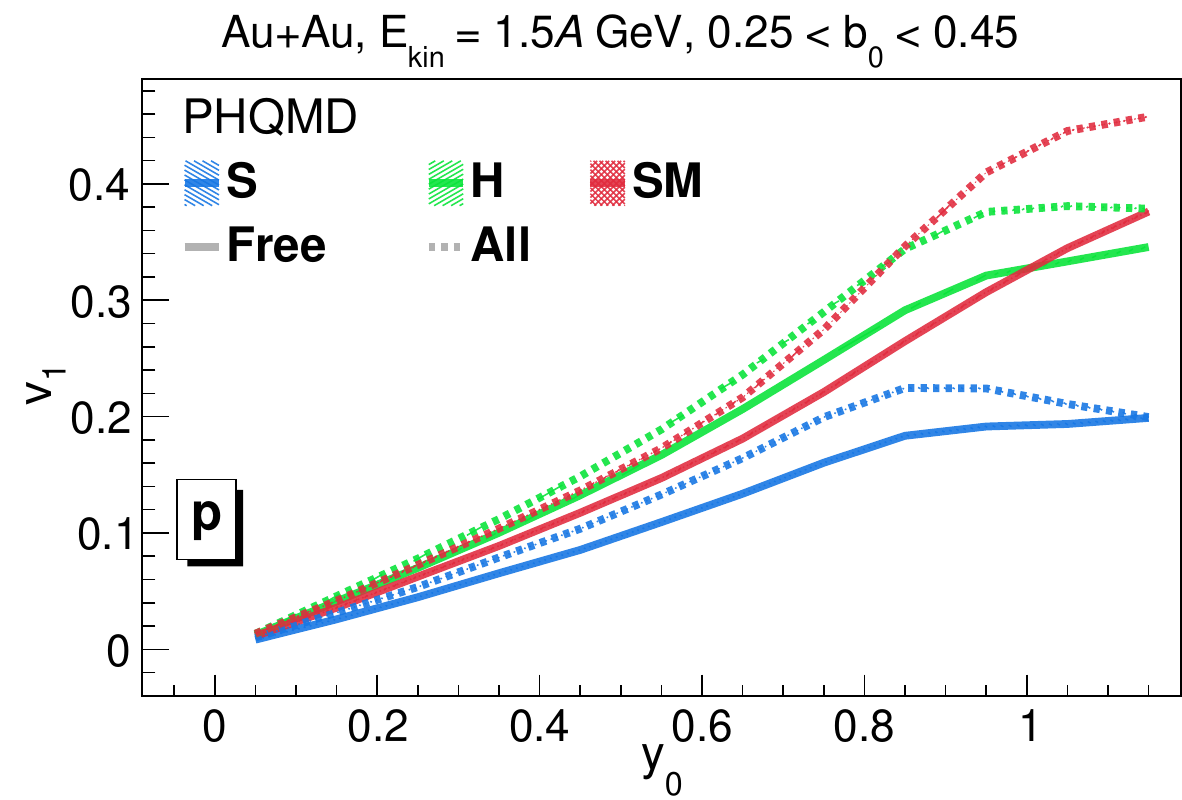}
    \caption{$v_1$ of protons  versus $y_0$ for Au+Au collisions at $E_{kin}=1.5$ A GeV for the impact parameter range $0.25<b_0<0.45$. The colour code is the same as in Fig. \ref{fig:dens}. The solid lines are for the free unbound protons, while the dashed lines show all (free + bound) protons.}
    \label{Fig:v1pV}     
\end{figure}

A second general observation concerns the directed flow of clusters:  
we observe that  the  $p_T$ integrated directed flow, $v_1$, of finally unbound protons is smaller than $v_1$ of all (free + bound in clusters) protons for all 3 EoS. 
This is shown, as a function of $y_0=y/y_{beam}$, in Fig. \ref{Fig:v1pV}, which displays $v_1$ for Au+Au collisions at $E_{kin}=1.5$ A GeV  in the impact parameter range $0.25<b_0<0.45$ .
This observation contradicts the assumption that the phase space distribution of clusters, which contain $N$ nucleons, is just a single proton distribution to the $N$'th power and taken at a momentum $p/N$:
$f^N(p) = [f(p/N)]^N.$ 

Already in Ref. \cite{Hartnack:1994ce} it has been shown that in QMD calculations clusters have a larger value of  $v_1$ than protons. $v_1$ of protons depends on their trajectory during the heavy-ion collision. Those protons, which pass the transition region between participant and spectator, where the density gradient in transverse direction is large,  have the highest value of $v_1$ whereas for those, which pass  the center of the participant region, $v_1$ is close to zero due to symmetry reasons \cite{David:1998qu}. In QMD calculations light clusters are preferably formed by nucleons whose trajectories pass close to the intersection between the participant and spectator region \cite{Aichelin:1988me} and have therefore on the average a larger $v_1$ than protons. A detailed comparison of these calculations with data by the FOPI   Collaboration \cite{FOPI:2011aa} found a very good agreement with data.  A  larger $v_1$ of clusters compared to protons is also expected in hydrodynamical calculations - cf. \cite{FOPI:2011aa}.

\begin{figure}[h!]
\includegraphics[scale=0.4]{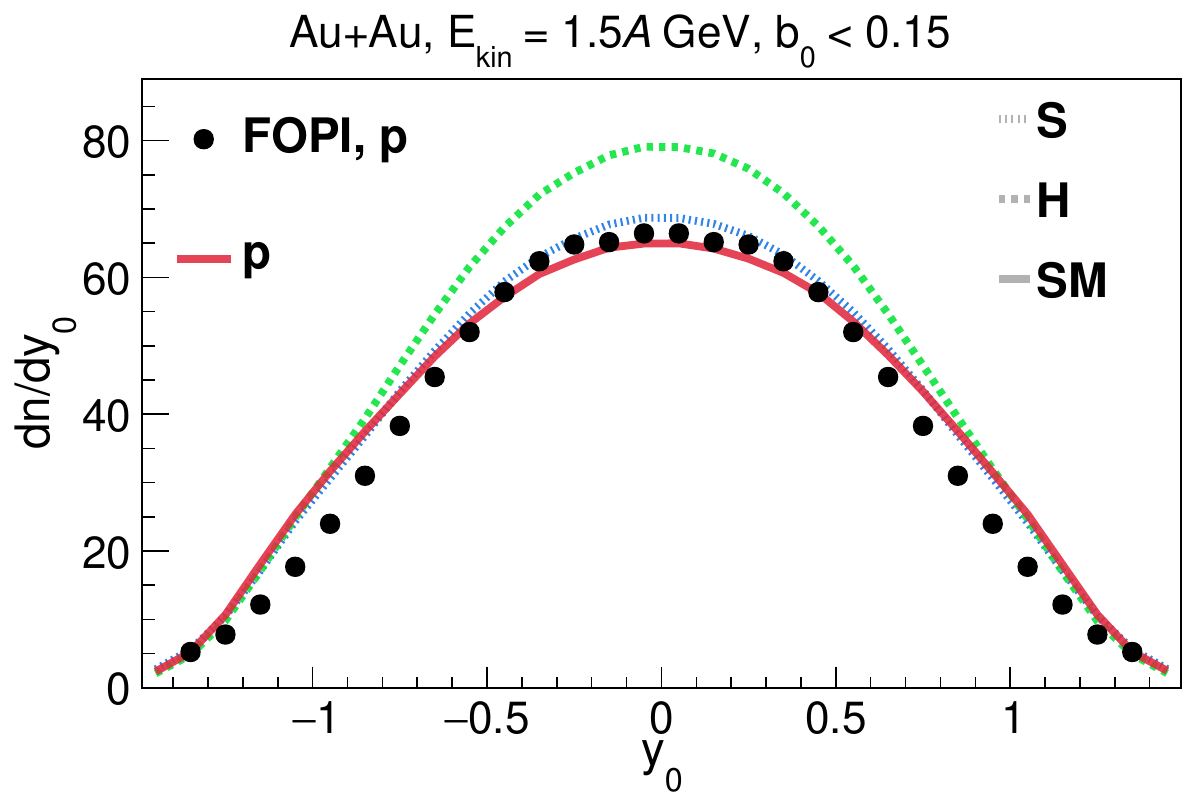}
\includegraphics[scale=0.4]{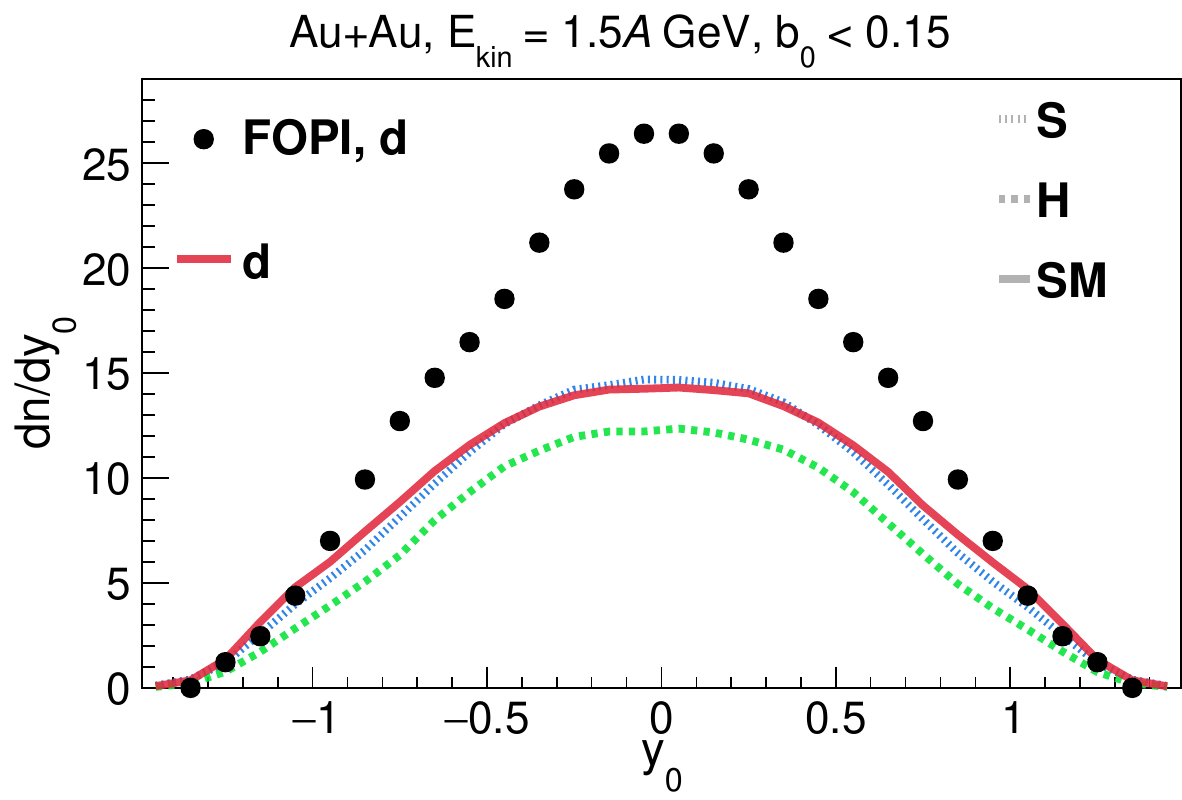}
\includegraphics[scale=0.4]{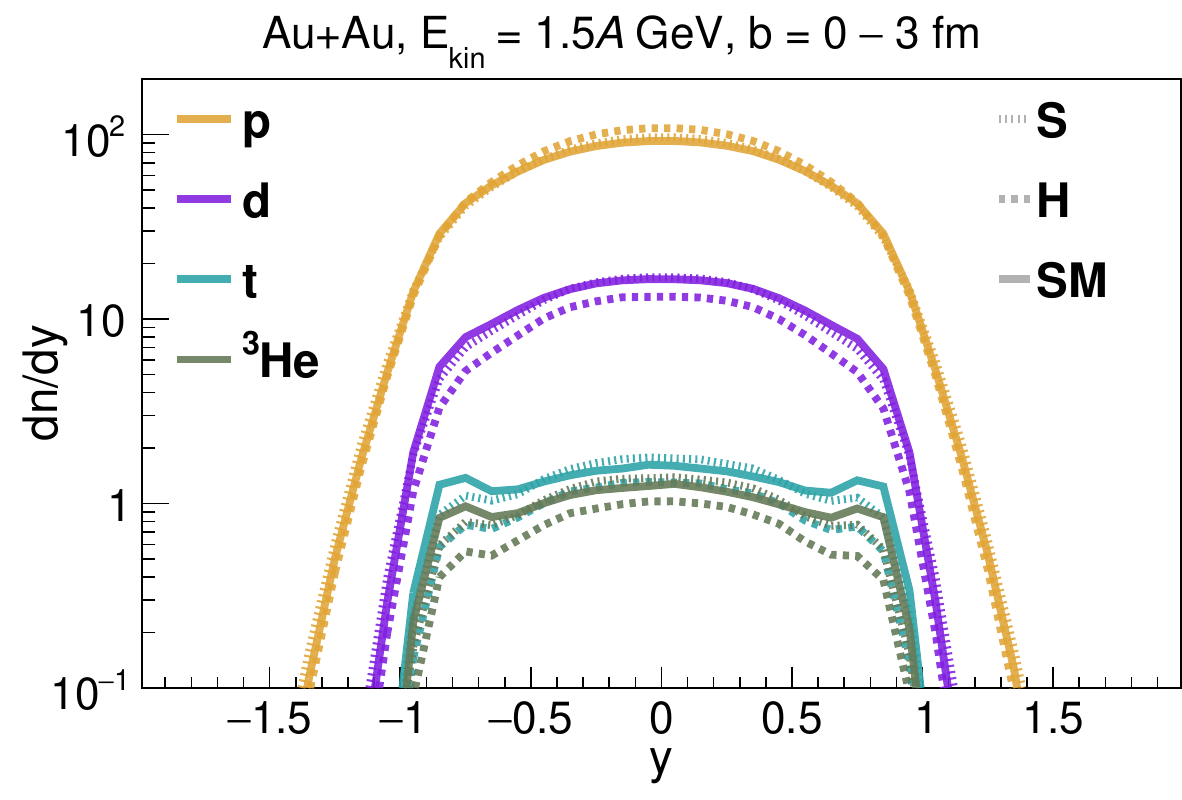}
    \caption{The PHQMD results for the scaled rapidity distribution $dN/dy_0$ of protons (upper), deuterons (middle)  for Au+Au collisions at $E_{kin}=1.5$ A GeV for the impact parameter $b_0 <0.15$ fm  in comparison to the  experimental data from the FOPI Collaboration \cite{Reisdorf:2010aa}.
    The lower plot shows the $dN/dy$ distributions of protons (yellow), deuterons (magenta), tritons (blue) and $^3He$ (olive) for Au+Au collisions at $E_{kin}=1.5$ A GeV for the impact parameter range $0< b <3$ fm. The PHQMD results are presented for S (dotted lines), H (dashed lines), SM (solid lines) EoS's. 
      }
    \label{Fig:dndy_all}     
\end{figure}

\begin{figure}[h!]
\includegraphics[scale=0.4]{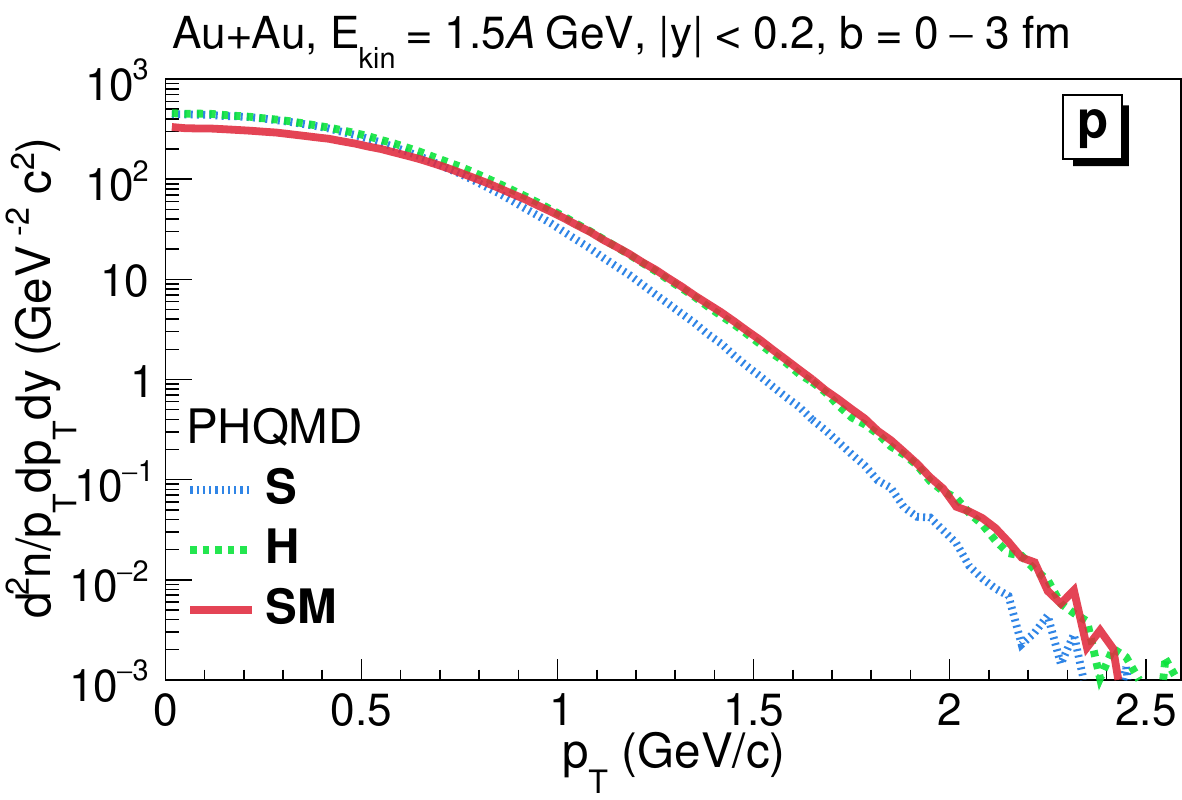}
\includegraphics[scale=0.4]{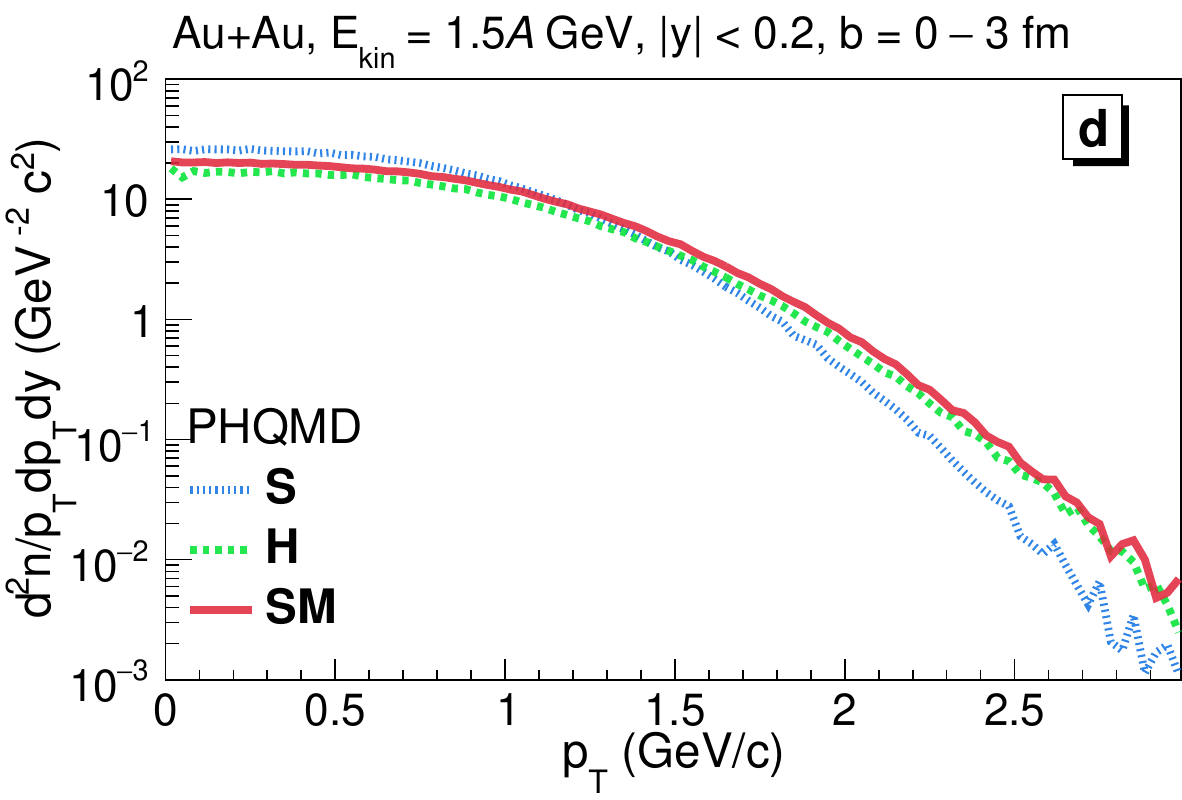}
\includegraphics[scale=0.4]{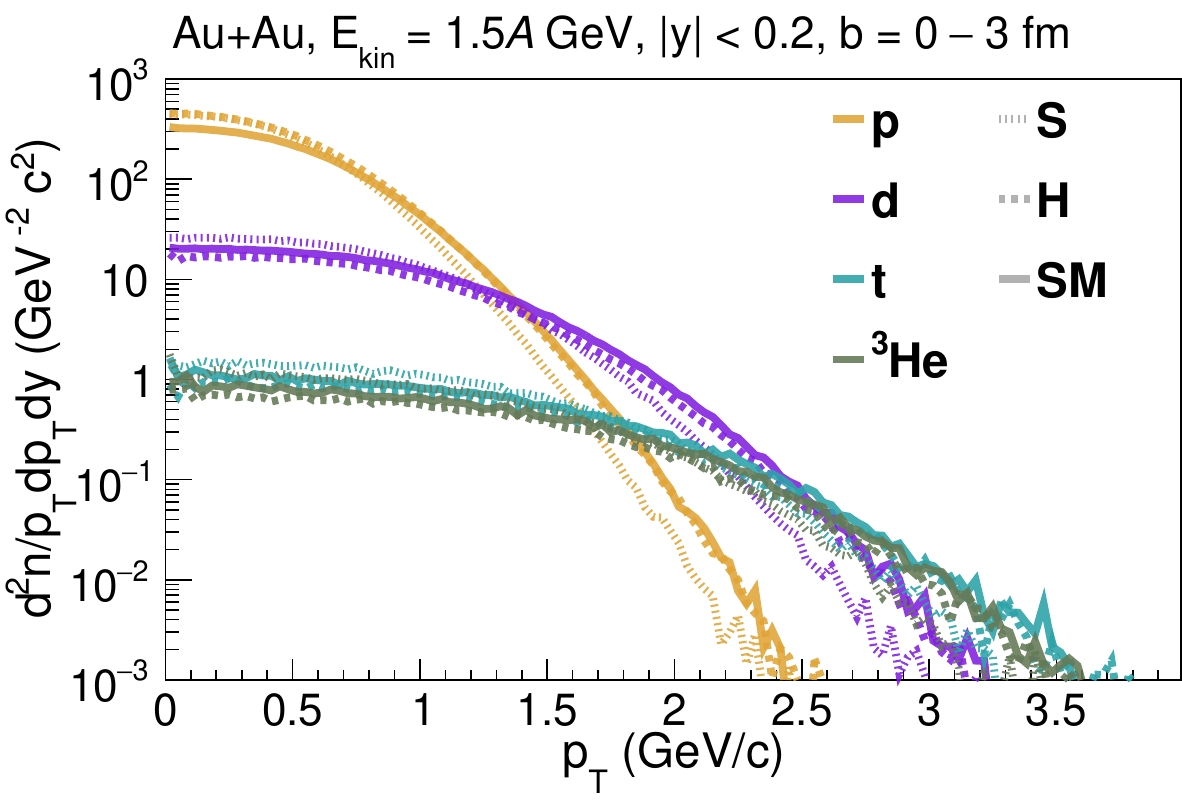}
    \caption{Invariant $p_{T}$ spectra of protons (upper), deuterons (middle) and compilation of protons, deuterons, tritons and $^3He$ for mid-rapidity ($|y|<0.2$) Au+Au collisions at $E_{kin}=1.5$ A GeV for the impact parameter range $0< b <3$ fm. The colour and line types scheme is the same as on the Fig. \ref{Fig:dndy_all}.}
    \label{Fig:pt_all}     
\end{figure}

A third general observation is that the rapidity spectra and and the slope of mid-rapidity $p_T$-spectra depend on the EoS. This is shown in Figs. \ref{Fig:dndy_all} and \ref{Fig:pt_all}, which display, from top to bottom, the rapidity distributions and $p_T$-spectra of free protons, deuterons for mid-rapidity Au+Au collisions at $E_{kin}=1.5$ A GeV and a compilation of free protons (yellow), deuterons (magenta), tritons (green) and $^3 He$ (blue) in a logarithmic scale. S, H and SM are presented by dotted, dashed and full lines, respectively. 

Moreover, the upper and middle plots of Figs. \ref{Fig:dndy_all} show the PHQMD results for the scaled rapidity distribution $dN/dy_0$ of protons (upper), deuterons (middle)  for Au+Au collisions at $E_{kin}=1.5$ A GeV for the impact parameter $b_0 <0.15$ fm  in comparison to the  experimental data from the FOPI Collaboration \cite{Reisdorf:2010aa}. While the proton rapidity distribution is well described by the SM EoS, the deuteron production is strongly underestimated mainly due to the double suppression scenario for kinetic deuterons applied for this calculations - by the excluded volume and momentum projections into the deuteron wave function. This suppression has been investigated in the previous PHQMD study in Ref. \cite{Coci:2023daq} (cf. Fig. 11). 
We further note that the relative importance of the MST and kinetic deuteron production mechanisms has been investigated in Refs. \cite{Coci:2023daq,Kireyeu:2024woo} as a function of bombarding energy and rapidity. These studies demonstrate that, within the PHQMD framework, the most of deuterons produced at midrapidity — from low BES up to top RHIC energies — originate from the MST mechanism, whereas the kinetic production channel remains subdominant - cf. Fig. 17 in Ref. \cite{Coci:2023daq}, while at SIS energies as here the contributions are more equal, cf. Fig. 7 in the same reference. A dedicated investigation of these effects would therefore be required but is beyond the scope of the present work, which focuses on the influence of the EoS on flow harmonics.

As follows from Figs. \ref{Fig:dndy_all} and \ref{Fig:pt_all} the larger the mass of the cluster the broader becomes the $p_T$ distribution whereas the width of the rapidity distribution decreases. This indicates that participant matter produces light clusters more frequently than spectator matter.
We see here again that the S and SM EoS produce about the same number of deuterons and about 30\% more than calculations with a hard EoS.  
As follows from  Fig. \ref{Fig:dndy_all}, the momentum-dependence of the EoS has small influence on particles emitted close to midrapidity, which originate from a fireball with a temperature of about $T \sim 100$ MeV and have therefore a small relative momentum as compared to the beam momentum. 
There, due to the same compressibility, S and SM behave similarly.
As shown in the bottom plot of Fig. \ref{Fig:dndy_all}, this lower multiplicity for a hard EoS is also visible for larger clusters. S and SM produce around 30\% more $A = 3$ clusters than H. 
The sensitivity to EoS is also visible in the slope of $p_T$-spectra as shown in Fig. \ref{Fig:pt_all}. The S EoS leads to much softer transverse spectra of protons and light clusters compared with the SM EoS and H EoS. We can conclude these general observations with the remark that, depending on the observables, the results of calculations with SM can be close to that of H or close to that of S. This offers the possibility, by comparing a multitude of observables, to determine experimentally the momentum and density dependence of the EoS.

\subsection{Directed flow \texorpdfstring{$v_1$}{v1}}

The first coefficient $v_1$, the directed flow, measures the deflection of the projectile and target nucleons in opposite direction towards a finite average value of $p_x$, where $x$ is the direction of the impact parameter. The value of $p_x$ is (per definition) positive for particles with a positive rapidity and negative for those with a negative rapidity.  As discussed, the finite $v_1$ can have two origins: either the high density overlap zone of projectile and target creates forces in transverse direction or the momentum-dependent interaction deviates projectile and target in transverse direction. The former sets in when a high density is reached, the second when projectile and target start to overlap and therefore the momentum-dependent potential is strongest.

\subsubsection{Comparison of the PHQMD \texorpdfstring{$v_1$}{v1} to the HADES data}

We start with the comparison of the PHQMD results for the directed flow $v_1$ of protons and light clusters with the experimental data from the HADES Collaboration \cite{HADES:2022osk} for Au+Au collisions at $E_{kin}=1.23$ A GeV ($\sqrt{s_{NN}}=2.4$~GeV).
We note that with "protons" we mean only "free" protons (if not specified explicitly), i.e. those which are not bound in clusters.

\begin{figure*}[h!]
\includegraphics[scale=0.4]{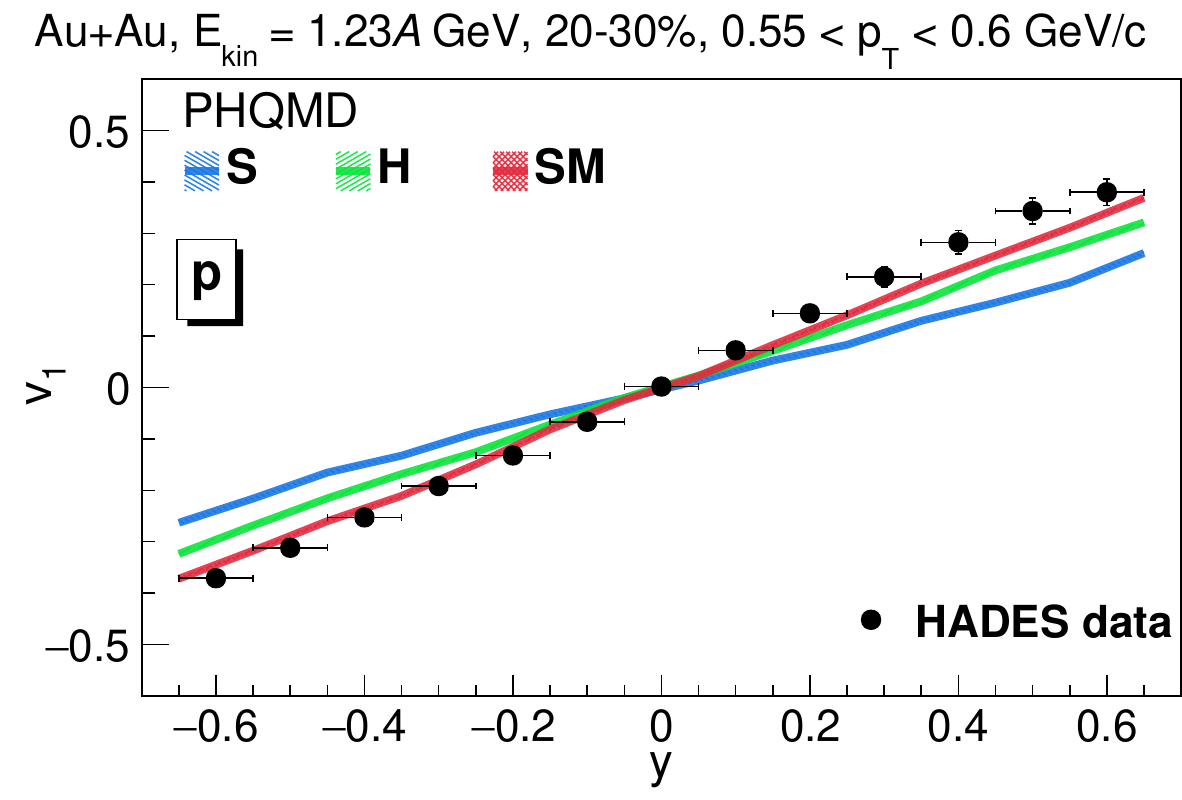}
\includegraphics[scale=0.4]{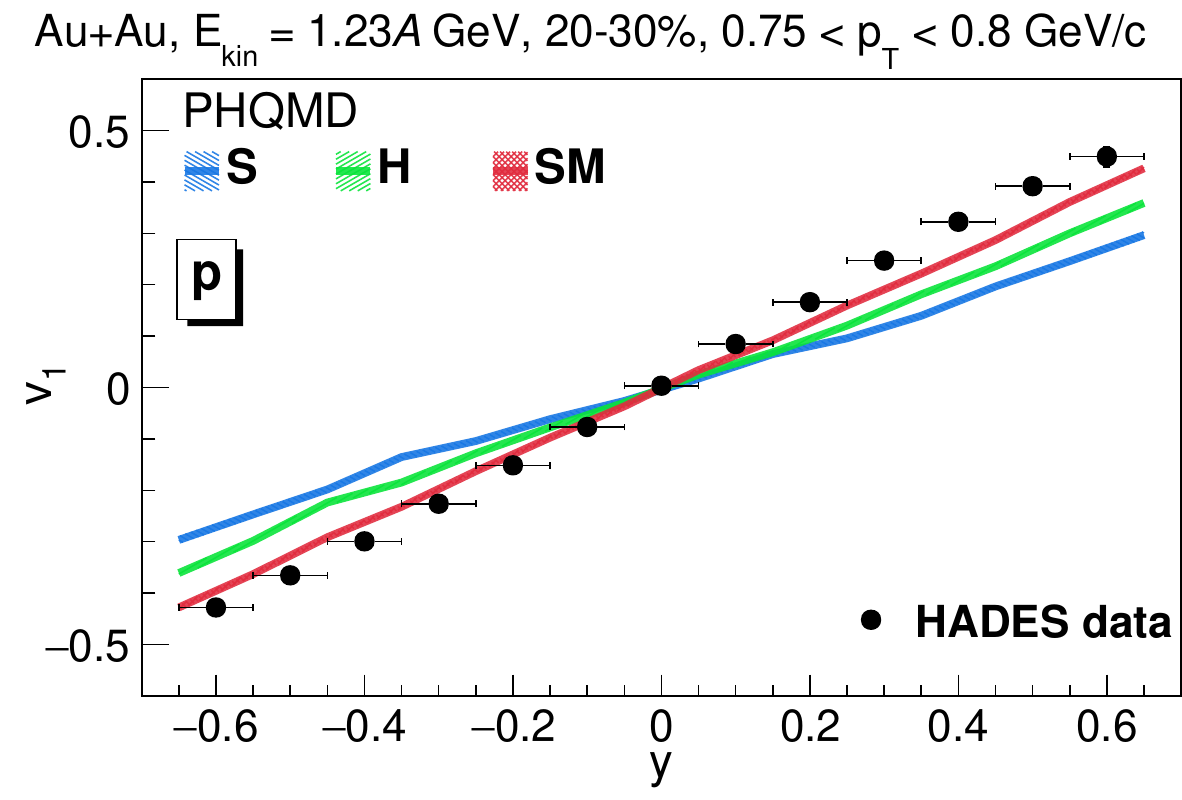}
\includegraphics[scale=0.4]{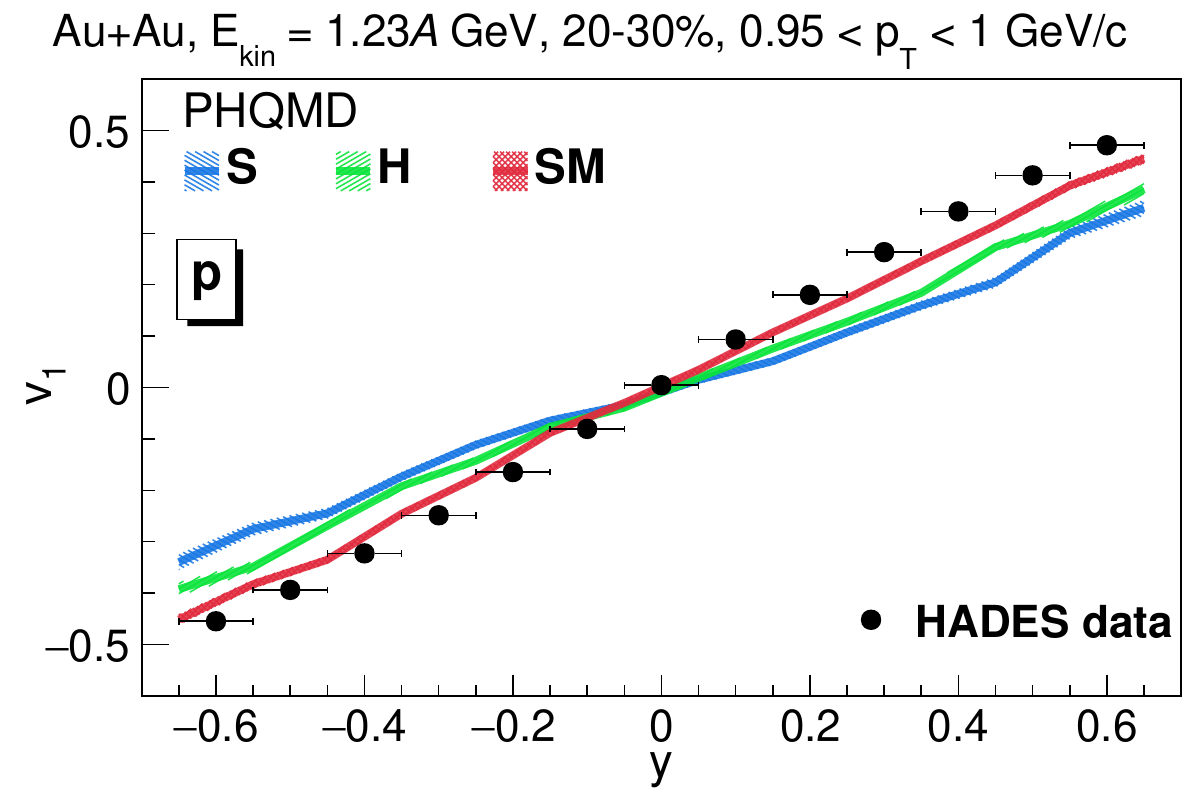}
\includegraphics[scale=0.4]{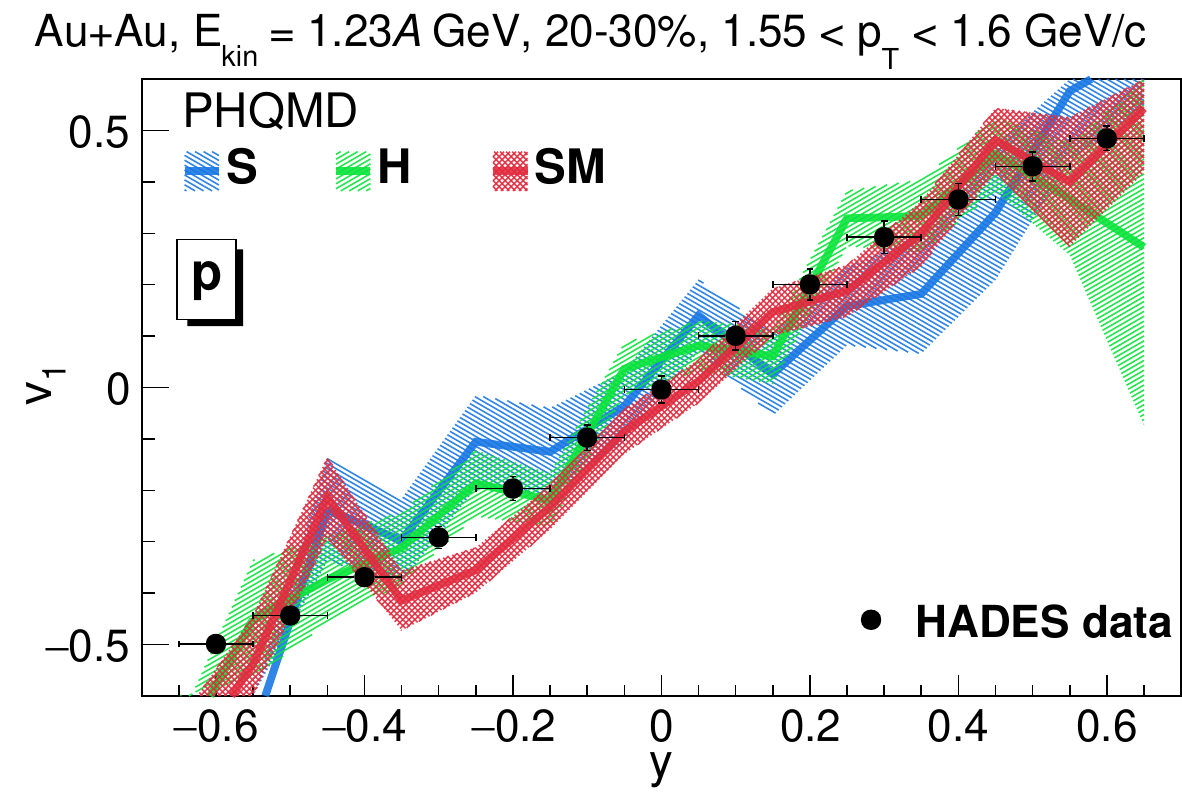} 
   \caption{The directed flow $v_1$ of protons as a function of rapidity $y$  for different $p_T$ intervals for  20-30\% central Au+Au collisions at $E_{kin}=1.23$ A GeV for different $p_T$ intervals:
   $0.55 < p_T < 0.6$ GeV/c (upper left),  $0.75 < p_T < 0.8$ GeV/c (upper right), 
    $0.95 < p_T < 1.0$ GeV/c (lower left),  $1.55 < p_T < 1.6$ GeV/c (lower right).
   The colour code is the same as in Fig. \ref{fig:dens}. The height of the filled areas represent the statistical errors.    
   The HADES experimental data are taken from Ref. \cite{HADES:2022osk}.} 
\label{fig:v1y_winn52} 
\end{figure*}

\begin{figure*}[h!]
\includegraphics[scale=0.4]{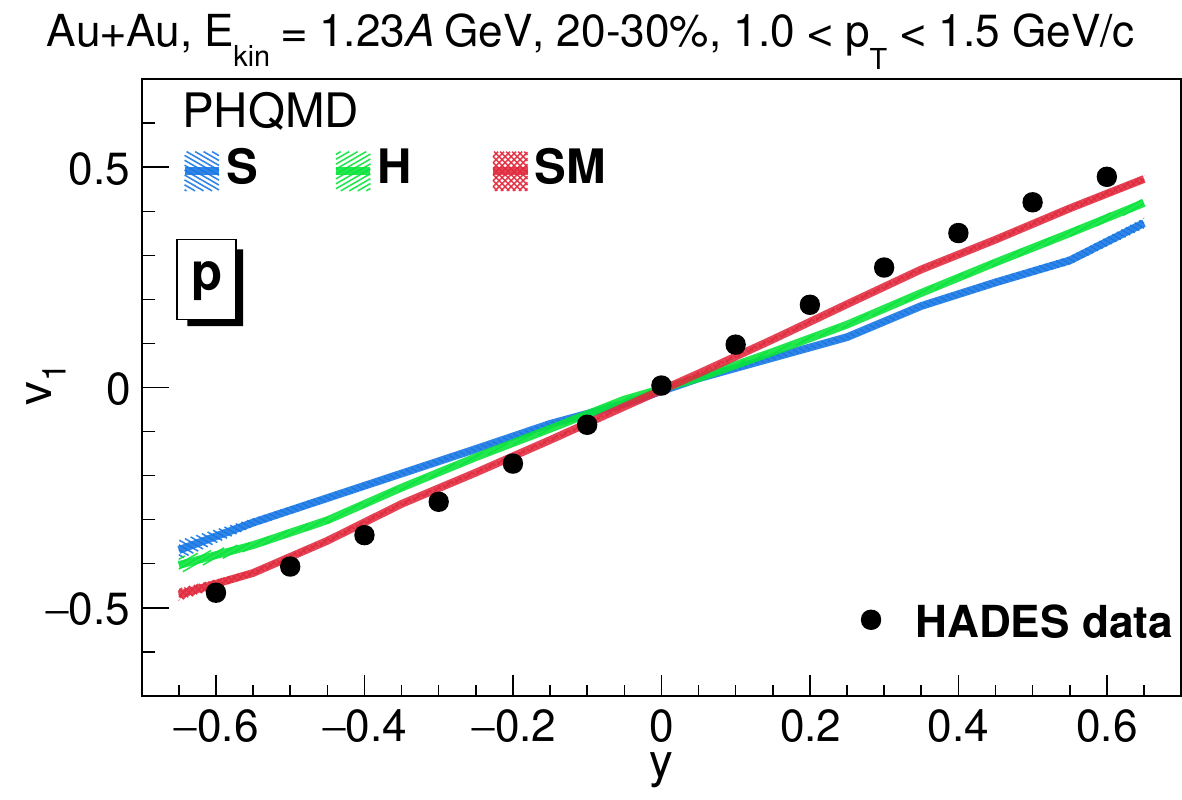}
\includegraphics[scale=0.4]{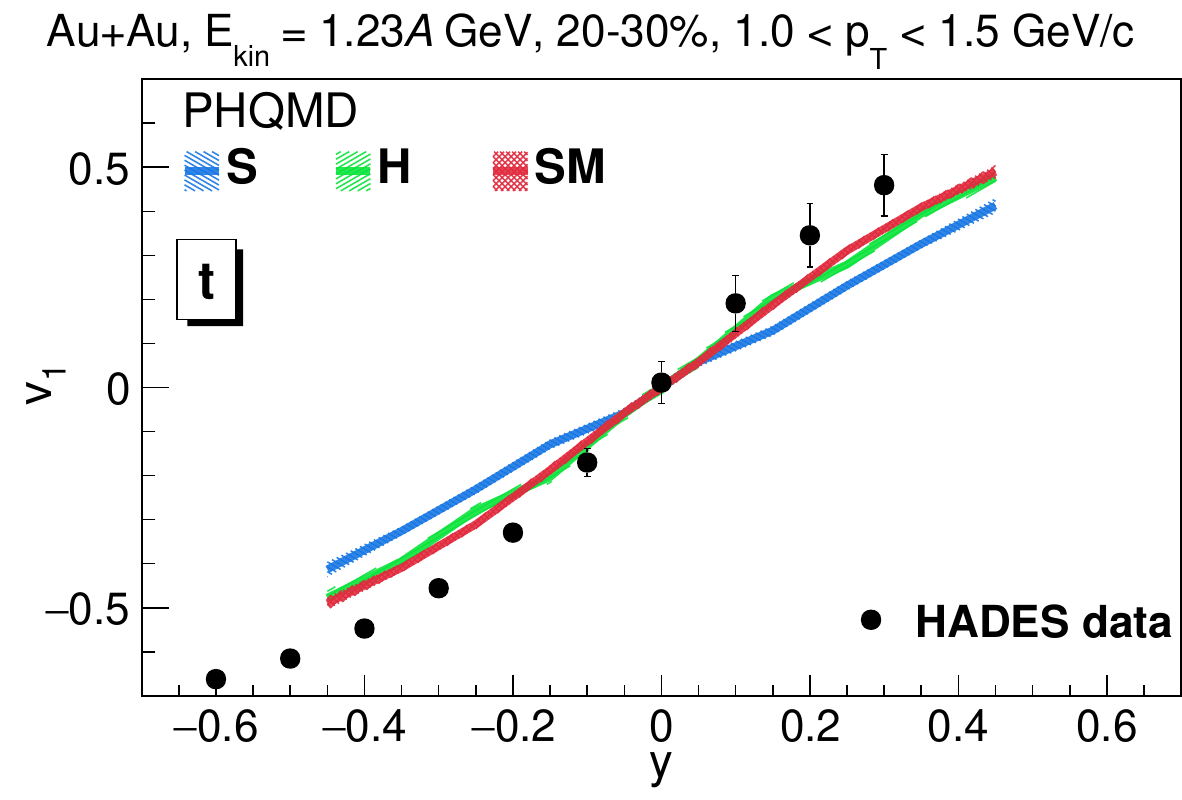}
\includegraphics[scale=0.4]{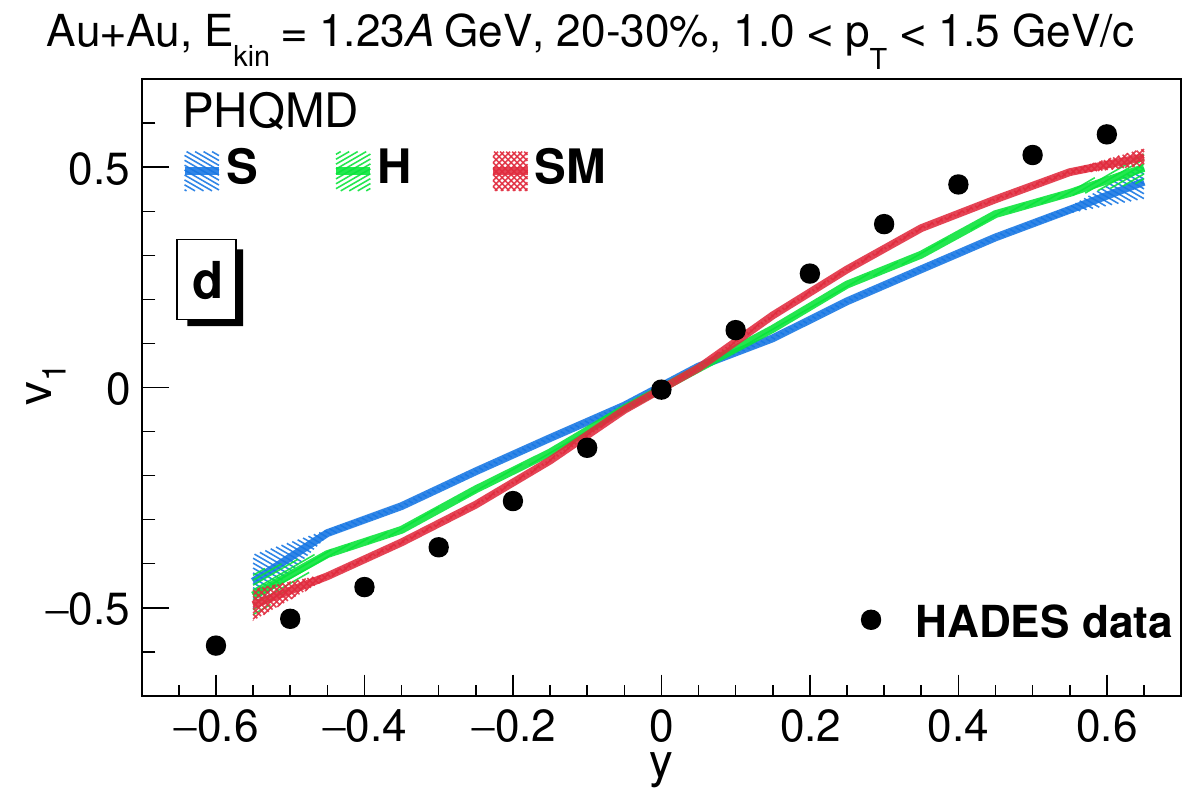}
\includegraphics[scale=0.4]{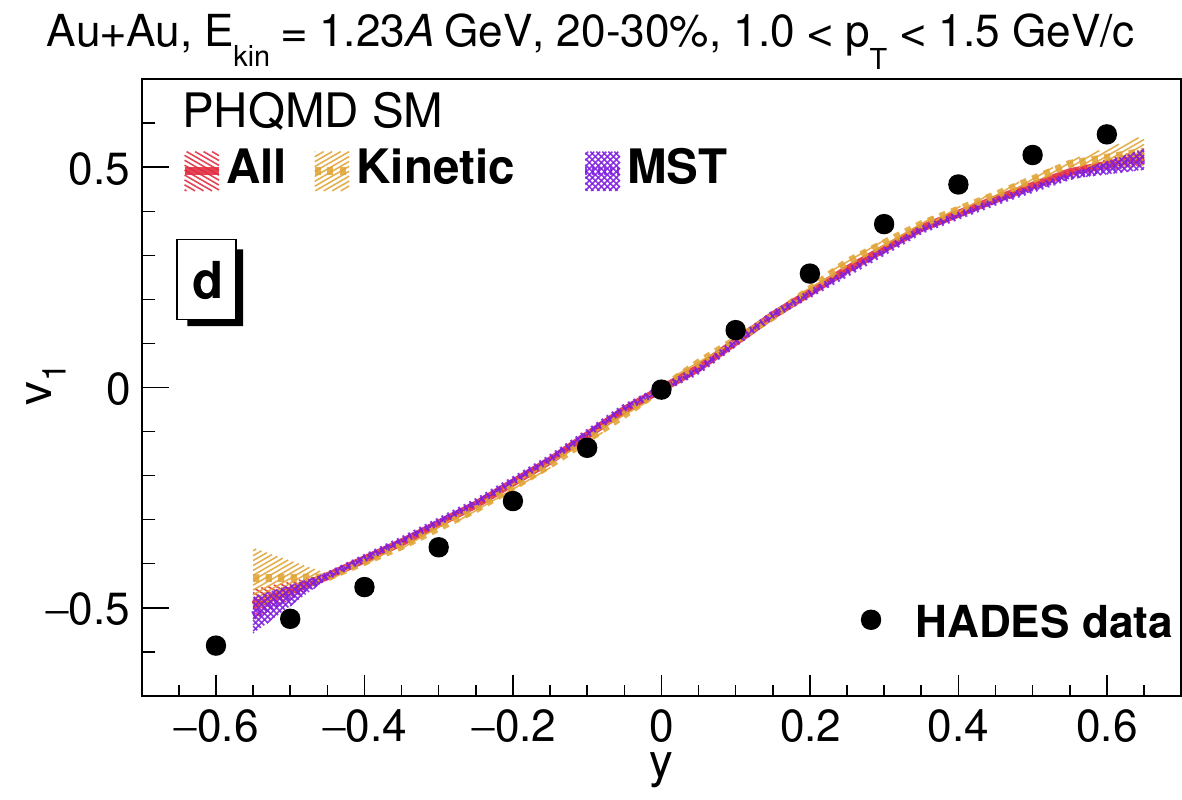}
    \caption{$v_1$ of protons (upper left) and deuterons (lower row) and tritons (upper right) as a function of rapidity for 20-30\% central Au+Au collisions at $E_{kin}=1.23$ A GeV for  $1.0 < p_T < 1.5$ GeV/c.
   The colour code is the same as in Fig. \ref{fig:dens}. The low right plot shows the $v_1(y)$ of deuterons for the SM EoS produced by different mechanisms: kinetic (yellow line), MST (violet line) as well as the $v_1(y)$ of all deuterons (red line, identical to the SM result on the lower left plot).    
    The HADES experimental data are taken from Ref. \cite{HADES:2020lob}.}
    \label{fig:hadesv1y}    
\end{figure*}

In Fig. \ref{fig:v1y_winn52} we show the PHQMD results for the directed flow $v_1$ of protons as a function of rapidity $y$  for different $p_T$ intervals for 20-30\% central Au+Au collisions at $E_{kin}=1.23$ A GeV  and for different $p_T$ cuts in comparison to the HADES experimental data from Ref. \cite{HADES:2022osk}.
The PHQMD calculations are done for three different equations-of-state summarized in table \ref{table_eos}.  In this plot and in all the following plots  the blue line "S" corresponds to the PHQMD calculations with the "soft" EoS, the green line "H" shows the "hard" EoS, the red line "SM" represents the momentum-dependent "soft" EoS; the colored areas represent the statistical errors.
Both, the theoretical as well as the experimental rapidity distributions of $v_1$, are rather insensitive to the chosen $p_T$ interval.  The soft momentum-dependent EoS gives even a steeper slope than the hard EoS and comes closest to the experimental data. The slope of the calculations with a soft EoS  is incompatible with the experimental data. 
We note that the soft and hard EoS have been chosen as a boundary of the values for the compressibility, which has been obtained from studies of monopole vibrations \cite{Mekjian:2011wut}, which are sensitive to densities around normal nuclear matter density, and early Plastic Ball data \cite{Gustafsson:1984ka}, which are sensitive to much higher densities and could be explained by a larger compressibility modulus \cite{Molitoris:1986pp}.

In Fig. \ref{fig:hadesv1y} we present the PHQMD results for $v_1(y)$ of protons (top left), deuterons (bottom left) and tritons (top right) in comparison to the HADES data \cite{HADES:2020lob} for Au+Au collisions at $E_{kin}=1.23$ A GeV for the $p_T$ interval $1.0<p_T<1.5$ GeV/c for hard, soft and momentum-dependent EoS. In the  bottom right figure we display $v_1(y)$ separately for  two mechanisms of deuteron production:  kinetic (yellow) and MST (violet).

Fig. \ref{fig:hadesv1y} shows that the increase of the slope of $v_1(y)$ with the size of the cluster, discussed in section \ref{conseq}, is confirmed experimentally.  This points towards a formation of the deuterons close to the border of the overlap region between projectile and target, where $v_1$ of nucleons is highest. Some of these  PHQMD results are in line with earlier findings in Ref. \cite{Aichelin:1987ti} and confirm recent calculations in Refs. \cite{Hillmann:2019wlt,Mohs:2020awg}. 
Also for clusters (left bottom and top right) the inclusion of the momentum-dependent interaction  in a S EoS gives even a steeper slope than a hard EoS and the results come closer to the HADES data. The slope of  $v_1(y)$ for tritons is still higher than that for deuterons and even a soft momentum-dependent interaction underpredicts noticeable the slope. The calculations reproduce also the slight non-linearity of $v_1(y)$ at large rapidities.   
$v_1(y)$  is rather similar for kinetic and MST deuterons (bottom right).

\begin{figure}
\includegraphics[scale=0.4]{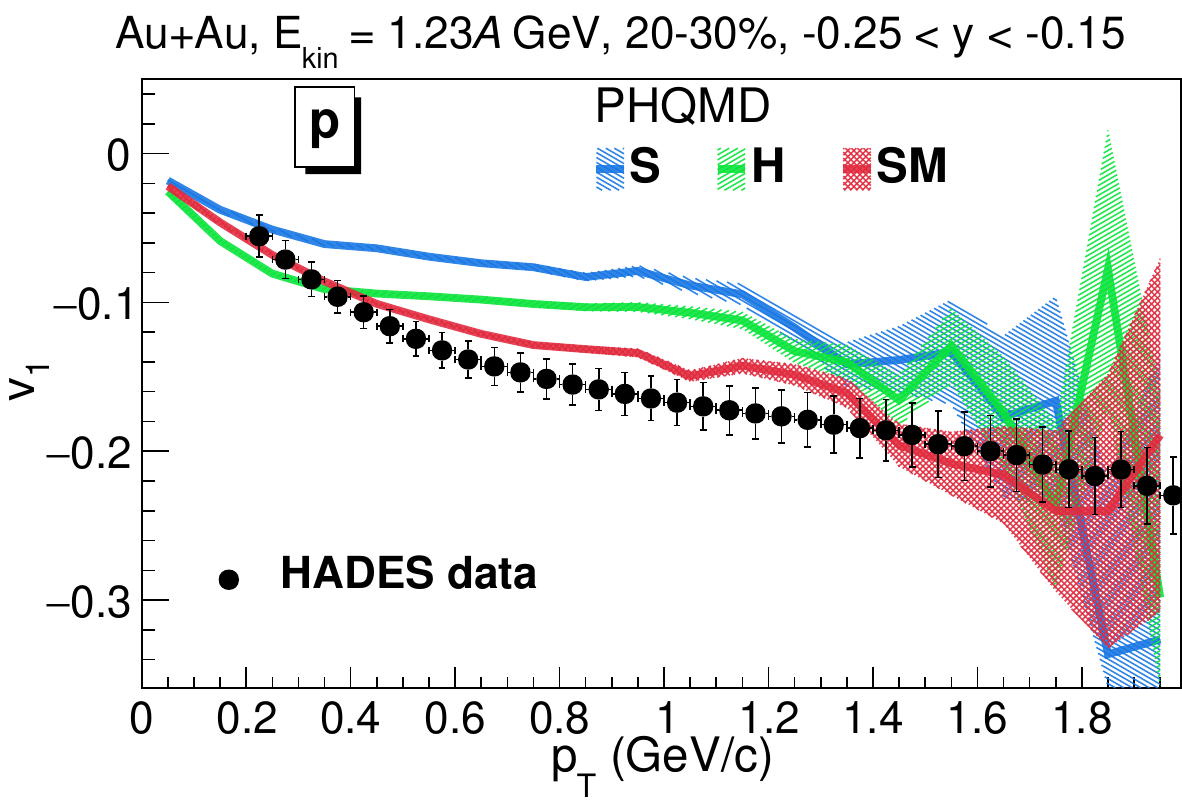}
\includegraphics[scale=0.4]{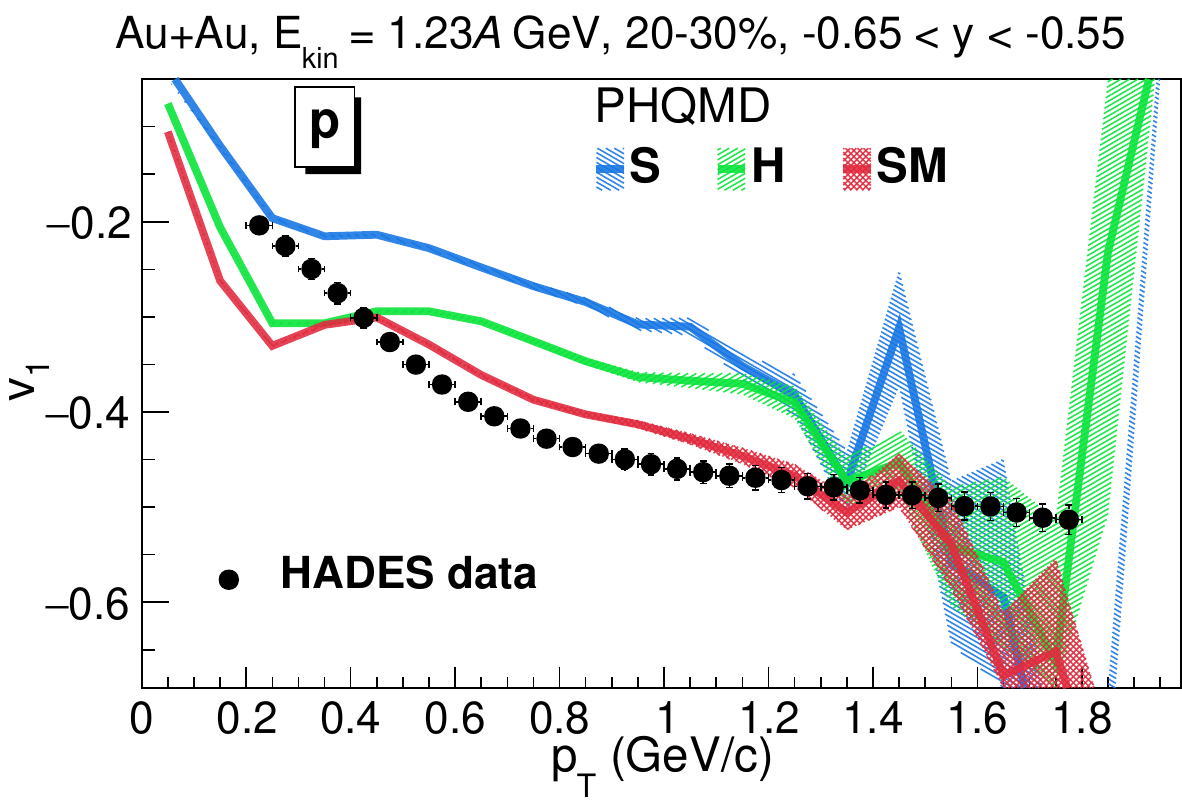}
    \caption{$v_1$ of protons as a function of $p_T$ for 20-30\% central Au+Au collisions at $E_{kin}=1.2$ A GeV for different rapidity bins:  
     $-0.25<y<-0.15$ (upper plot) and $-0.65<y<-0.55$ (lower plot). 
    The colour code is the same as in Fig. \ref{fig:dens}. 
    The experimental data are taken from Ref. \cite{HADES:2022osk}.}    \label{protonptv1}   
\end{figure}

\begin{figure}[h!]
\includegraphics[scale=0.4]{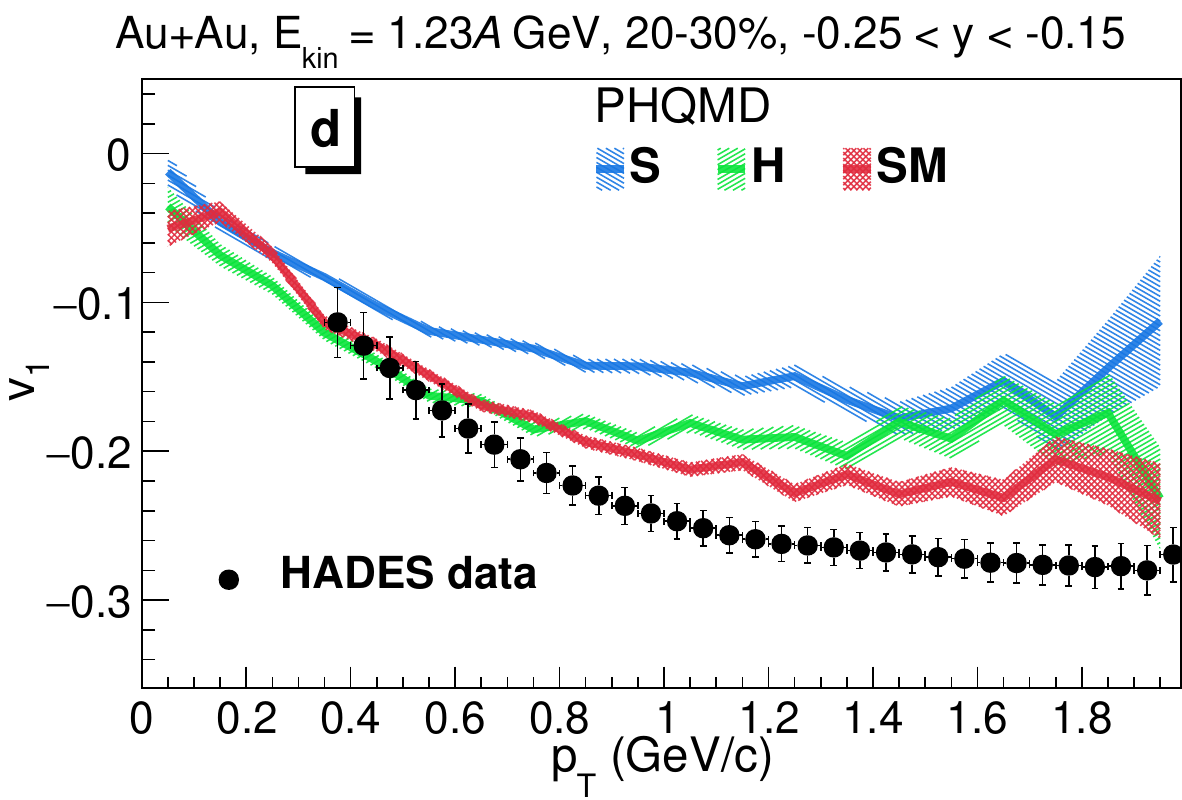}
\includegraphics[scale=0.4]{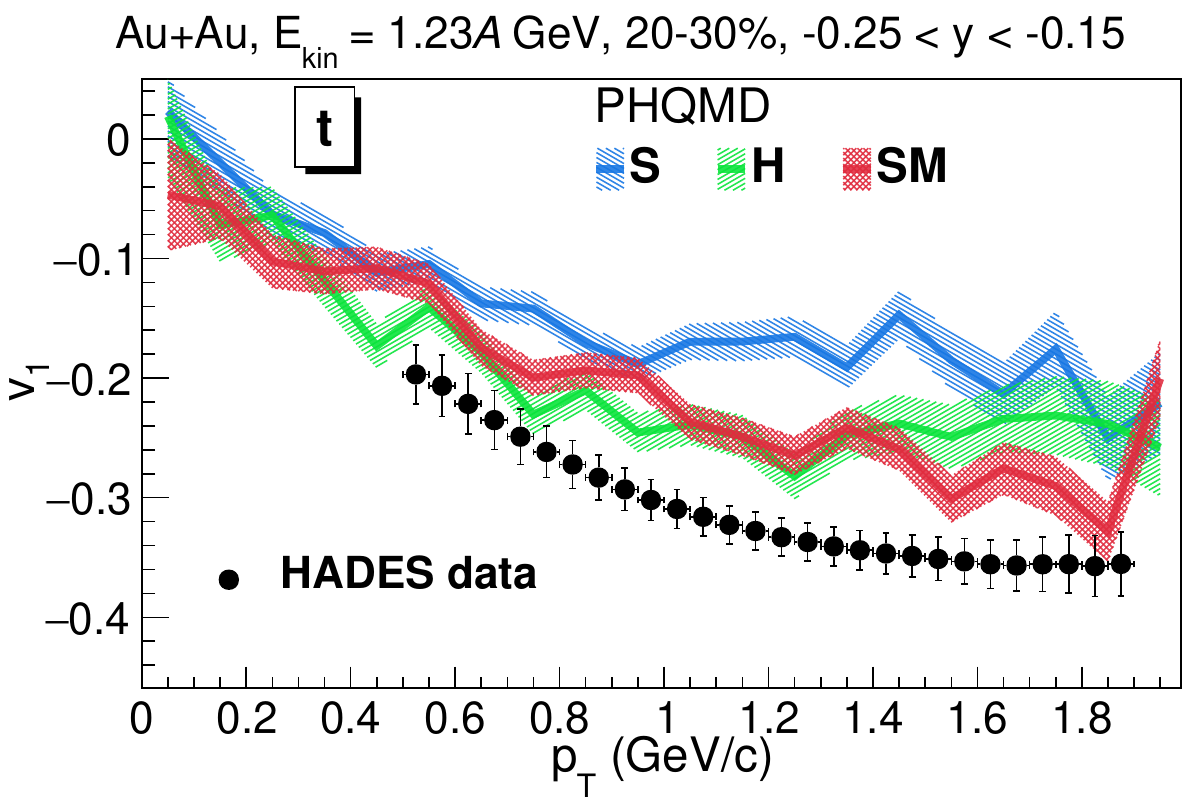}
    \caption{$v_1$ of deuterons (upper plot) and tritons (lower plot) as a function of $p_T$ for 20-30\% central Au+Au collisions at $E_{kin}=1.2$ A GeV  in the rapidity bin $-0.25<y<-0.15$. 
    The colour code is the same as in Fig. \ref{fig:dens}. 
    The experimental data are taken from Ref. \cite{HADES:2022osk}.
    \label{pdtv1pt}}   
\end{figure}

In Fig. \ref{protonptv1} we show the $p_T$ dependence of $v_1$ of protons for 20-30\% central  Au+Au collisions at $E_{kin}=1.23$ A GeV for different rapidity bins - $-0.25<y<-0.15$ (top) and $-0.65<y<-0.55$ (bottom) - calculated for three EoS. 
One can see that the value of $v_1(p_T)$ increases with increasing rapidity for all EoS, keeping the hierarchy - soft, hard, momentum-dependent soft EoS. One can also see that the SM EOS provides the best description of the HADES experimental data. 

A similar  behaviour of $v_1(p_T)$ holds for the light clusters as demonstrated in Fig. \ref{pdtv1pt}, which shows the $p_T$ dependence of the directed flow of deuterons (top) and tritons (bottom) for the rapidity bin $-0.25<y<-0.15$. One can see that the directed flow $v_1(p_T)$ for the hard EoS and SM EoS  are closer to each other as compared to the corresponding proton case in Fig. \ref{protonptv1}  , however, well below that for a soft EoS, which substantially underestimates the experimental data.

\subsubsection{Comparison of the PHQMD \texorpdfstring{$v_1$}{v1} to the FOPI data}

Now we step to a comparison of the PHQMD results for the directed flow for Au+Au collisions at $E_{kin}=1.2$ and 1.5 A GeV with the FOPI data \cite{FOPI:2011aa}.  The FOPI Collaboration presents the results using the following kinematic variables: \\
- scaled rapidity $y_{0}$:
   $$y_{0} = \frac{y}{y_{proj}}, \ \ {\rm where} \ y = \frac{1}{2}\log(\frac{E+p_z}{E-p_z})$$ 
is the rapidity of a particle with energy $E$ and 3-momentum ${\bf p}\equiv p$,
   $$y_{proj} = \frac{1}{2}\log(\frac{E_P+P_P}{E_P-P_P})$$ 
is the projectile rapidity and $E_P, \ P_P=P_{P,z}$ are the energy and momentum of the initial projectile nucleon in center-of-mass system; \\
- scaled  transverse momentum $u_{t0}$: 
  $$u_{t0} = \frac{u_T}{u_{proj}},$$
 where $u_T = \beta_T \gamma$ is the transverse velocity of a particle with velocity $\beta=p/E$, $\gamma = 1/\sqrt{1 - \beta^2}$  and $\beta_T = p_T / E$, 
while the projectile velocity is $u_{proj} = \beta_P \gamma_P$ with  $\beta_P = P_{P} / E_{P}$. \\
-  the centrality is defined in terms of the scaled impact parameter $b_0=b/b_{max}$
with $b_{max} = 1.15 (A_P^{1/3} + A_T^{1/3})$~fm, where $A_P, A_T$ are the atomic numbers of projectile (P) and target (T).

The FOPI measurement of $v_1$ covers different sub-regions in the $(y_0,u_{t0})$ plane.
Here we present the PHQMD results for $v_1(y_0)$ for  $u_{t0} >0.4$ and for $v_1(u_{t0})$ for $0.4<|y_0|<0.8$ for the centrality bin $0.25<b_0<0.45$  ($3.34 < b < 6.022$ fm).

\begin{figure*}
\includegraphics[scale=0.4]{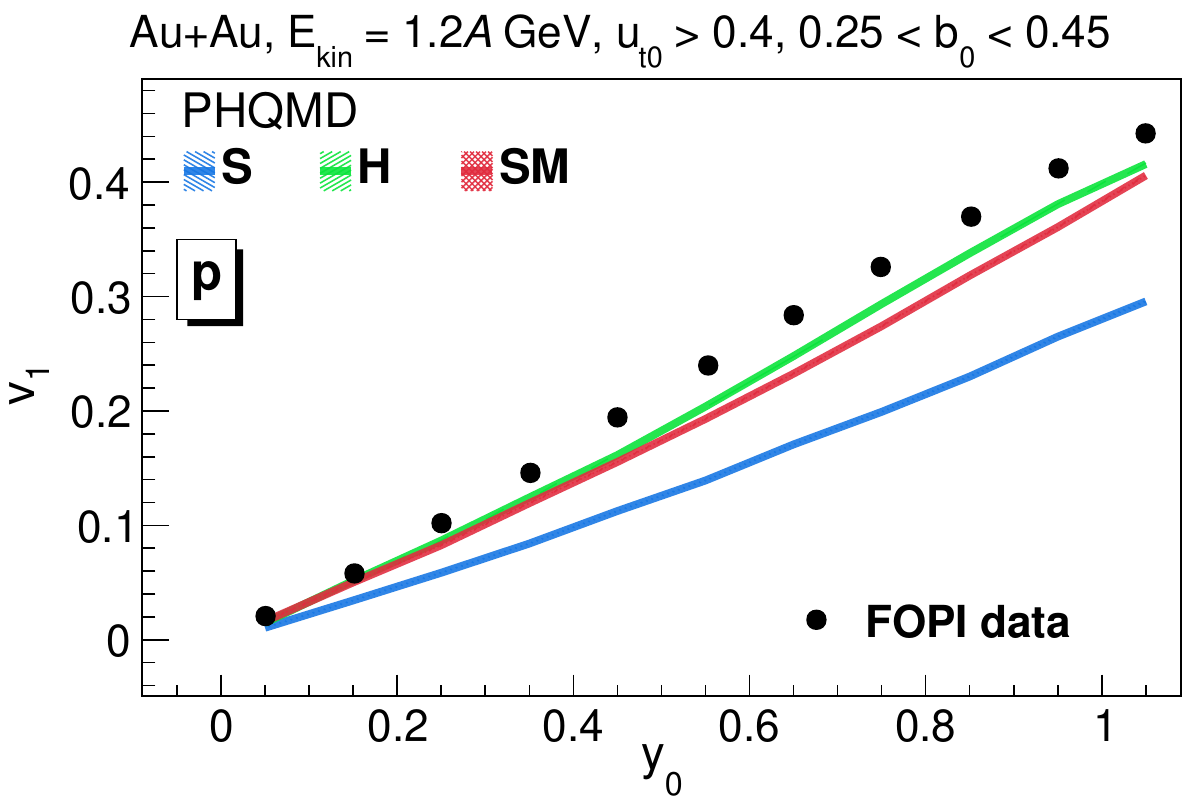}
\includegraphics[scale=0.4]{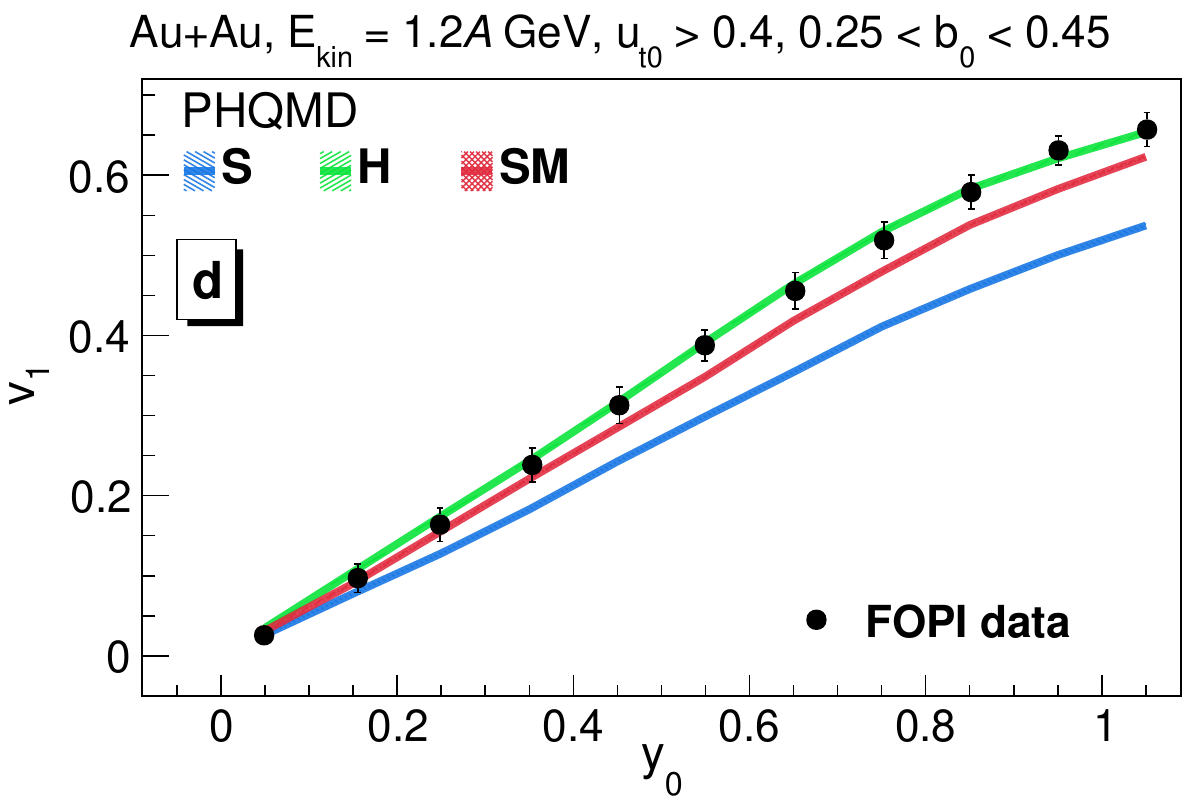}
\includegraphics[scale=0.4]{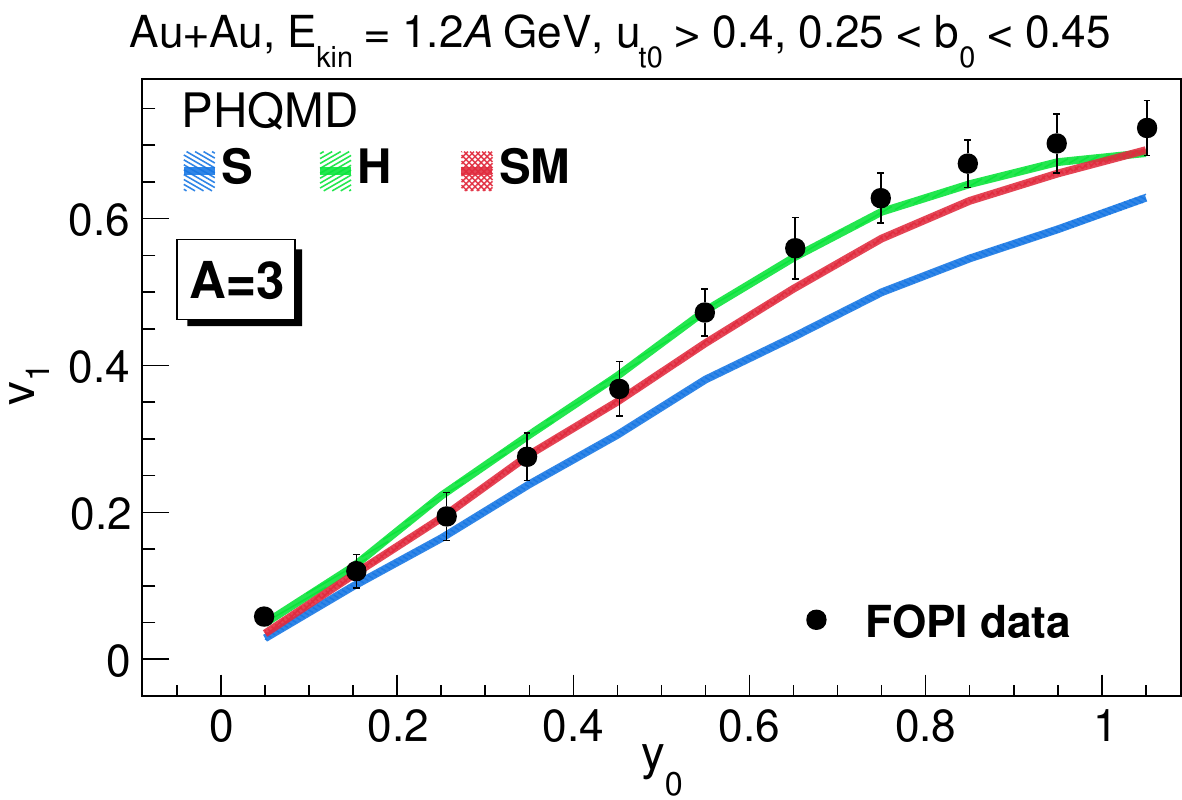}
\includegraphics[scale=0.4]{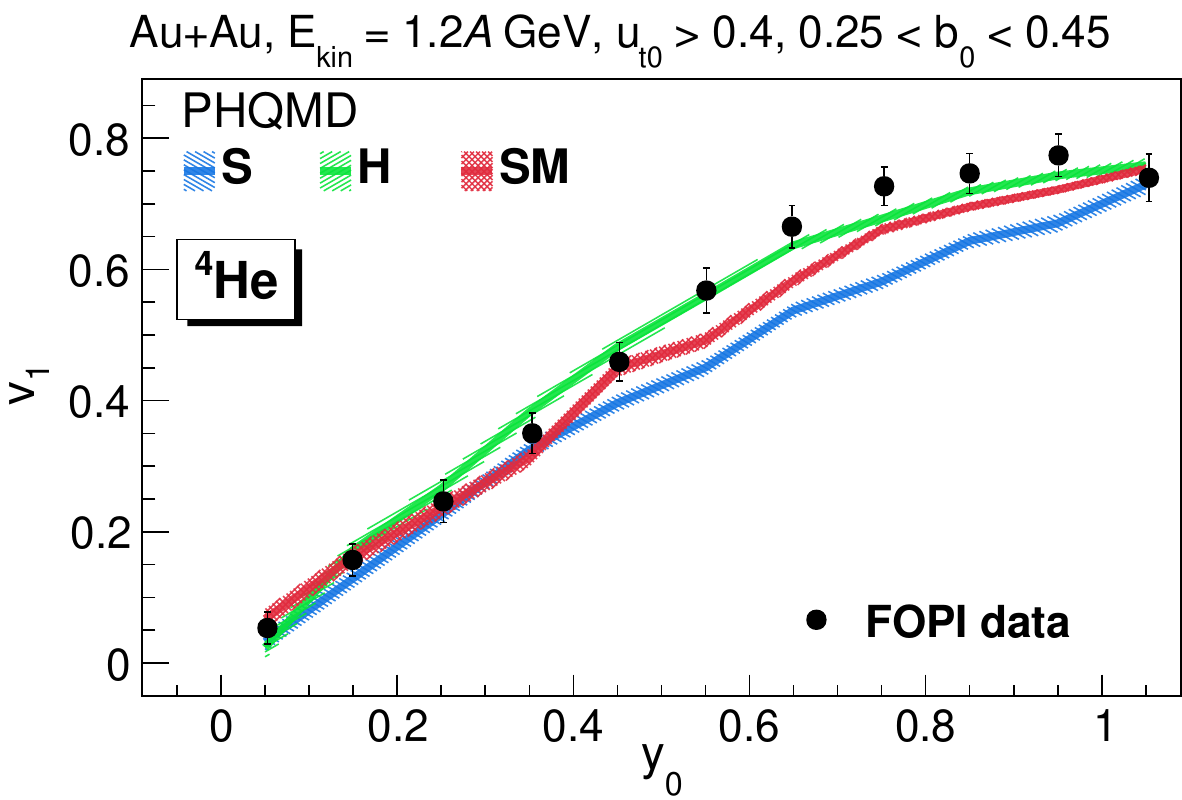} \\
\includegraphics[scale=0.4]{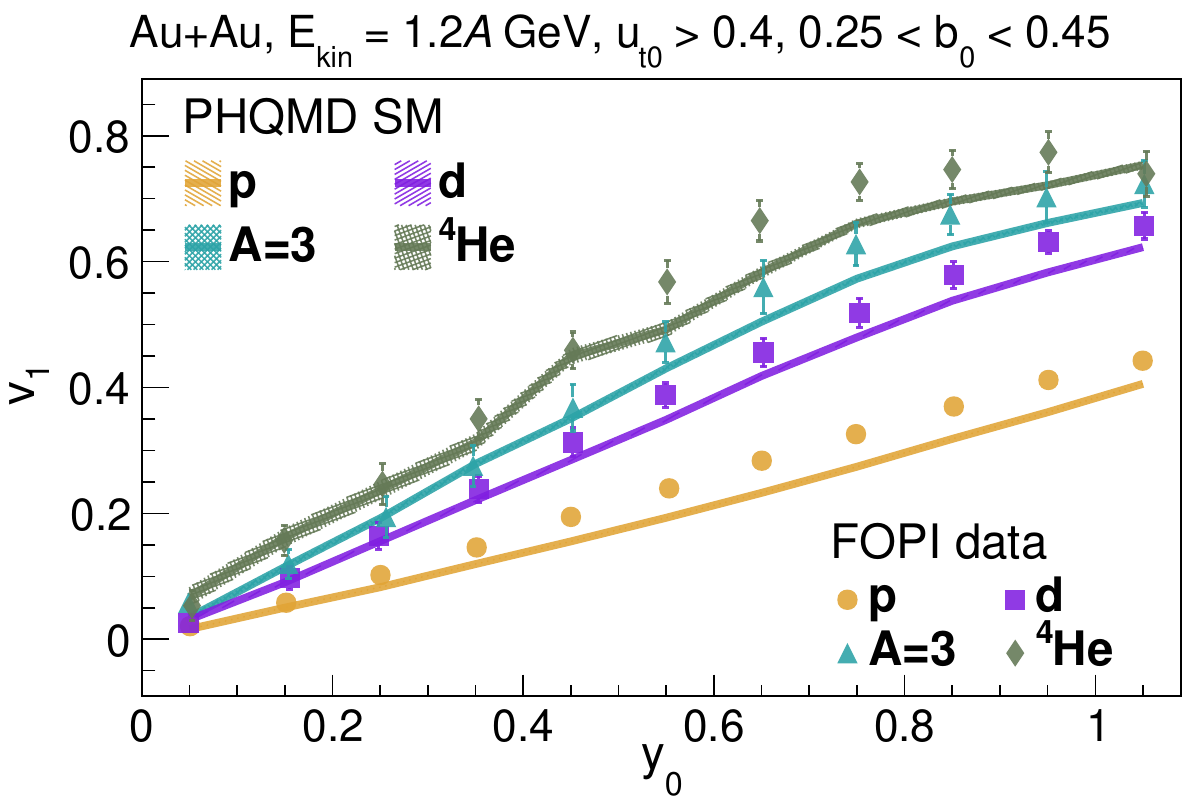}
    \caption{$v_1$ of protons (upper left), deuterons (upper right), $A=3$ (middle left) and $^4$He (middle right) as a function of $y_0$ for Au+Au collisions at $E_{kin}=1.2$ A GeV for $u_{t0} >0.4$ and the impact parameter range $0.25<b_0<0.45$. The colour code is the same as in Fig. \ref{fig:dens}. 
    The plot on the lower row shows the compilation of $v_1(y_0)$ for protons (yellow), deuterons (magenta), $A=3$ clusters (blue) and $^4$He (olive) for the SM EoS. 
    The FOPI experimental data are taken from Ref. \cite{FOPI:2011aa}.}
    \label{Fig:FOPI12y0}       
\end{figure*}

\begin{figure*}
\includegraphics[scale=0.4]{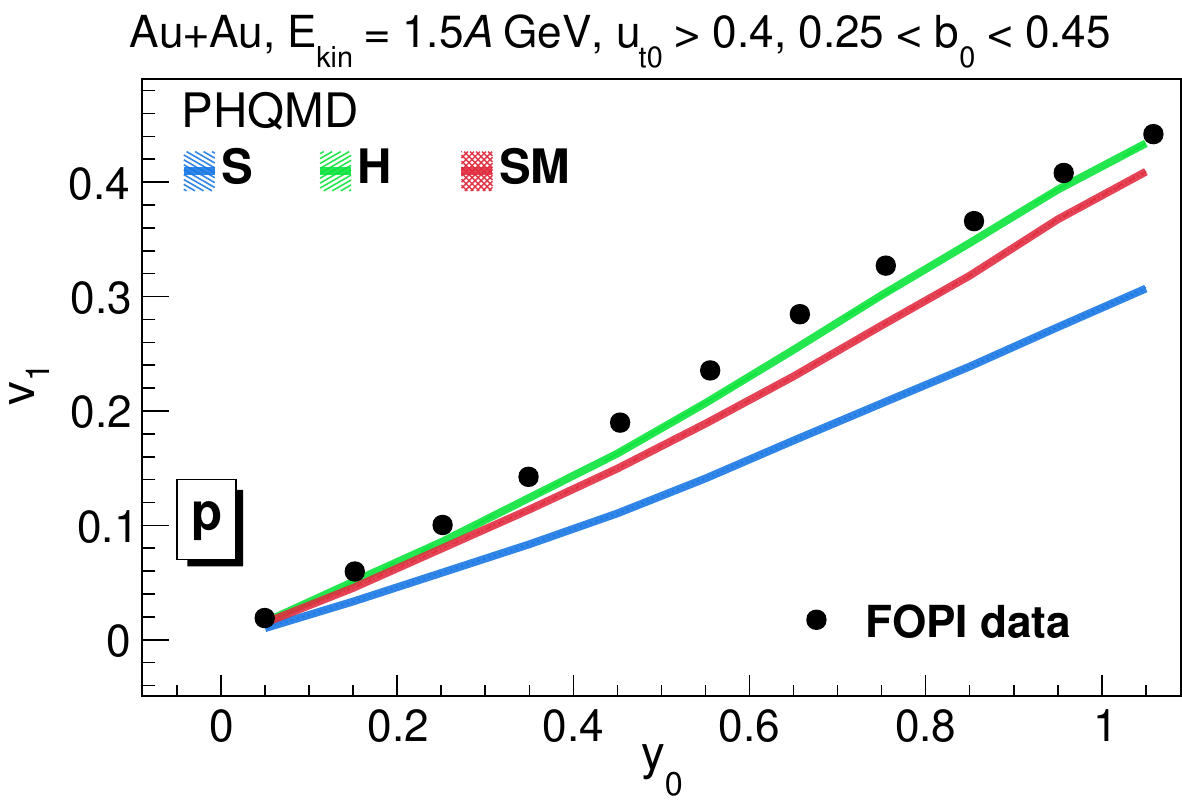}
\includegraphics[scale=0.4]{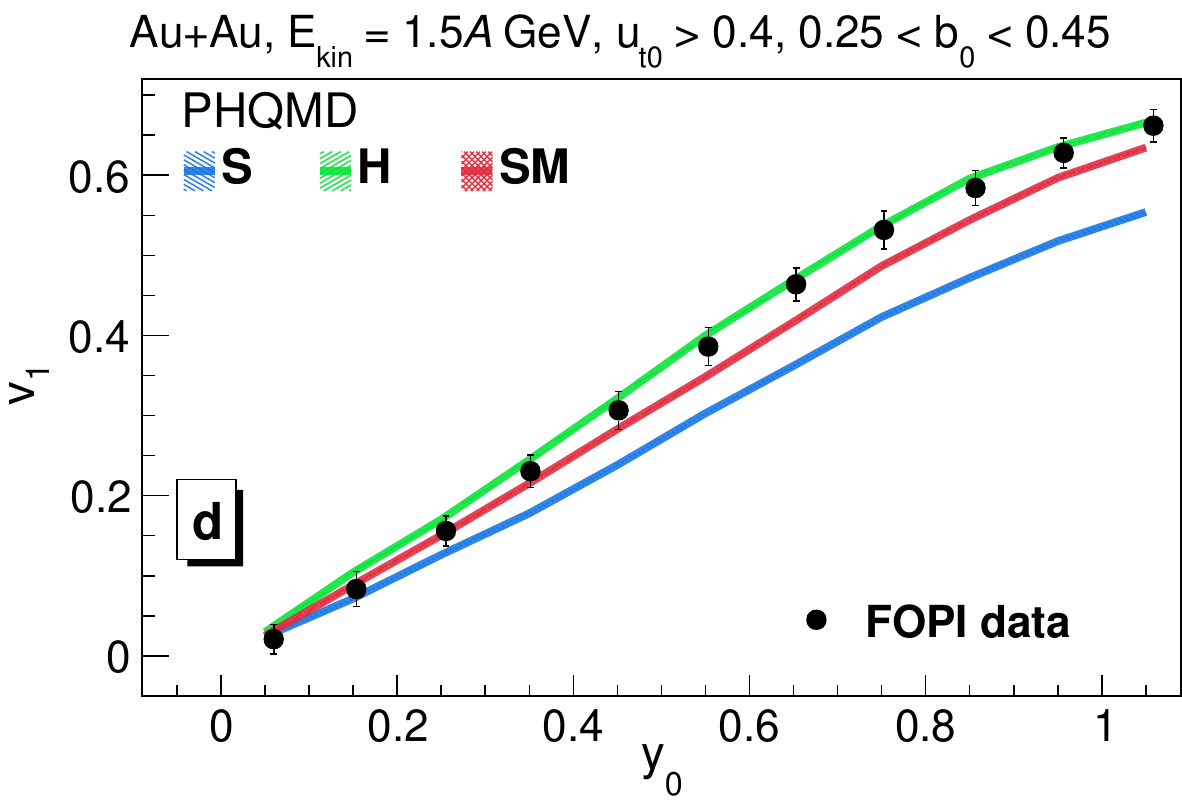}
\includegraphics[scale=0.4]{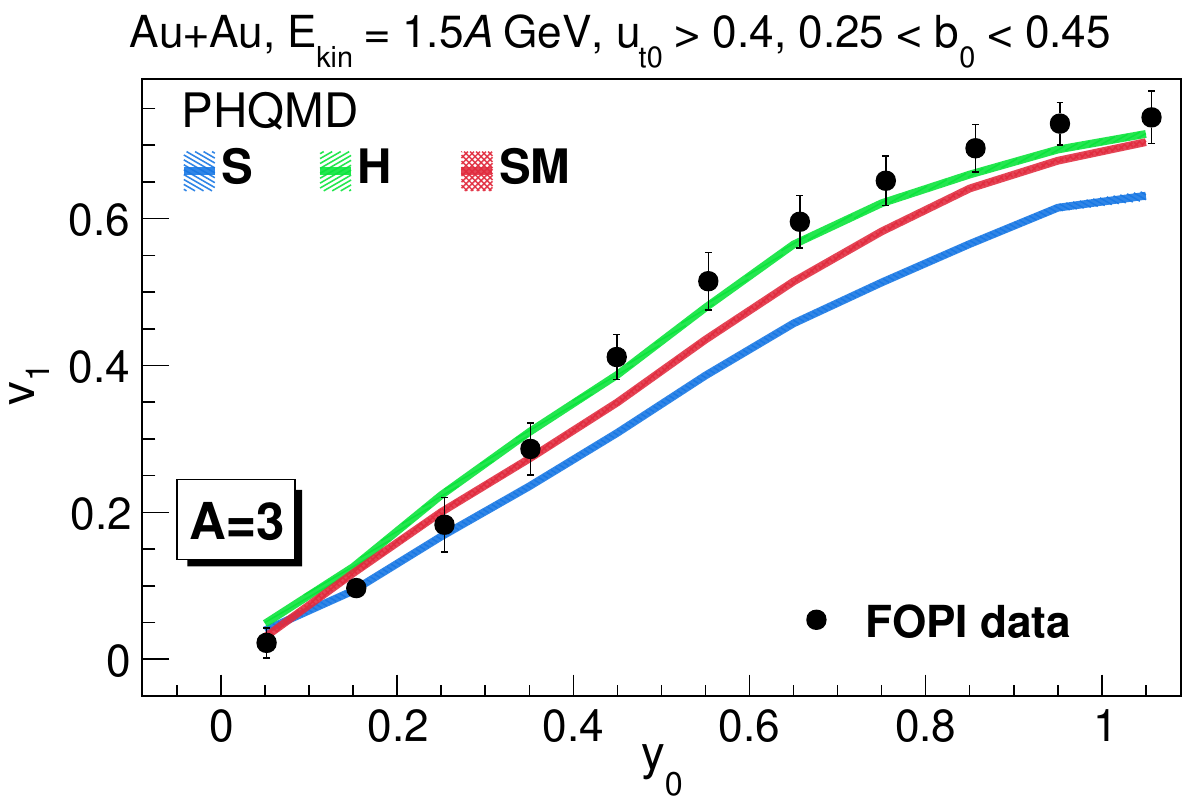}
\includegraphics[scale=0.4]{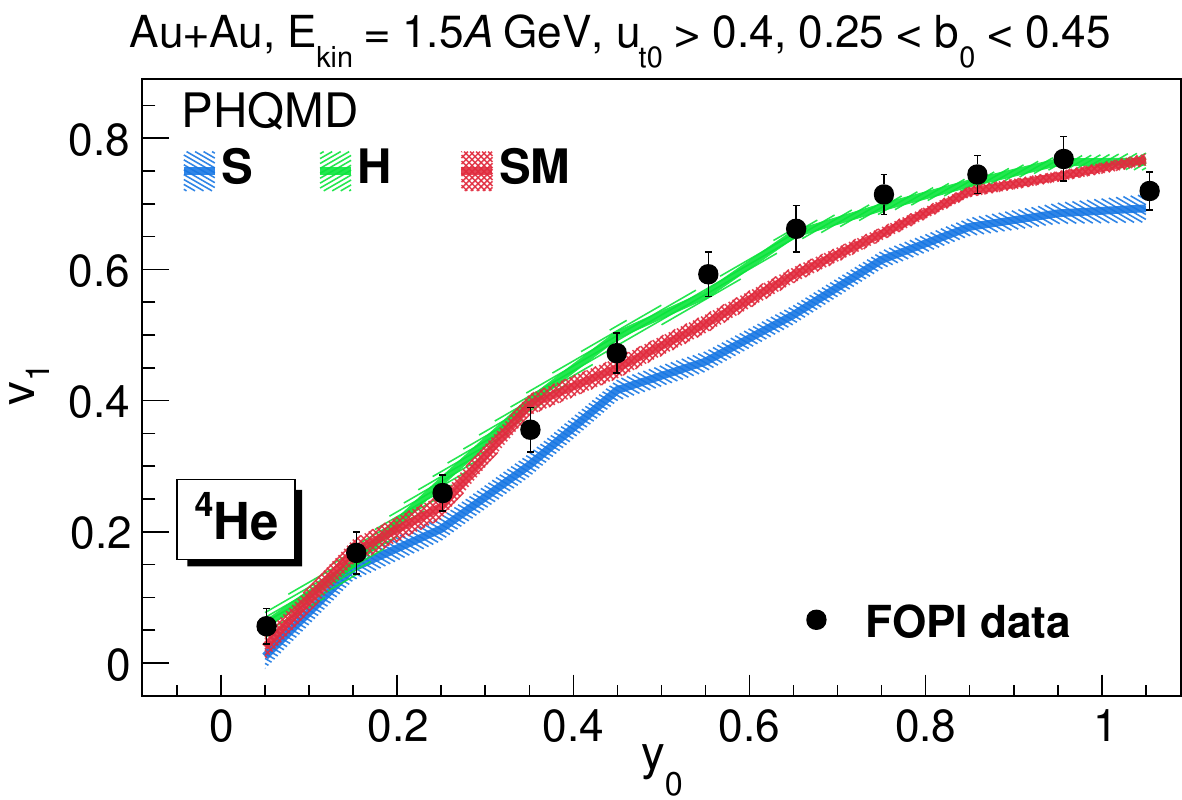} 
\includegraphics[scale=0.4]{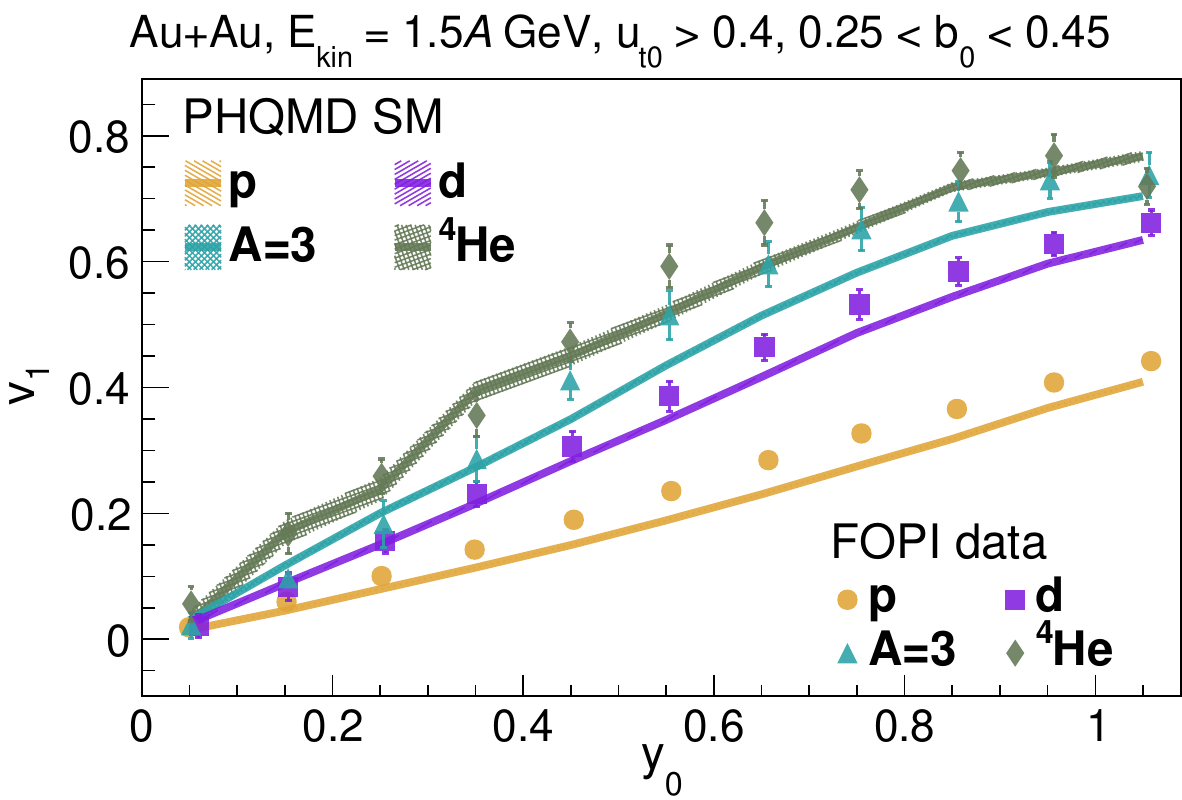}
    \caption{$v_1$ of protons (upper left), deuterons (upper right), $A=3$ (middle left) and $^4$He (middle right) as a function of $y_0$ for Au+Au collisions at $E_{kin}=1.5$ A GeV for $u_{t0} >0.4$ and the impact parameter range $0.25<b_0<0.45$.  
    The plot on lower row shows the compilation of  $v_1(y_0)$ for protons, deuterons, $A=3$ clusters and $^4$He for the SM EoS. 
    The colour code is the same as in Fig. \ref{Fig:FOPI12y0}.
    The FOPI experimental data are taken from Ref. \cite{FOPI:2011aa}. }
    \label{Fig:FOPI15y0}           
\end{figure*}

Figs. \ref{Fig:FOPI12y0} and \ref{Fig:FOPI15y0} show the directed flow $v_1$ of protons, deuterons, $A=3$ clusters and $^4$He as a function of $y_0$ for Au+Au collisions at $E_{kin}=1.2$ and 1.5 A GeV, respectively, for $u_0 >0.4$ and the impact parameter range $0.25<b_0<0.45$ in comparison to the FOPI  data \cite{FOPI:2011aa}.
One can see that for all clusters as well as for protons the soft EoS underestimates also the FOPI data. The difference between the hard and the soft momentum-dependent EoS is small for  $u_{t0} >0.4$ and the hard EoS leads  to a slightly larger $v_1$, in contradistinction to the calculation for the HADES data 
Fig. \ref{fig:hadesv1y} (which cover a partially different kinematical range)

That these findings are not trivial is shown in the bottom panel of Figs. \ref{Fig:FOPI12y0} and \ref{Fig:FOPI15y0}, which display $v_1(y_0)$ for a SM EoS for different cluster sizes in comparison with the experimental results. 
We observe  a strong dependence of the slope of $v_1(p_T)$ on the cluster size, which has been discussed in section \ref{conseq} and which is indeed seen in the experimental data.

\begin{figure*}
\includegraphics[scale=0.4]{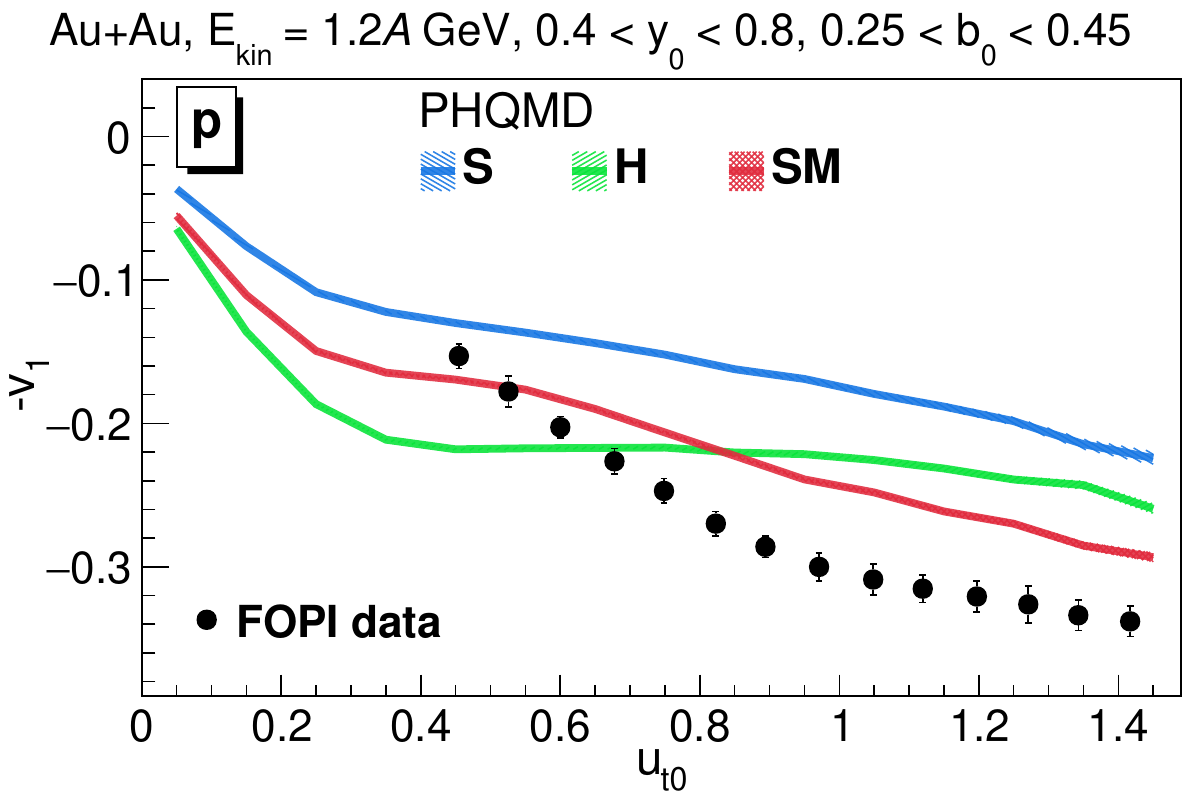}
\includegraphics[scale=0.4]{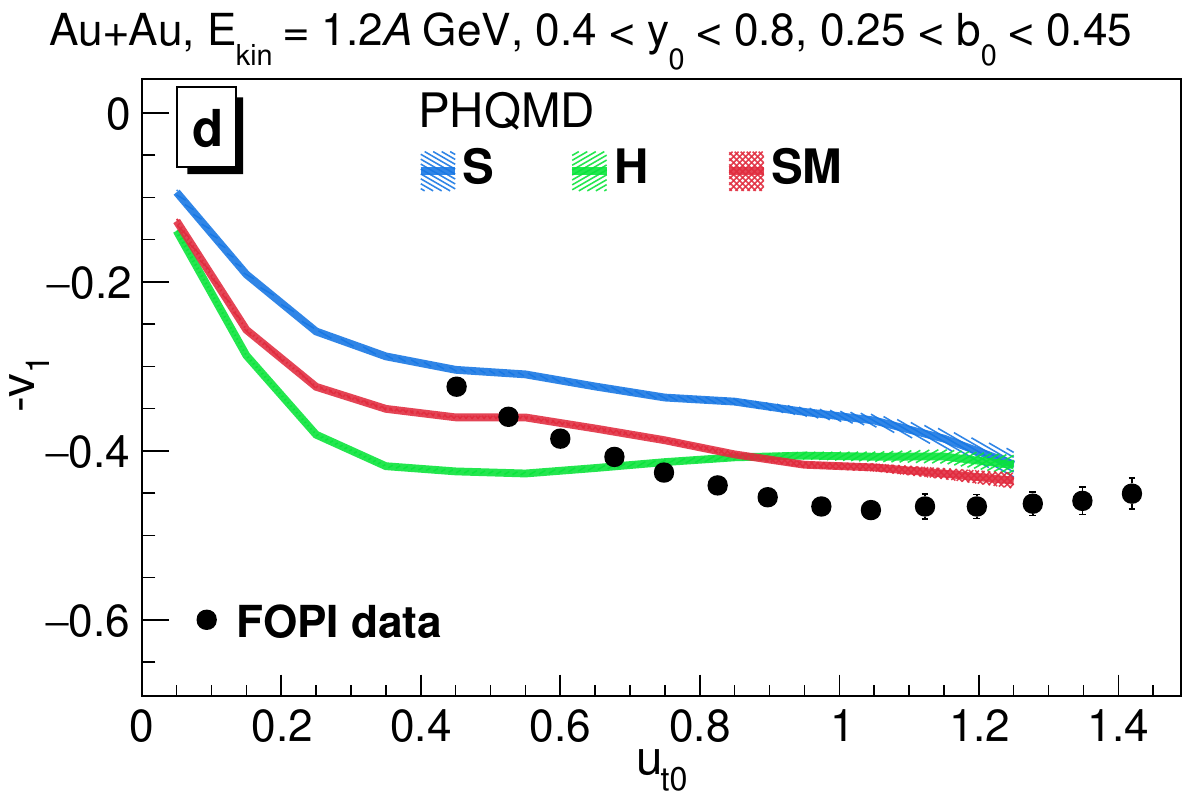}
\includegraphics[scale=0.4]{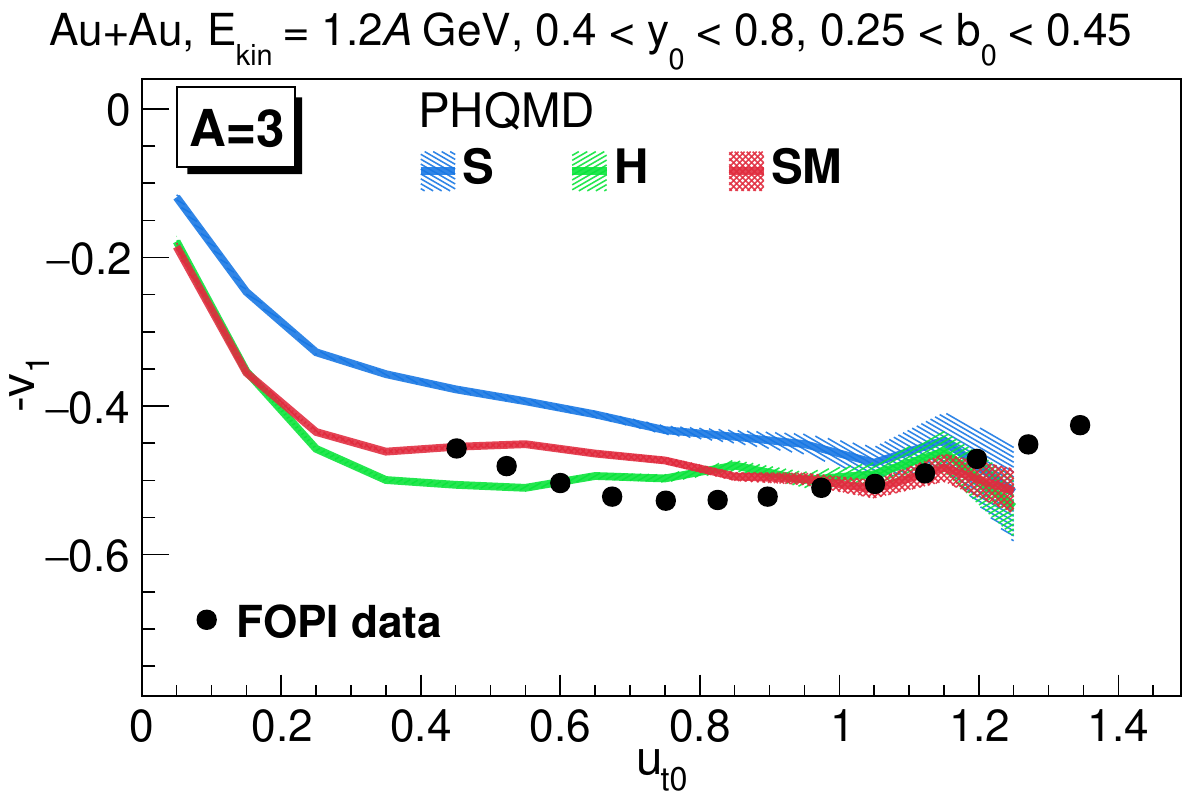}
\includegraphics[scale=0.4]{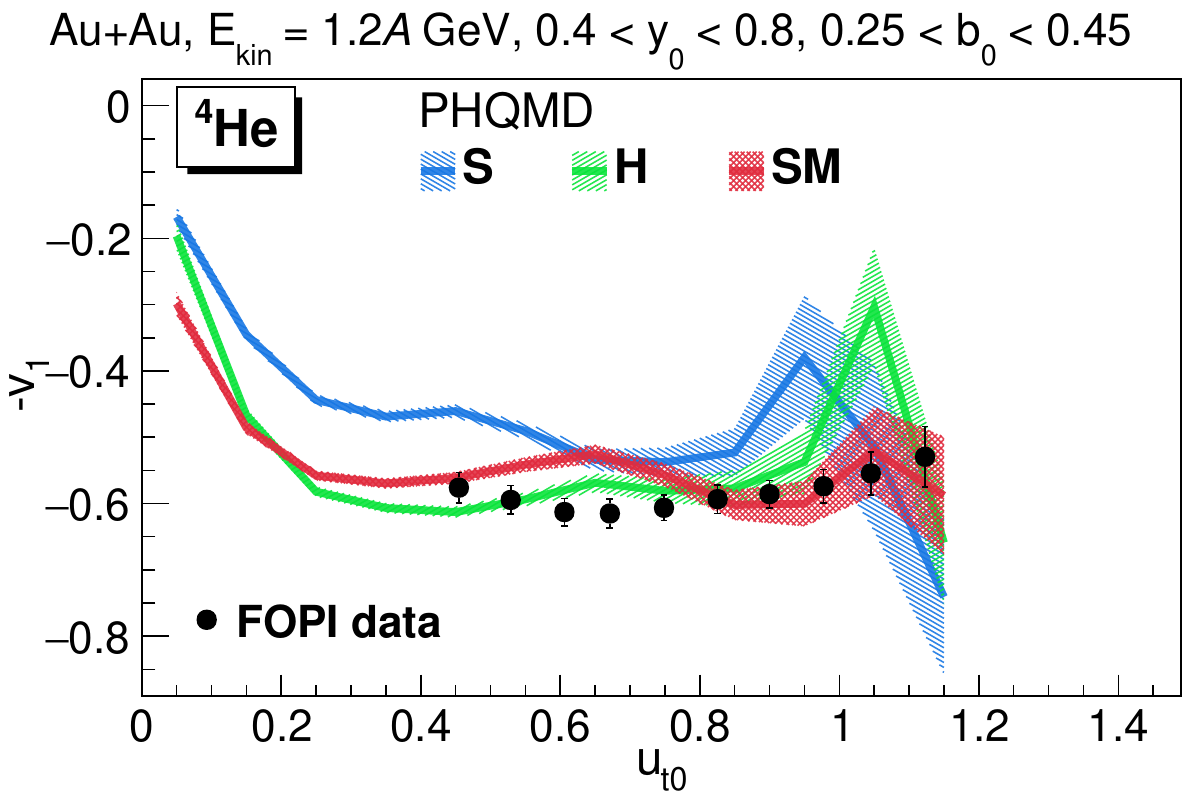}
    \caption{$v_1$ of protons (upper left), deuterons (upper right), $A=3$ (lower right) and $^4$He (lower left) as a function of $u_{t0}$ for Au+Au collisions at $E_{kin}=1.2$ A GeV for $0.4<y_0<0.8$ and the impact parameter range $0.25<b_0<0.45$.
   The colour code is the same as in Fig. \ref{fig:dens}. 
   The FOPI experimental data are taken from  Ref. \cite{FOPI:2011aa}.}
    \label{Fig:FOP12u0}           
\end{figure*}

\begin{figure*}
\includegraphics[scale=0.4]{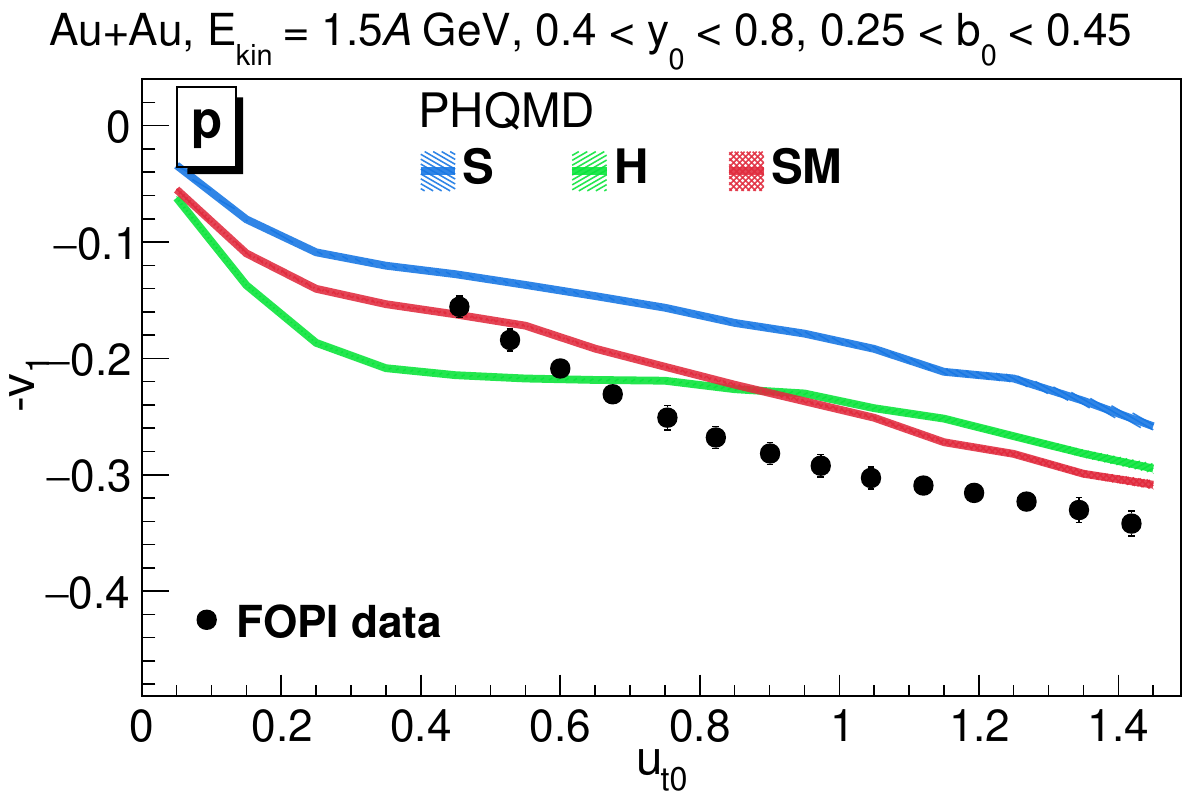}
\includegraphics[scale=0.4]{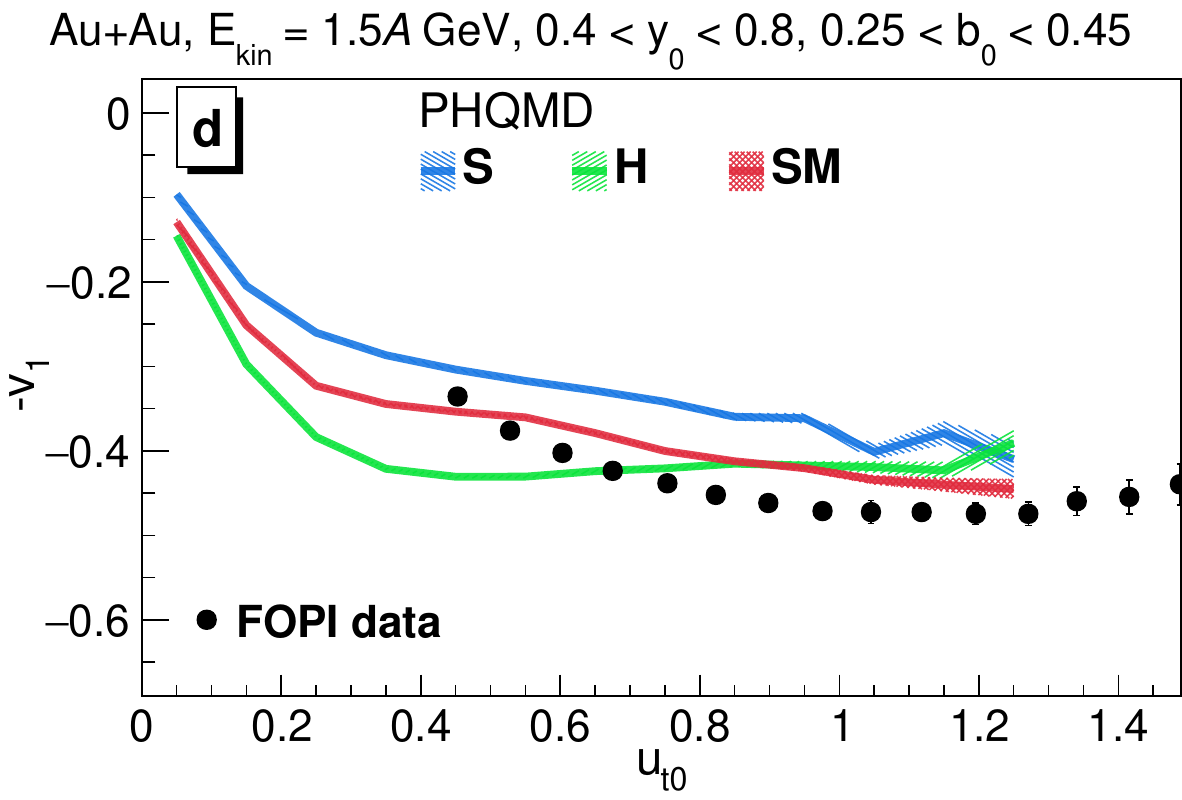} 
\includegraphics[scale=0.4]{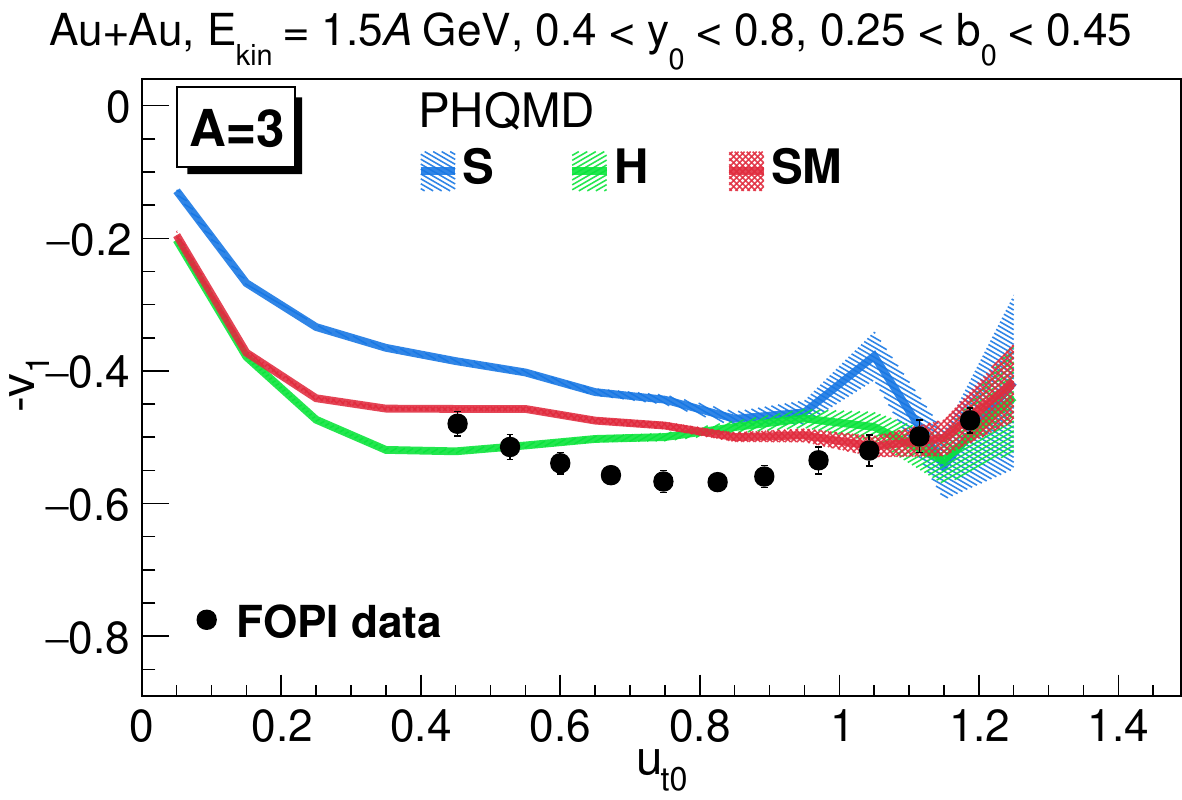} 
\includegraphics[scale=0.4]{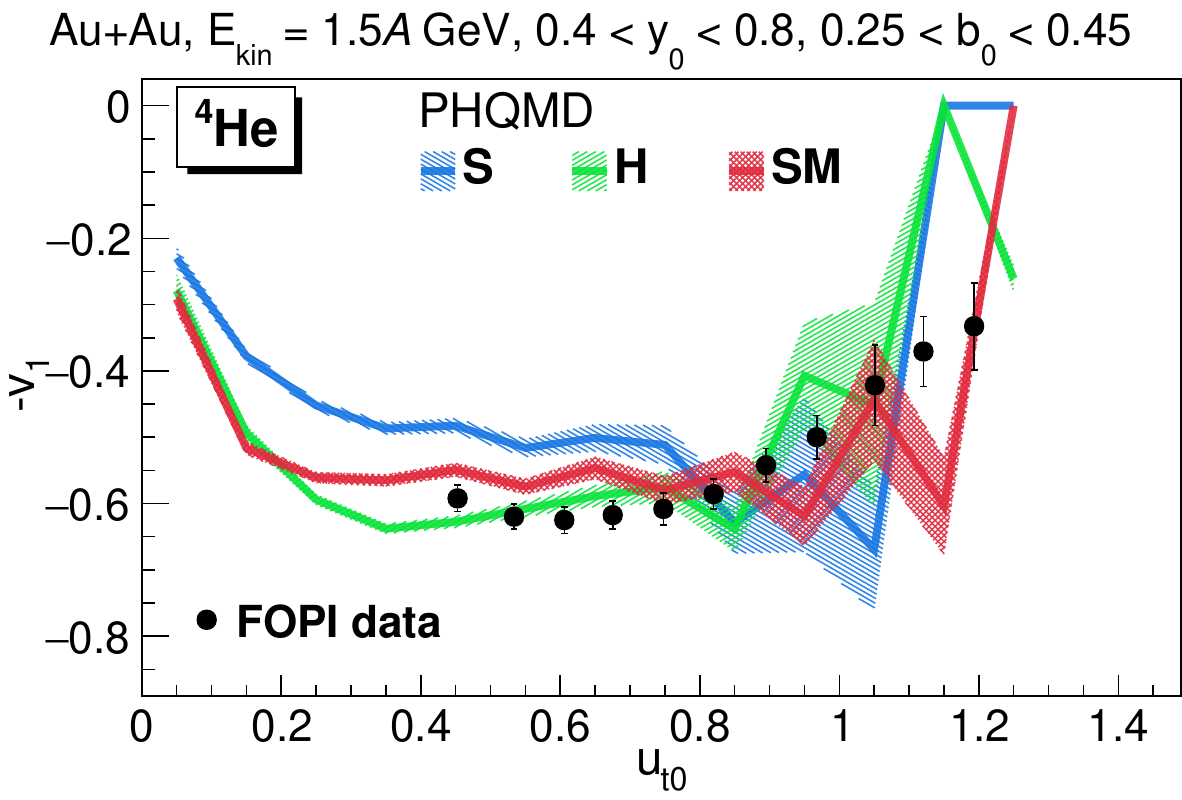}
\includegraphics[scale=0.4]{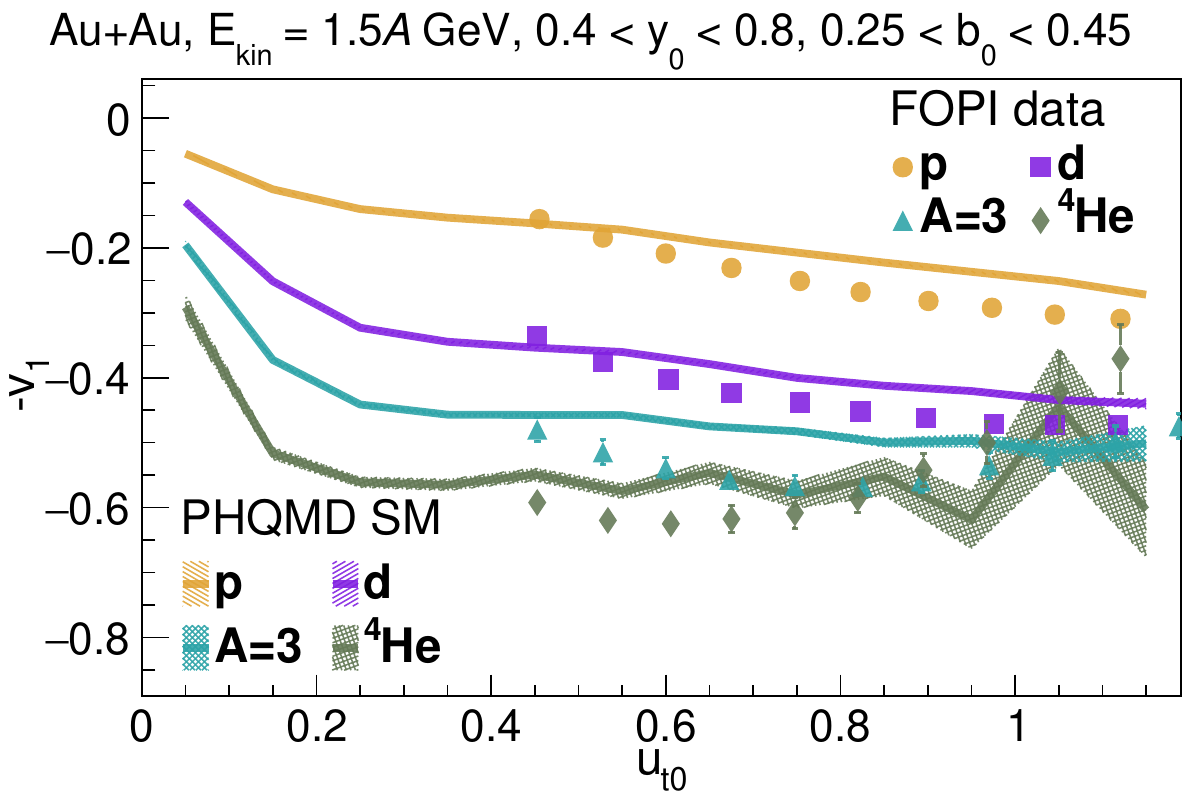}
    \caption{$v_1$ of protons (upper left), deuterons (upper right), $A=3$ (middle left) and $^4$He (middle right) as a function of $u_{t0}$ for Au+Au collisions at $E_{kin}=1.5$ A GeV  for $0.4<y_0<0.8$ and the impact parameter range $0.25<b_0<0.45$. 
    The plot on the lower row shows the compilation of $v_1(u_{t0})$ for protons, deuterons, $A=3$ clusters and $^4$He for the SM EoS. 
    The colour code is the same as in Fig. \ref{Fig:FOPI12y0}.
    The FOPI experimental data are taken from Ref. \cite{FOPI:2011aa}. }    
    \label{Fig:FOP15u0}     
\end{figure*}

In Figures \ref{Fig:FOP12u0} and \ref{Fig:FOP15u0} we present the PHQMD results for $v_1$ of protons, deuterons, $A=3$ clusters and $^4$He as a function of $u_{t0}$ for Au+Au collisions at $E_{kin}=1.2$ and 1.5 A GeV, respectively, for $0.4<y_0<0.8$ and the impact parameter range $0.25<b_0<0.45$ in comparison to the FOPI data from \cite{FOPI:2011aa}.
We note that the FOPI data were measured in a positive rapidity hemisphere. However, we have reflected the FOPI data $v_1(u_{t0}) \to -v_1(u_{t0})$ for the better comparison to the HADES results. 
One can see  that different EoS yield quite different $v_1(u_{t0})$ curves:  at low $u_{t0}$, $v_1$ of a hard EoS is larger while at large $u_{t0}$ the SM EoS gives larger $v_1$ for protons and joins the results for a H EoS for clusters. For a SM EoS we summarize in the bottom figures the results for the different clusters.

As follows from Figs. \ref{Fig:FOPI12y0} - \ref{Fig:FOP15u0}, the PHQMD calculations provide a qualitative description of the experimental data also for the FOPI data in rapidity as well as in $u_{to}$ . The form of the different $v_1(u_{to})$  and  $v_1(y_0)$ spectra at $E_{kin}$ = 1.2 and 1.5 GeV are similar, in theory as well as in experiment, and the deviations between theory and experiment are also almost energy independent. 
Also the FOPI data show that $v_1(u_{to})$  and  $v_1(y_0)$ depend substantially on the cluster size.  This is as well reproduced by PHQMD calculations. 
Despite of the different centralities, the different observables and the different acceptance ranges, one might compare the calculations for the HADES and FOPI data and come to the conclusions  that  for all cluster sizes the theoretical  $v_1(y), v_1(y_0)$  and $v_1(p_T), v_1(u_{t0})$  calculations with a SM EoS are slightly below the data in all these experiments. 

\clearpage
\subsection{ The elliptic flow \texorpdfstring{$v_2$}{v2}}

The second Fourier coefficient in the azimuthal distribution (\ref{eq:azim}) , the elliptic flow $v_2$, measures whether matter is preferably emitted in the reaction plane (positive $v_2$) or perpendicular to the reaction plane (negative $v_2$). At SIS energies, considered here, $v_2$ is negative. In former times this has been interpreted either as the absorption of participant nucleons by the spectator matter or as squeezing of the nucleons perpendicular to the reaction plane. 

More detailed studies have revealed, however, that the  origin of $v_2$ is more complicated \cite{LeFevre:2016vpp} and that in reality  3 different processes contribute to the building of the  $v_2$ \cite{reichert}.  Participant nucleons entering the spectator matter have to have a positive $p_x$ component (if the spectator is located in positive $x$-direction). This positive $p_x$ component gets lowered  by finite angle collisions with spectator nucleons, creating an overall negative $v_2$. In addition, nucleons sitting in $y$-direction at the tip of the overlap region, feel a very strong density gradient in $y$-direction (from about twice nuclear matter density to vacuum) and get therefore considerably accelerated in $y$-direction. This yields also a negative $v_2$. Finally, the compressed matter of the overlap region, which is not spherical in transverse direction  and approaches with increasing beam energy an almond shaped form, develops an outward non-isotropic pressure, which accelerates  the participant nucleons outwards, preferably in x-direction due to the geometry. The $v_2$ is then an image of the initial eccentricity in coordinate space and is positive. This is the mechanism, which dominates $v_2$  at higher energies but it is already present at the energies considered here.  Both, collisions and potential interactions contribute to $v_2$ and therefore it is a challenge for every transport approach to reproduce the experimentally observed $v_2$ values quantitatively.

\subsubsection{Comparison of the PHQMD \texorpdfstring{$v_2$}{v2} to the HADES data}

\begin{figure}
\includegraphics[scale=0.4]{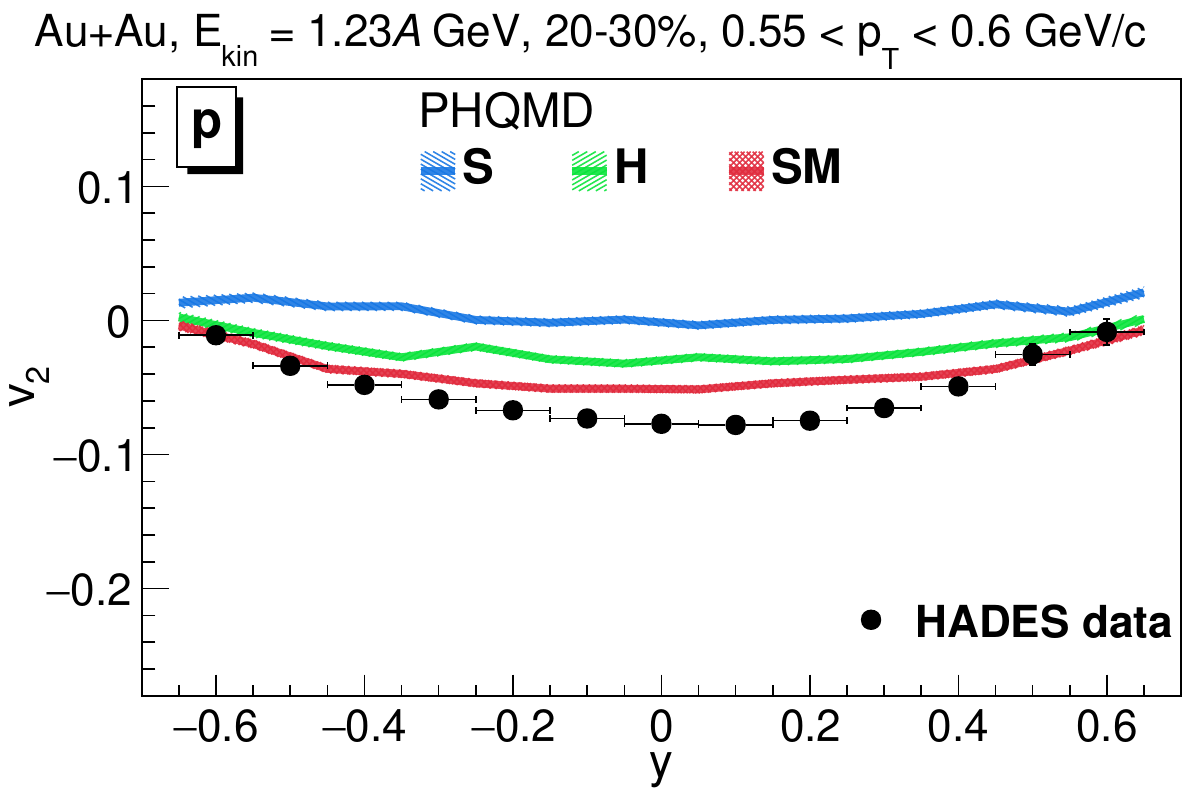}
\includegraphics[scale=0.4]{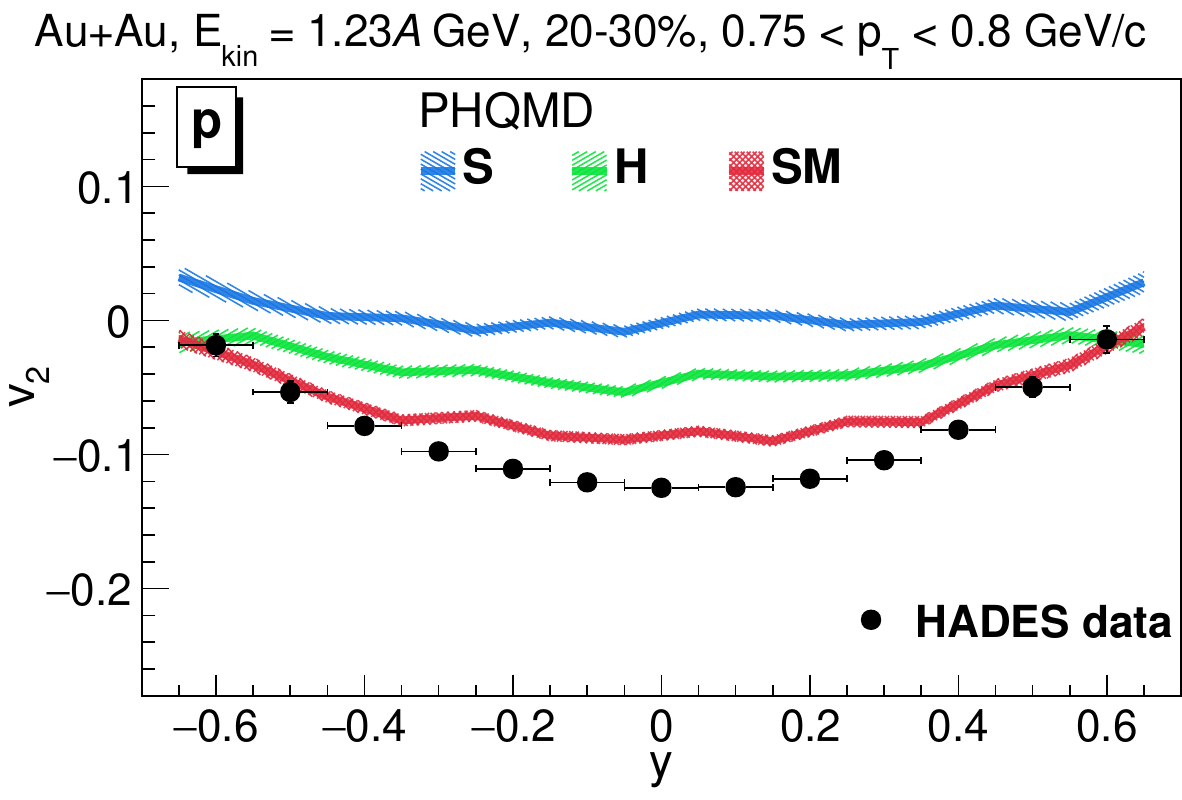}
\includegraphics[scale=0.4]{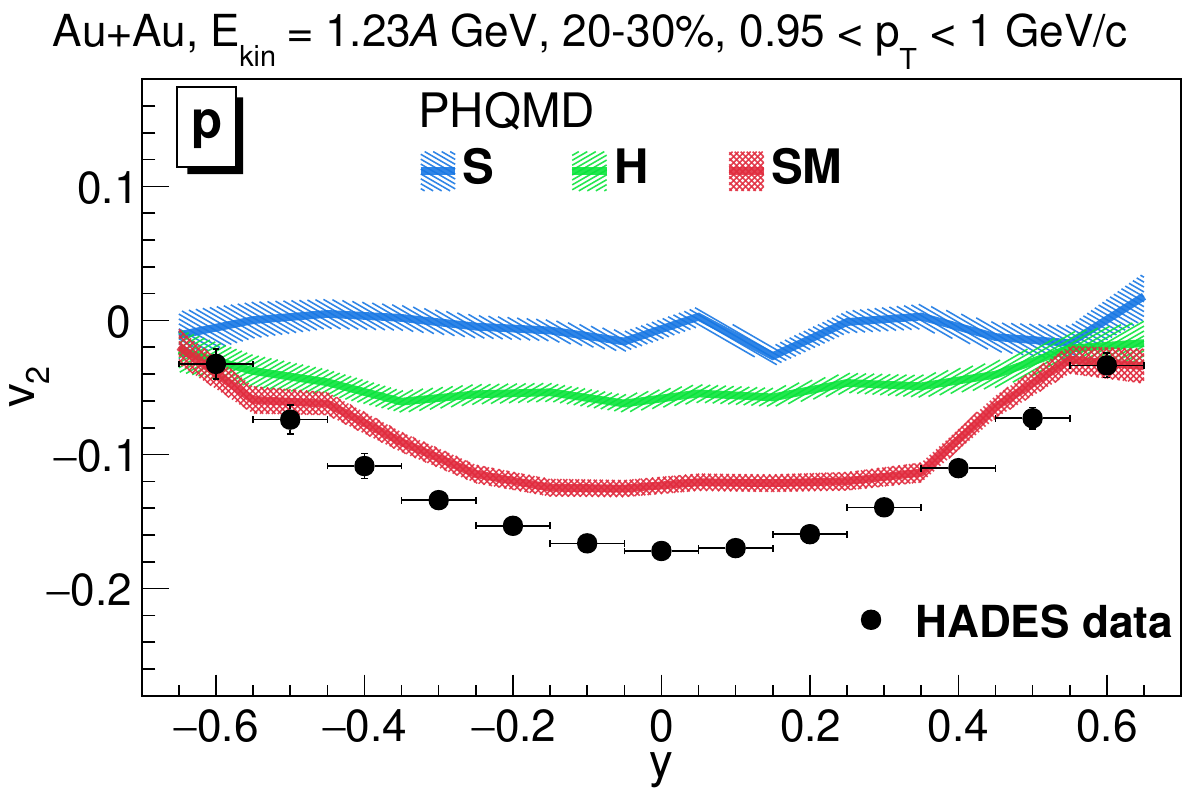}
\caption{$v_2$ of protons as a function of rapidity $y$  for 20-30\% central Au+Au collisions at $E_{kin}=1.23$ A GeV for different $p_T$ intervals: $0.55 < p_T < 0.6$ GeV/c (upper),  $0.75 < p_T < 0.8$ GeV/c (middle) and $0.95 < p_T < 1.0$ GeV/c (lower). The colour code is the same as in Fig. \ref{fig:dens}. The HADES experimental data are taken from Ref. \cite{HADES:2022osk}.} 
\label{fig:HADESv2yp}    
\end{figure}

\begin{figure*}[h!]
    \centering
\includegraphics[scale=0.4]{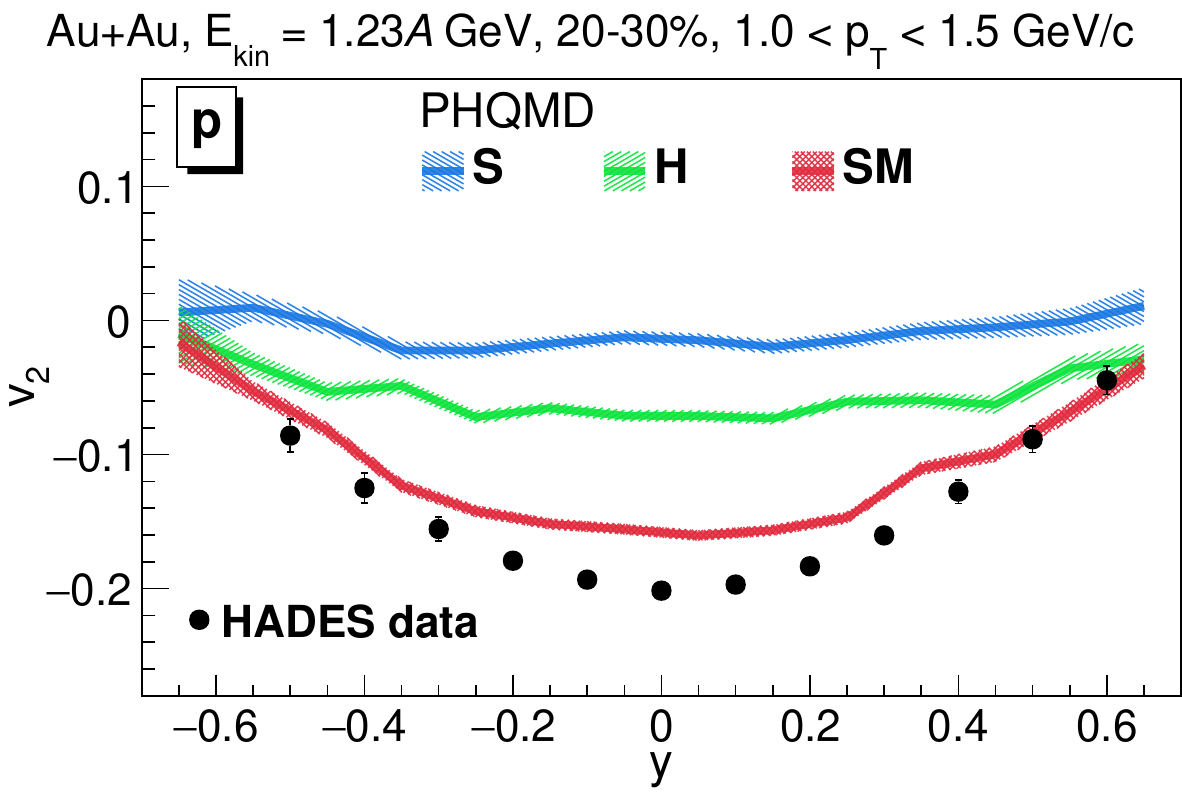}
\includegraphics[scale=0.4]{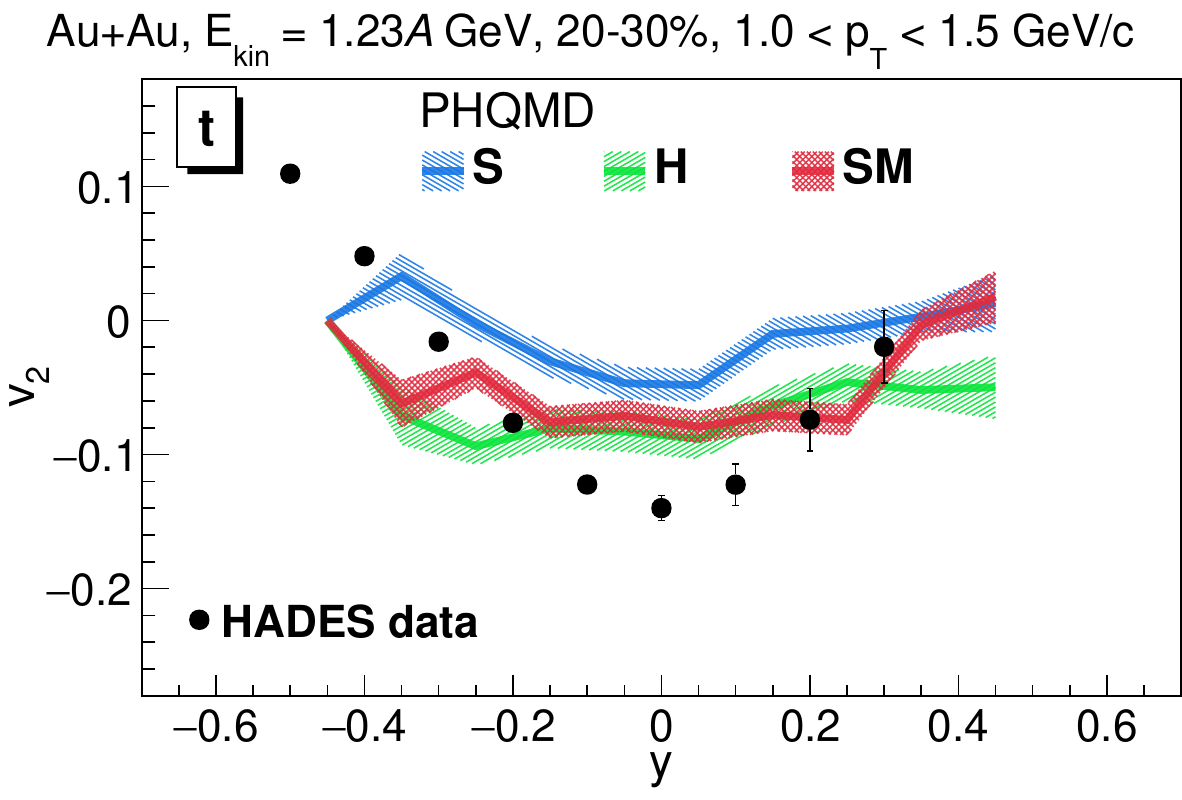}
\includegraphics[scale=0.4]{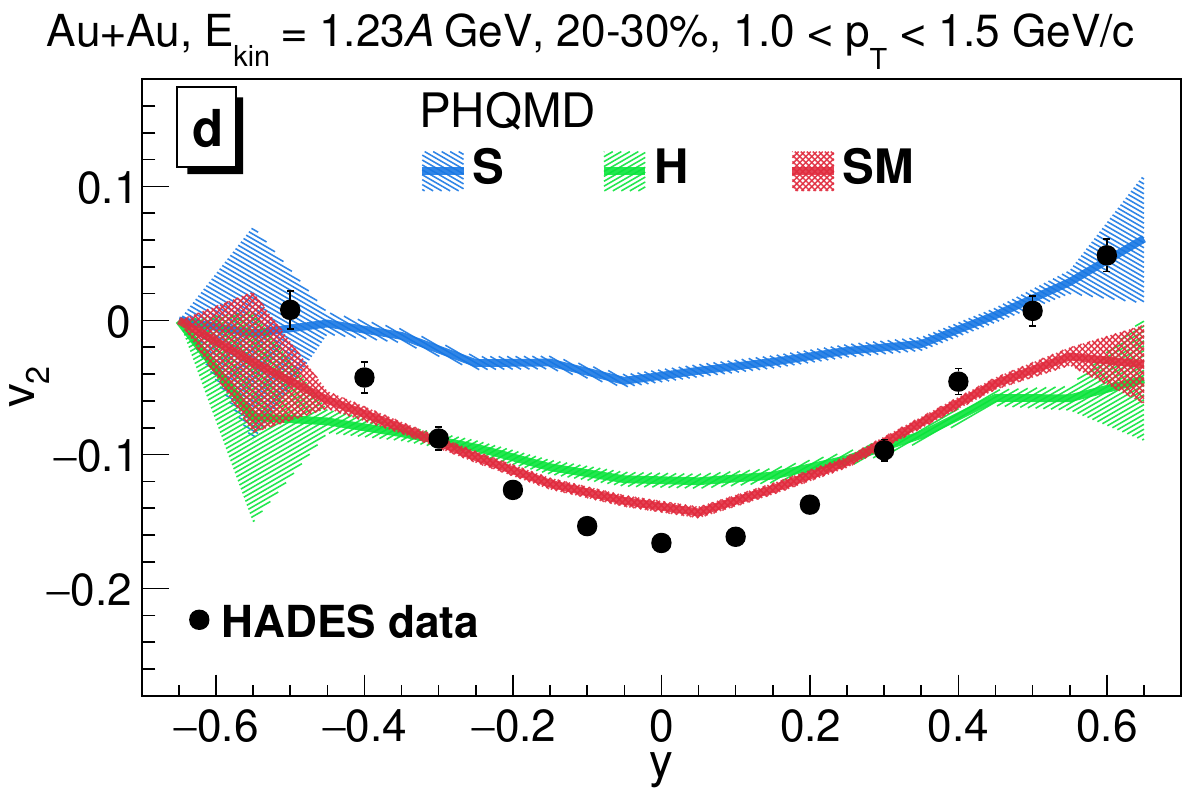}
\includegraphics[scale=0.4]{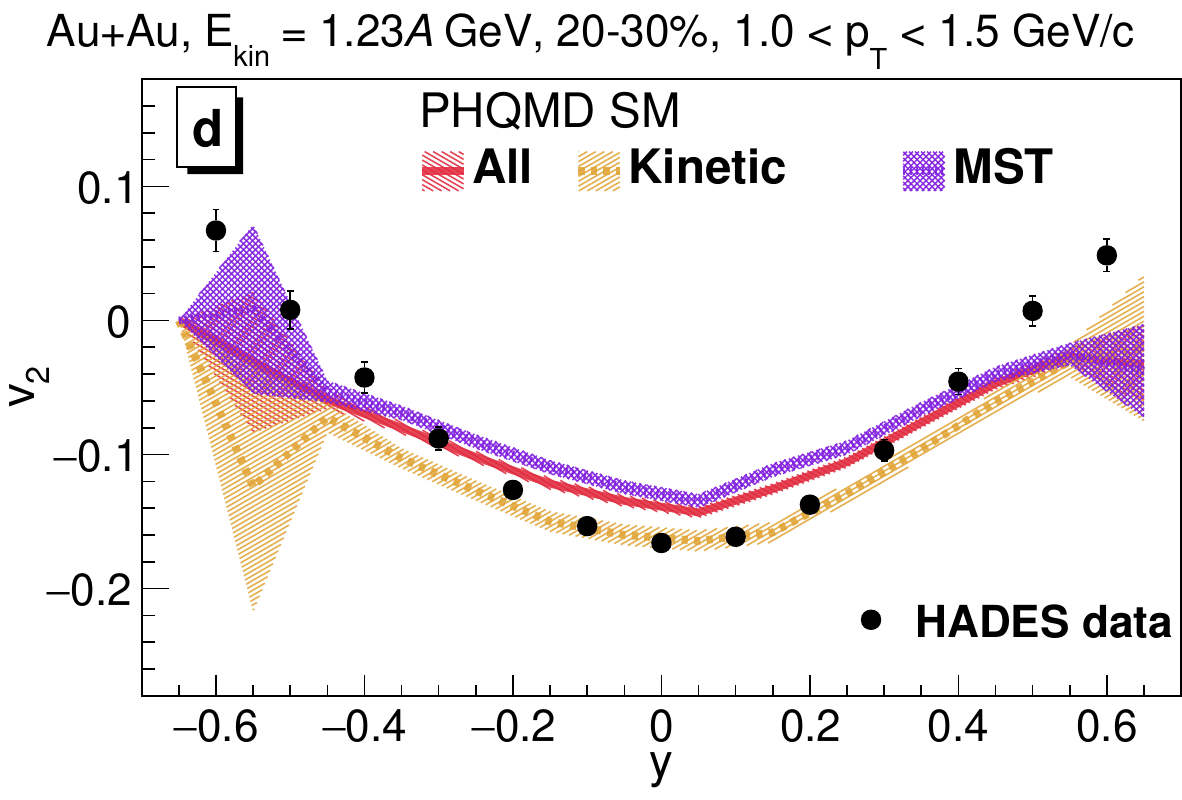}
   \caption{$v_2$ of protons (upper left) and deuterons (lower row) and tritons (upper right) as a function of rapidity for 20-30\% central Au+Au collisions at $E_{kin}=1.23$ A GeV for  $1.0 < p_T < 1.5$ GeV/c. The colour code is the same as in Fig. \ref{fig:dens}. The low right plot shows the $v_2(y)$ of deutrons for the SM EoS produced by different mechanisms: kinetic (yellow line), MST (violet line) as well as the $v_1(y)$ of all deuterons (red line, identical to the SM result on the lower left plot). The HADES experimental data are taken from Ref. \cite{HADES:2020lob}.}
    \label{fig:HADESv2yclast}      
\end{figure*}

The PHQMD results for $v_2(y)$ in comparison with HADES data \cite{HADES:2022osk} are presented in Fig. \ref{fig:HADESv2yp} for protons for 20-30\% central Au+Au collisions at $E_{kin}$=1.23 A GeV for different $p_T$ intervals: $0.55 < p_T < 0.6$ GeV/c (top),  $0.75 < p_T < 0.8$ GeV/c (middle) and  $0.95 < p_T < 1.0$ GeV/c (bottom). We obtain, in agreement with the SMASH results \cite{Mohs:2020awg}, that a soft and a hard EoS cannot reproduce the observed proton elliptic flow, whereas the momentum-dependent  interaction gives results, which come close to the HADES data  \cite{HADES:2022osk}, but still under-predict $v_2$ close to midrapidity, independent of the $p_T$ bin.

In Fig. \ref{fig:HADESv2yclast} we show  $v_2$ of protons (top left), deuterons (bottom left) and tritons (top right) as a function of rapidity for 20-30\% central Au+Au collisions at $E_{kin}=1.23$ A GeV for  $1.0 < p_T < 1.5$ GeV/c in comparison to the HADES data \cite{HADES:2020lob}. The bottom right plot shows for the SM EoS  $v_2(y)$ of deuterons produced by different mechanisms: kinetic (yellow line), MST (violet line)  as well as the combination of MST and kinetic deuterons, which is employed in the standard PHQMD calculation.  One can see that the $v_2(y)$ from kinetic and MST deuterons is rather similar (as has been already observed for  $v_1(y)$ - cf. Fig. \ref{fig:hadesv1y} ).

As seen from Fig. \ref{fig:HADESv2yclast},  the PHQMD results for protons, deuterons and tritons with the soft EoS substantially underestimate $v_2(y)$ at this interval of large $p_T$.
The hard EoS leads to an underestimation of the proton $v_2$ (similar to the results in Fig. \ref{fig:HADESv2yp} for the intervals of smaller $p_T$, while the SM EoS gives a  good description of the HADES data for the proton $v_2$, deuterons and tritons, which agree within 10\% with the experimental data.  For the latter two  $v_2(y)$ are rather similar for the SM and hard EoS. 
This result agrees with the UrQMD calculations with a static hard EoS in Ref. \cite{Hillmann:2019wlt}. This is not very surprising because it is known in literature that a static hard and a momentum dependent soft equation-of-state give similar results for the flow observables \cite{Aichelin:1987ti}.  It is in tension with the SMASH results \cite{Mohs:2020awg,Mohs:2024gyc,Tarasovicova:2024isp}, where a hard EoS gives a larger $v_1$ than observed experimentally. We note that in UrQMD and SMASH deuterons are produced by coalescence, however in a different way, see section II.B of Ref. \cite{Mohs:2024gyc}.

\begin{figure*}
\includegraphics[scale=0.4]{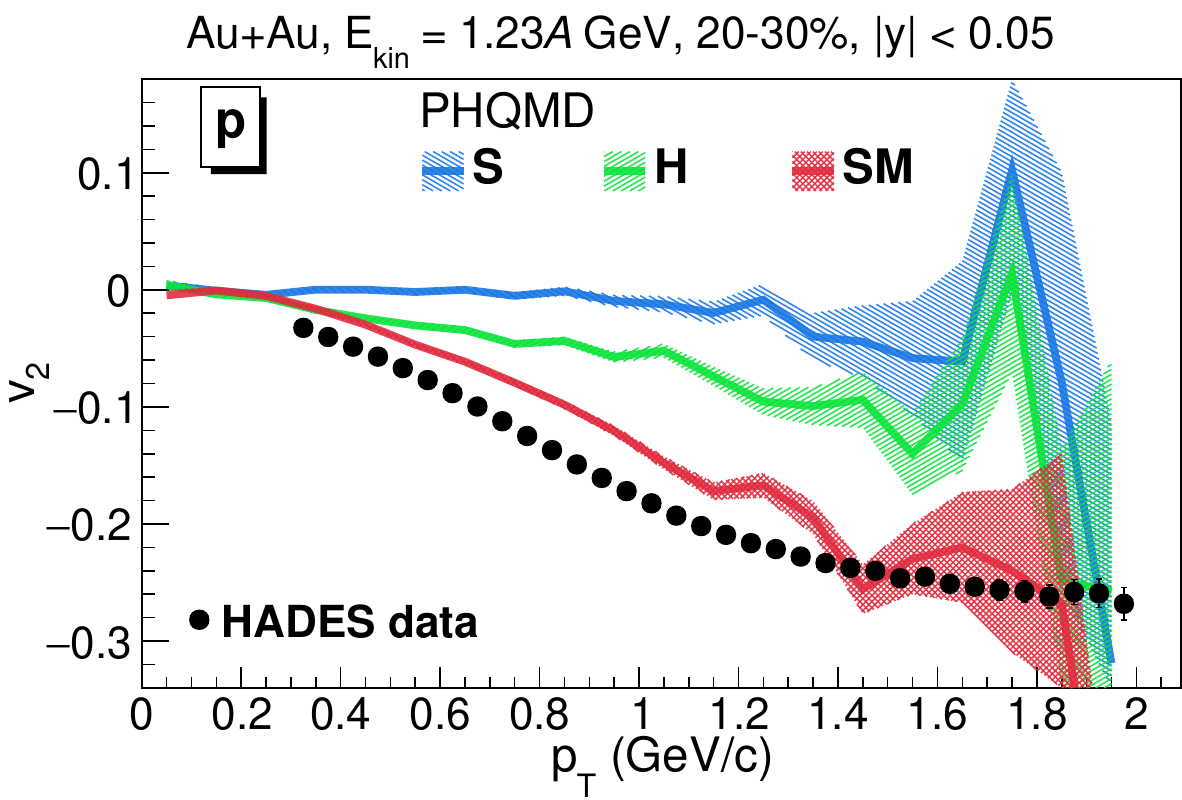}
\includegraphics[scale=0.4]{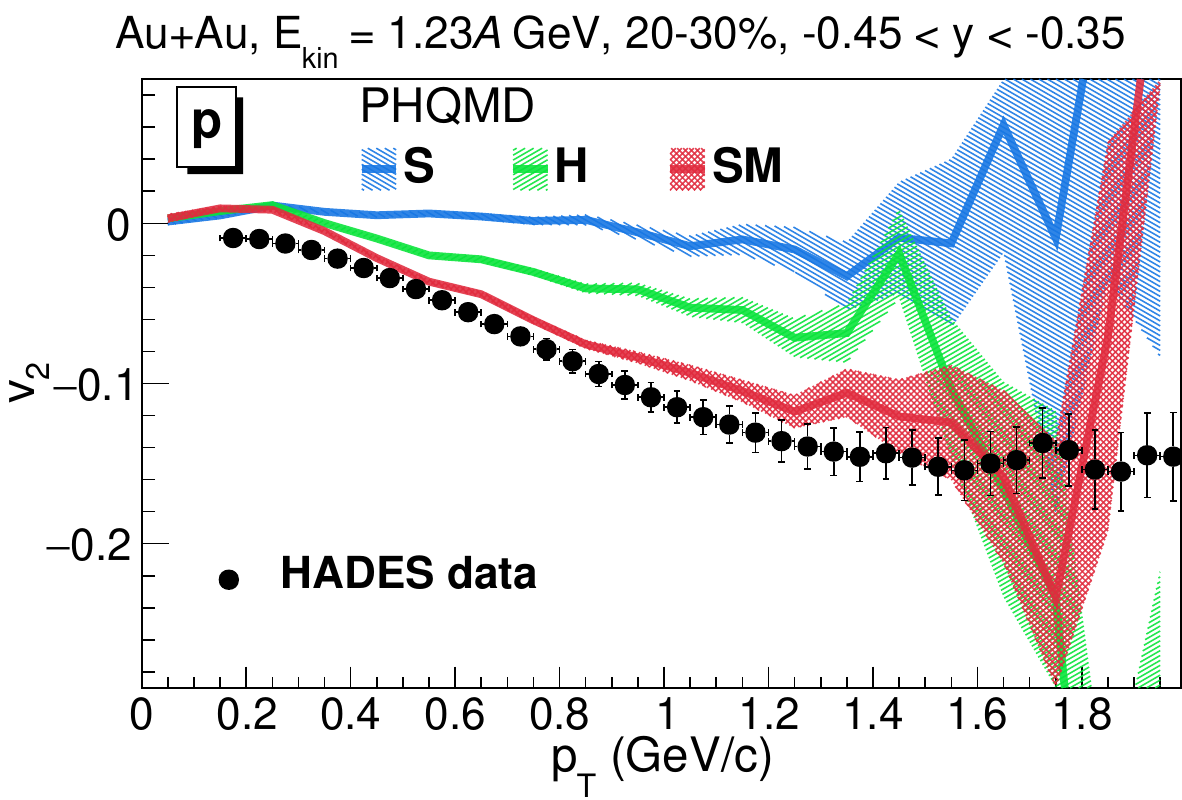}
\includegraphics[scale=0.4]{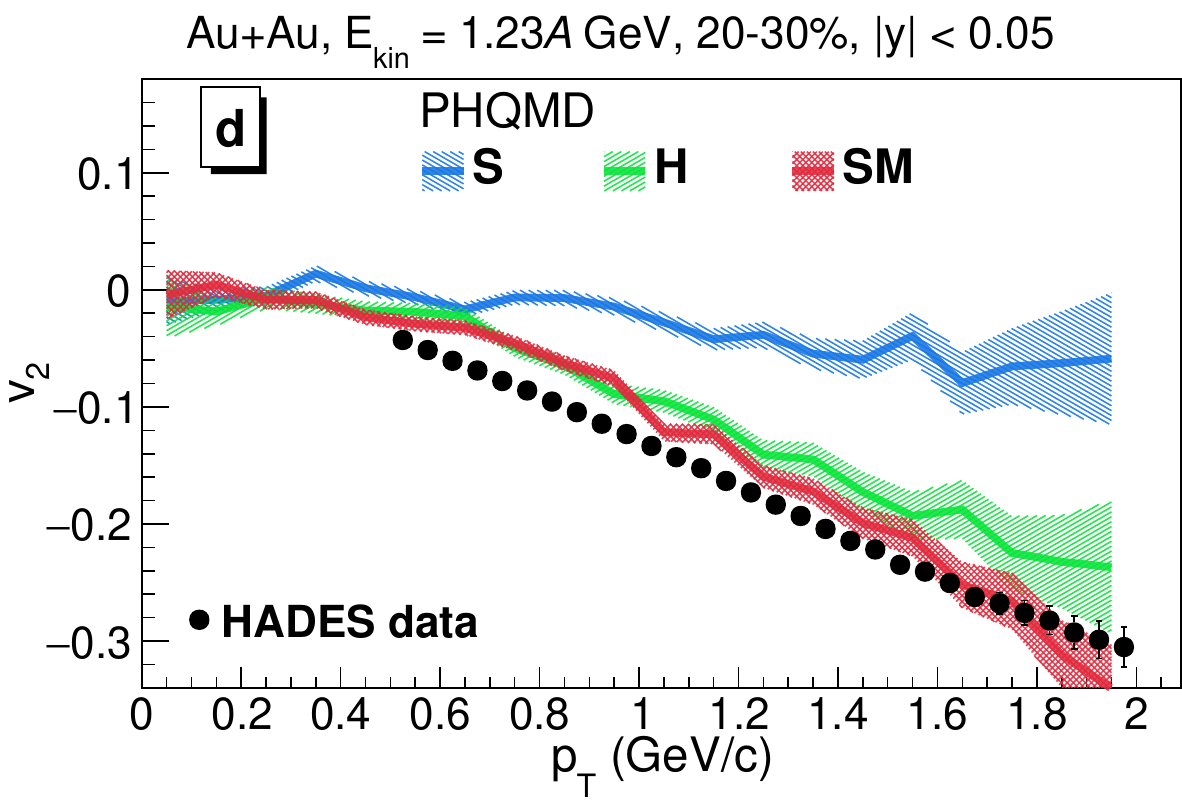}
\includegraphics[scale=0.4]{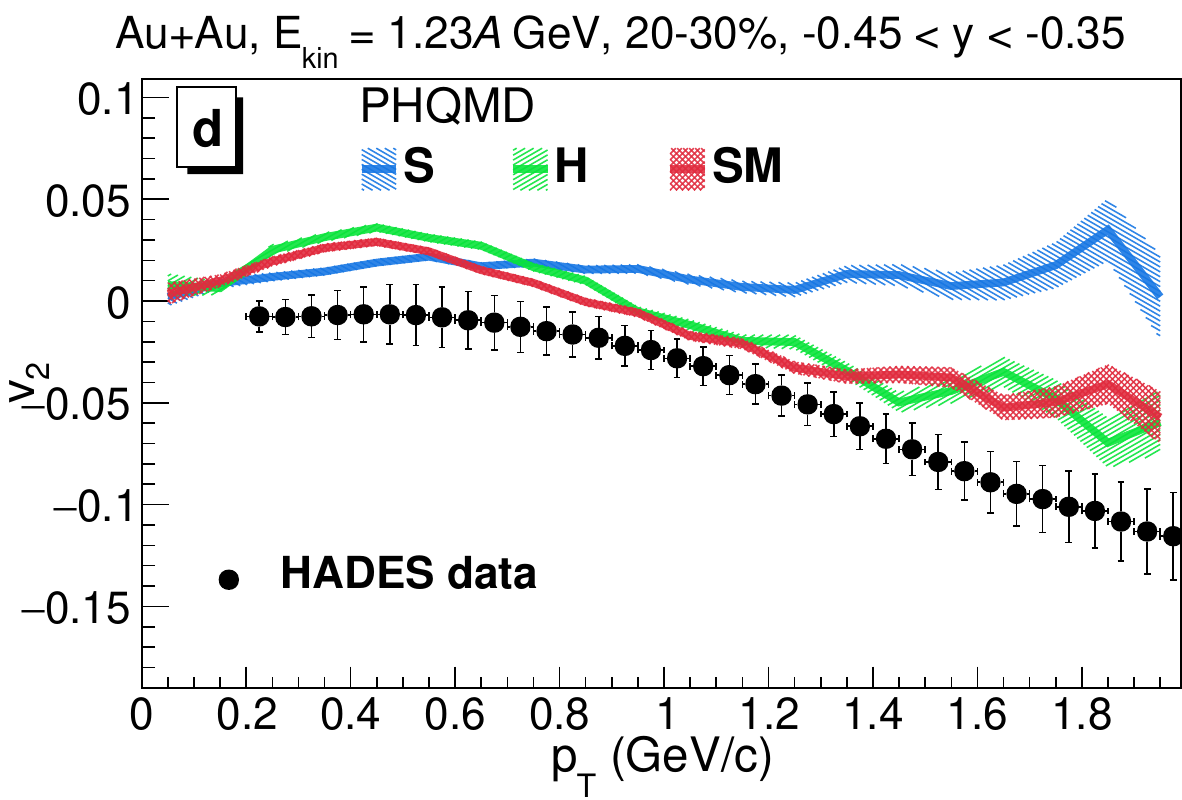}
\includegraphics[scale=0.4]{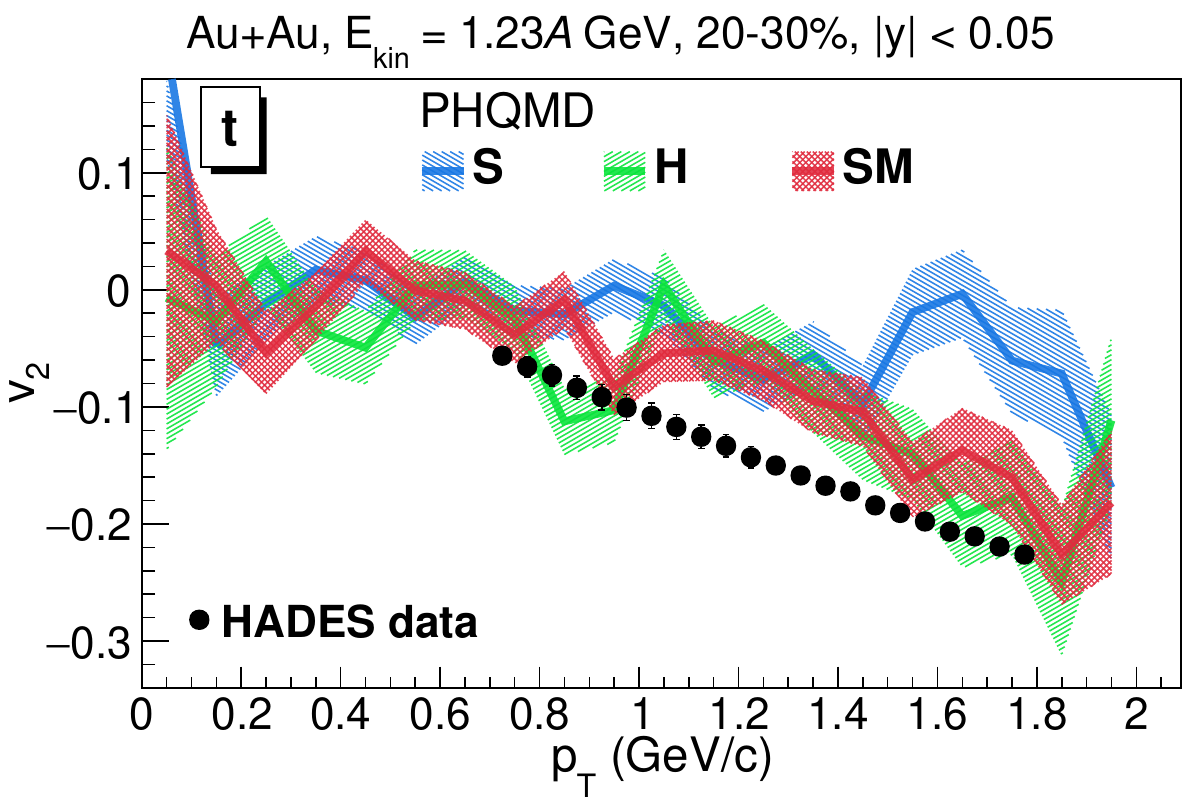}  
\includegraphics[scale=0.4]{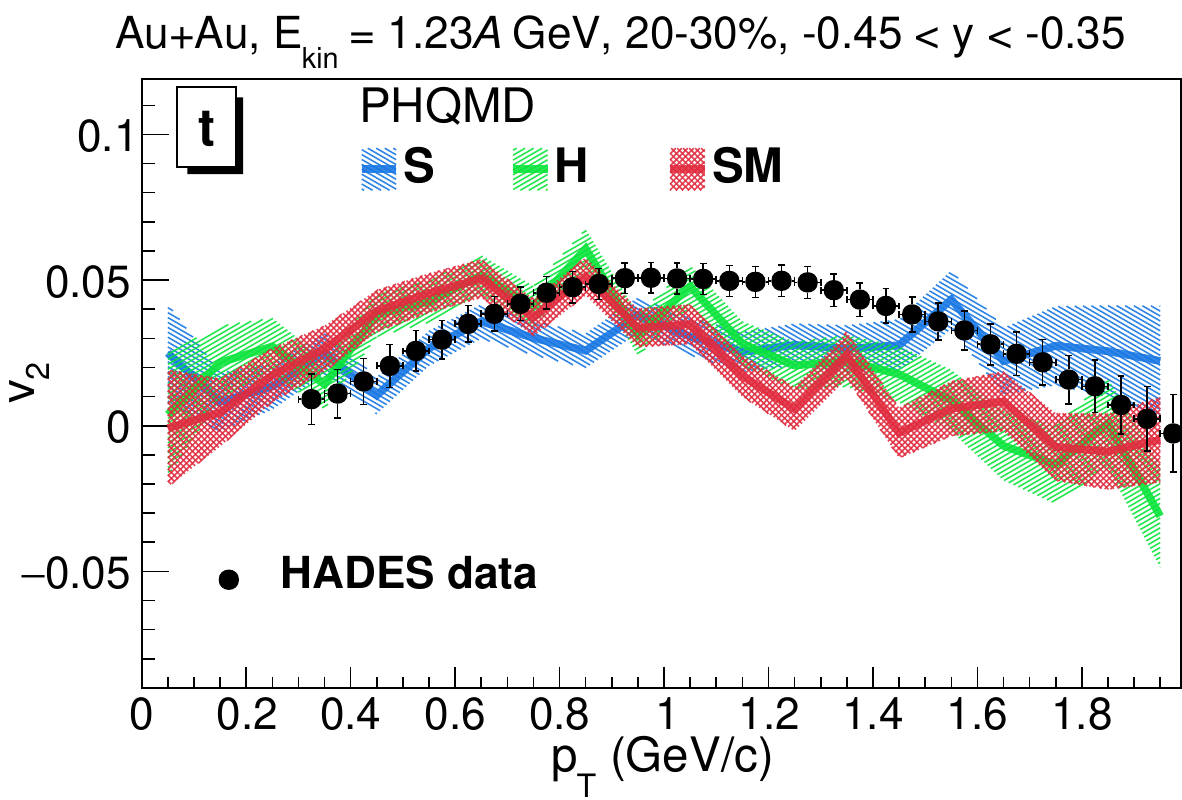}
\caption{ $v_2$ of protons (upper row), deuterons (middle row) and triton (lower row) as a function of $p_T$ for rapidity intervals $|y|<0.05$ (left column) and  $-0.45 < y <-0.35$ (right column) for 20-30\% central Au+Au collisions at $E_{kin}$=1.23 A GeV. The colour code is the same as in Fig. \ref{fig:dens}. The HADES experimental data are taken from Ref. \cite{HADES:2020lob,HADES:2022osk}.}
\label{fig:v2pt_HADES} 
\end{figure*}

The transverse momentum distribution of $v_2$ is presented in Fig. \ref{fig:v2pt_HADES} for protons (top), deuterons (middle) and tritons (bottom) as a function of $p_T$ for  rapidity intervals $|y|<0.05$ (left column) and  $-0.45 < y <-0.35$ (right column) for  20-30\% central Au+Au collisions at $E_{kin}$=1.23 A GeV. First of all we observe a strong $p_T$ dependence  of $v_2(p_T)$ and a strong dependence on EoS for all clusters and both rapidity intervals. For  both rapidity intervals  the soft EoS leads to a substantial underestimation of the HADES data on $v_2(p_T)$ for protons,  deuterons and tritons and the deviation grows with increasing $p_T$.  For the hard EoS the HADES data of the proton $v_2$ at high $p_T$ are not reproduced, what is in line with  SMASH results \cite{Mohs:2020awg, Mohs:2024gyc}. The soft momentum-dependent EoS brings, however, $v_2$ close to the experimental data.  Whereas for protons $v_2(p_T)$ is different for a H and a SM EoS, for  deuterons and tritons they yield almost the same result, at forward as well as at mid rapidity.  A hard EoS gives deuterons more $v_2$ than protons and brings the calculations closer to the experimental data. This has also been observed in UrQMD calculations in Ref. \cite{Hillmann:2019wlt}.
It is interesting to note that the functional form of the experimental $v_2(p_T)$ for protons, deuterons and tritons is different, especially in the $-0.45 < y < -0.35$ interval. The PHQMD calculations reproduce qualitatively the forms as well as the values of the measured $v_2(p_T)$.

\begin{figure}
\includegraphics[scale=0.4]{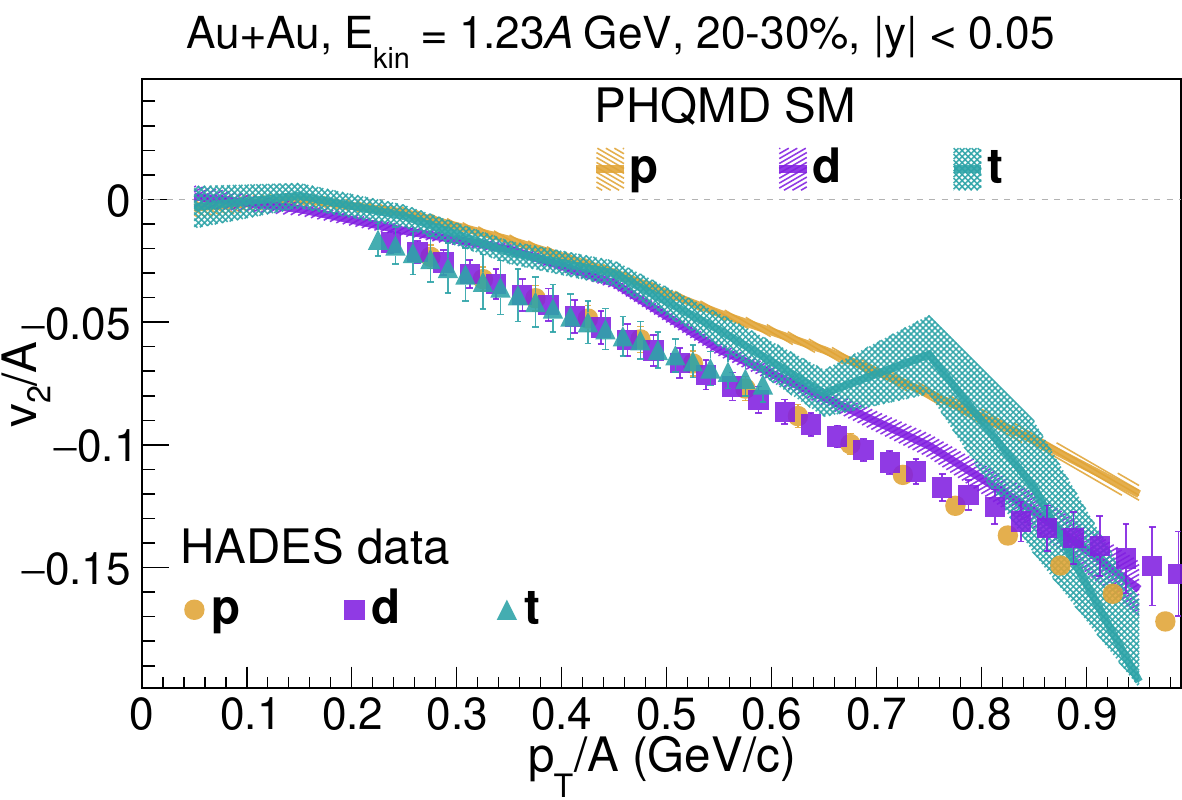}
\caption{The elliptic flow scaled with atomic number $v_2/A$ versus scaled $p_T/A$ for protons (ochre), deuterons (violet) and tritons (dark celeste) for 20-30\% central Au+Au collisions for $|y|<0.05$ at $E_{kin}$=1.23 A GeV. The HADES experimental data are taken from Ref. \cite{HADES:2020lob} and scaled with $A$.}
\label{fig:v2pt_HADES_scaling} 
\end{figure}

Fig. \ref{fig:v2pt_HADES_scaling} presents the elliptic flow for a SM EoS, scaled by the atomic number - $v_2/A$ , as a function of the transverse momentum of a nucleon in the cluster  $p_T/A$ for protons, deuterons and tritons at midrapidity $|y|<0.05$.  The scaled $v_2/A(p_T/A)$ is almost independent of $A$. This scaling behavior of $v_2/A (p_T/A)$ has been observed by the HADES Collaboration \cite{BKardan,HADES:2022osk}. For the deuterons or tritons this means that their azimuthal distribution  can be obtained  by assuming that two or three nucleons, with an azimuthal distribution as seen for the protons, combine without any further assumptions to a deuteron and, correspondingly, to larger clusters. We note that a similar scaling behavior of $v_2$ with the number of "constituents" has been predicted  in Ref. \cite{Fries:2003kq}. There $v_2$ of mesons and baryons scales with the number of valence quarks (i.e. 2 for mesons and 3 for baryons). This 'constituent quark number scaling' has been observed experimentally in central Au+Au collisions at RHIC \cite{STAR:2003wqp,PHENIX:2006dpn}. 
As follows from Fig. \ref{fig:v2pt_HADES_scaling}, the PHQMD calculations show that in this scaled presentation of $v_2$ at midrapidity the difference between the $v_2(p_T)$ distributions for protons and clusters is reduced substantially but we do not observe a perfect scaling, as the data show for $p_T<0.8$ GeV$/c$ (we note also that we have a limited statistics in our simulations). Moreover, the deviation from perfect scaling increases for large $p_T$. 
Our $v_2$ for the SM EoS is also lower than that observed experimentally  - in line with
Fig. \ref{fig:v2pt_HADES}.

\subsubsection{Comparison of the PHQMD \texorpdfstring{$v_2$}{v2} to the FOPI data}

\begin{figure}
    \centering
\includegraphics[scale=0.4]{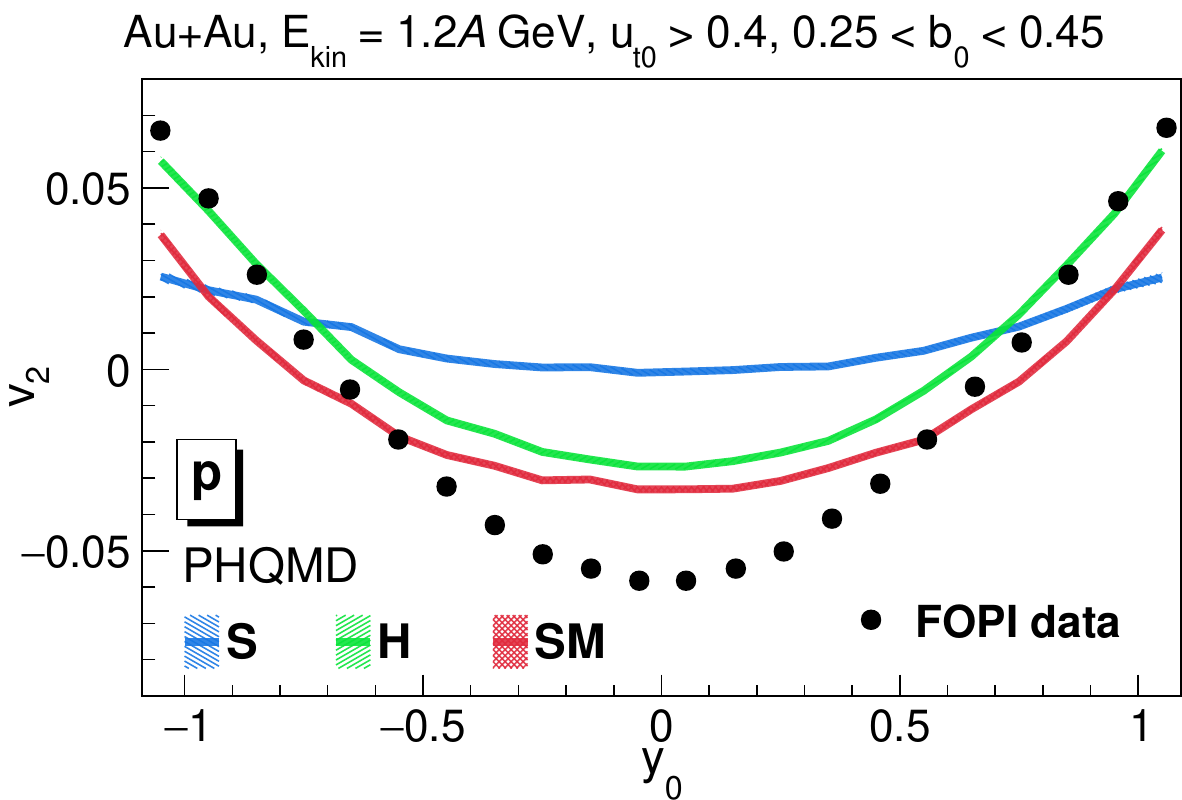}
\includegraphics[scale=0.4]{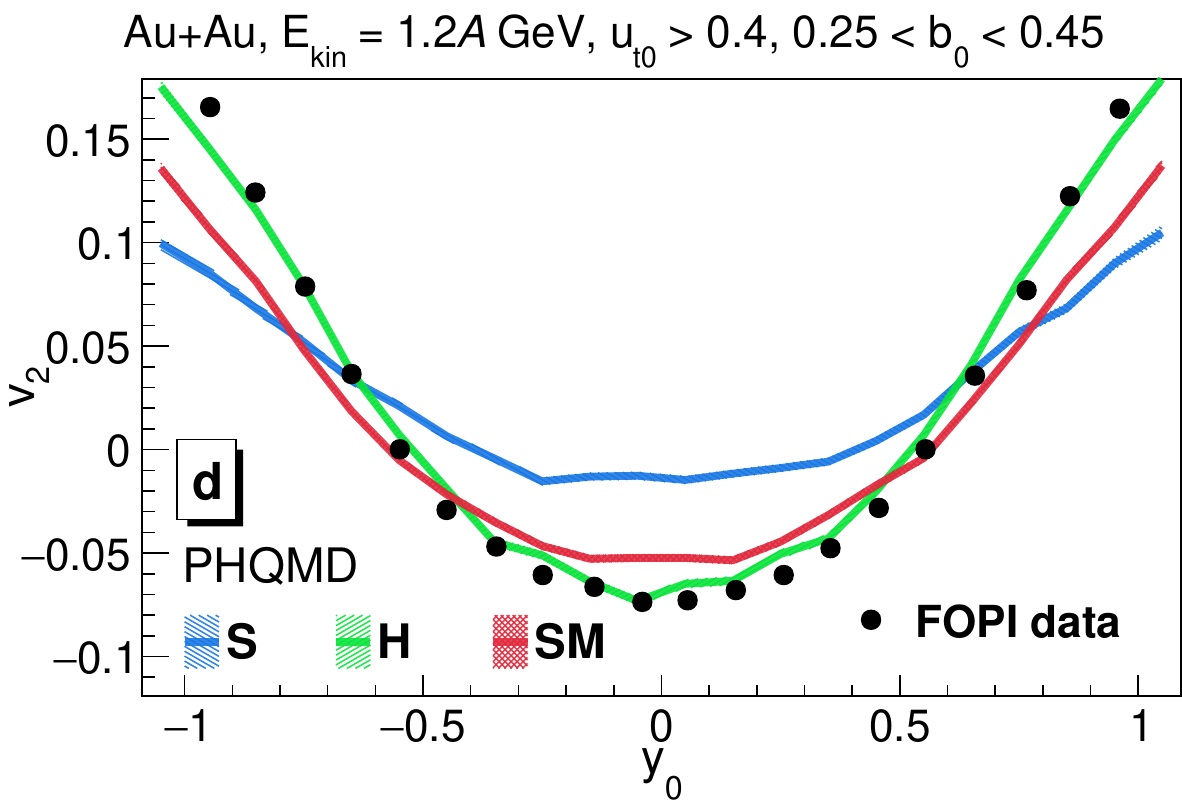}
\includegraphics[scale=0.4]{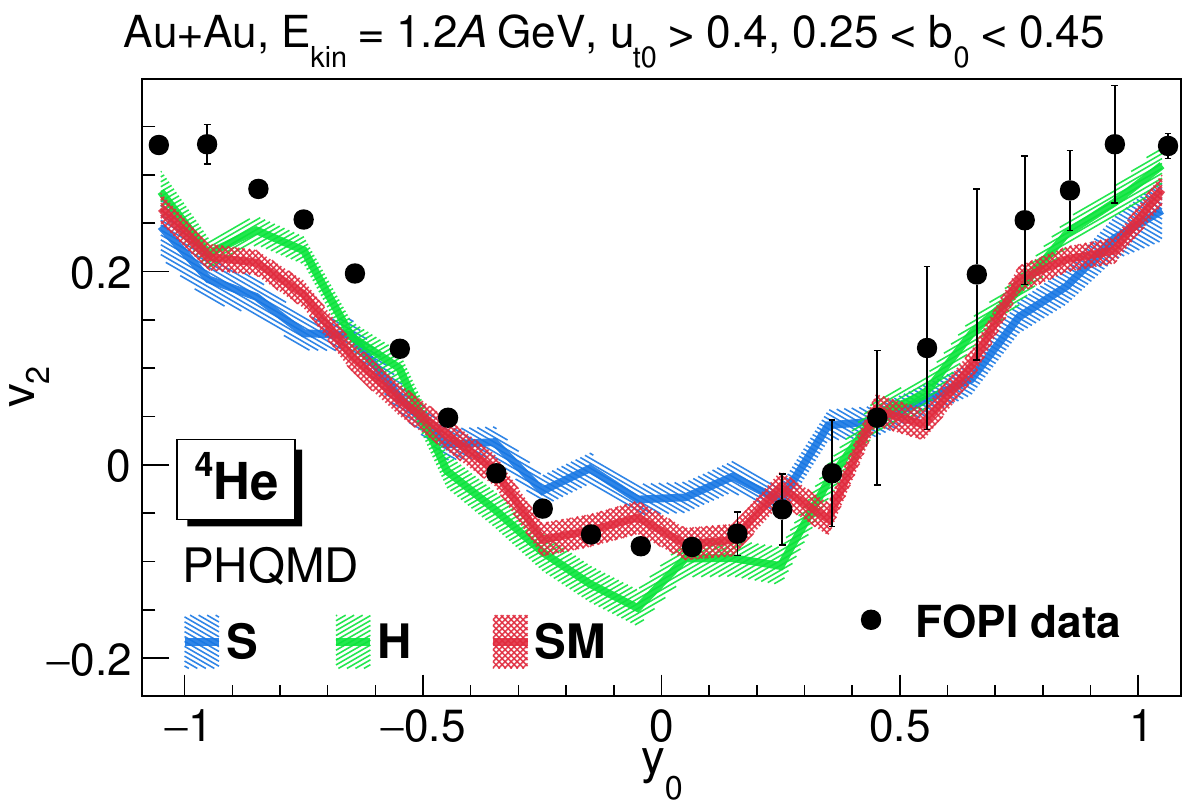}
\caption{$v_2$ of protons (upper), deuterons (middle) and $^4$He (lower) as a function of the scaled rapidity $y_0$ for  Au+Au collisions at $E_{kin}$=1.2 A GeV for $u_{t0} >0.4$ and the impact parameter range $0.25<b_0<0.45$. The colour code is the same as in Fig. \ref{fig:dens}. The FOPI experimental data are taken from Ref. \cite{FOPI:2011aa}.}
\label{fig:FOPI12v2y0}
\end{figure}

\begin{figure*}
\includegraphics[scale=0.4]{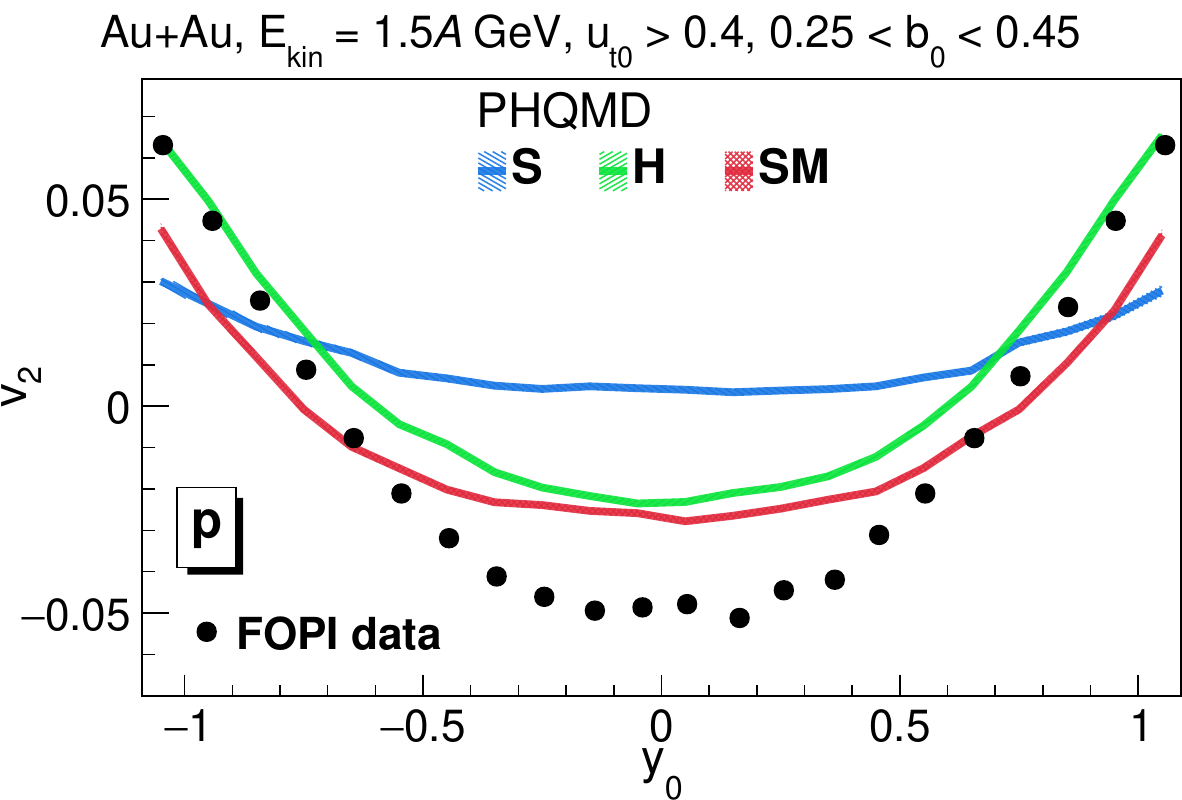}
\includegraphics[scale=0.4]{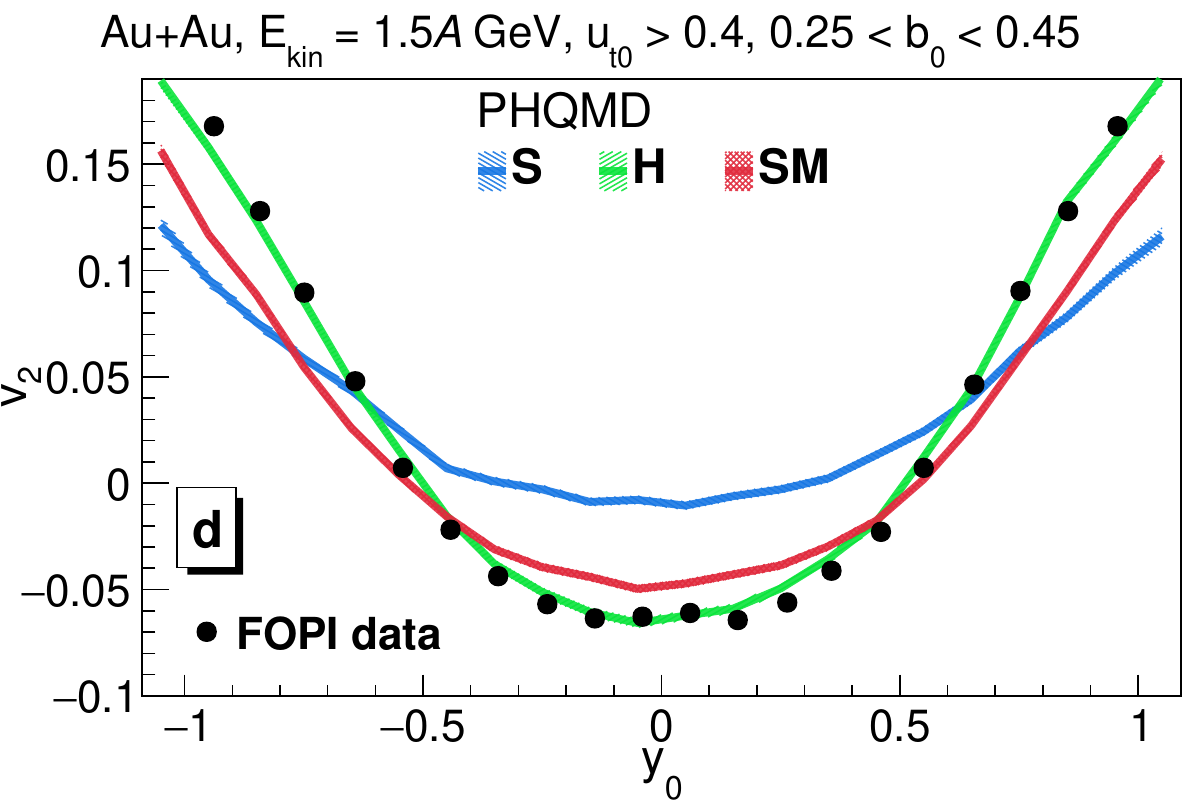}
\includegraphics[scale=0.4]{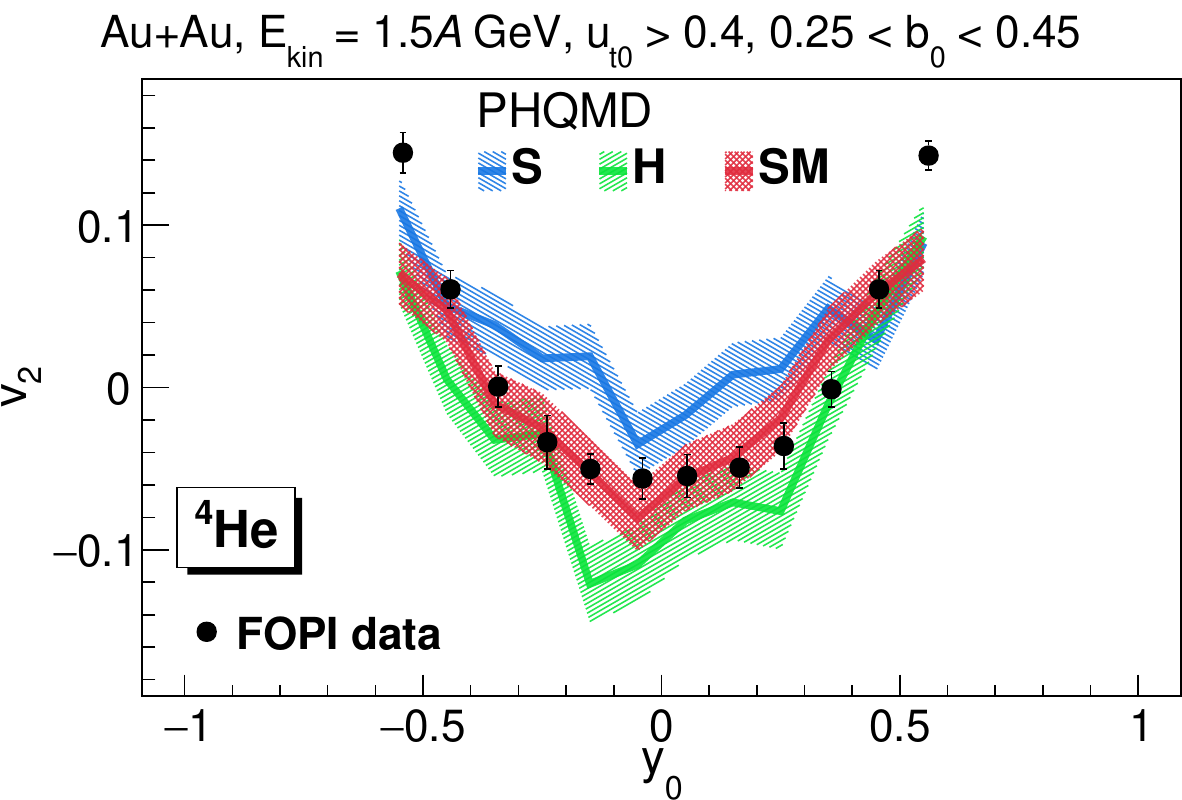}
\includegraphics[scale=0.4]{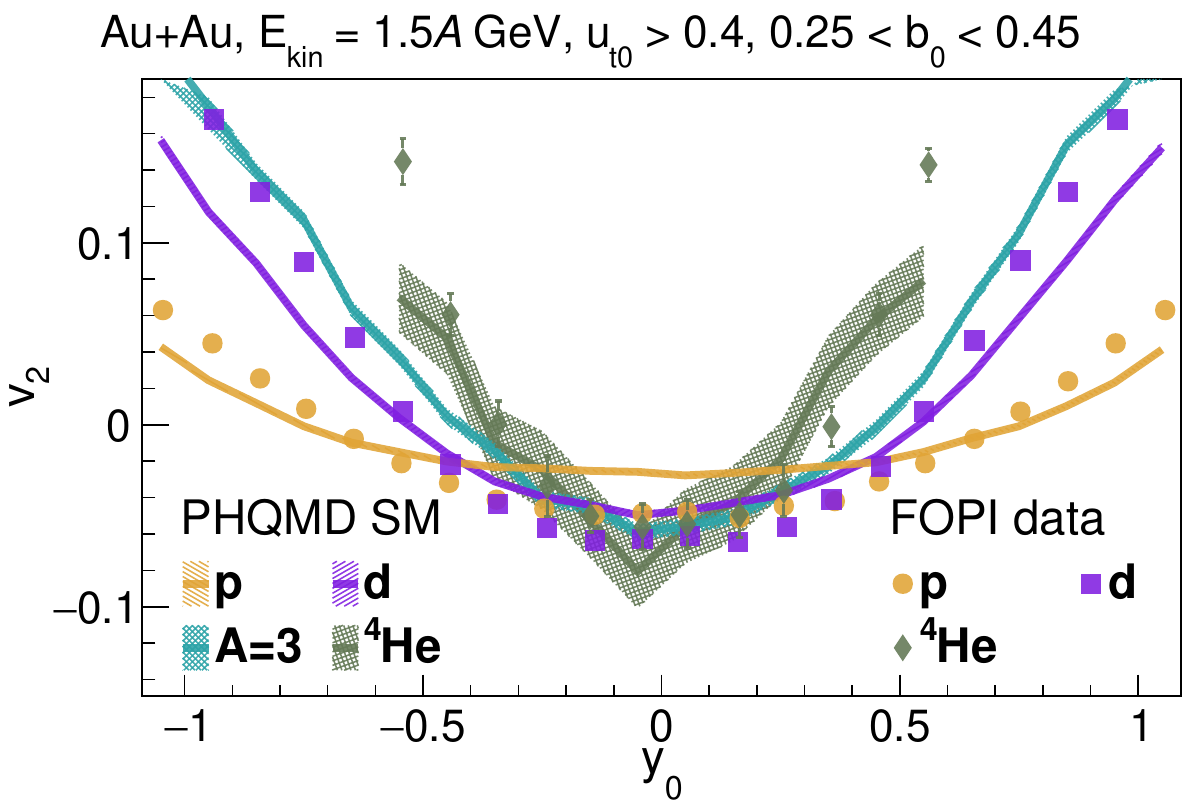}
\caption{$v_2$ of protons (upper left), deuterons  (upper right) and $^4$He (lower left) as a function of the scaled  rapidity $y_0$ for Au+Au collisions at $E_{kin}$=1.5 A GeV for $u_{t0} >0.4$ and the impact parameter range $0.25<b_0<0.45$. The lower right plot  shows the compilation of $v_2(y_0)$ for protons, deuterons, $A=3$ clusters and $^4$He for the SM EoS. The colour code is the same as in Fig. \ref{Fig:FOPI12y0}. The FOPI experimental data are taken from Ref. \cite{FOPI:2011aa}.}
  \label{fig:FOPI15v2y0}
\end{figure*}

We step now to a comparison of the PHQMD results with the experimental data from the FOPI Collaboration \cite{FOPI:2011aa}.
To allow for a better comparison with the HADES data we have changed the representation of $v_2$ from the FOPI Collaboration by plotting $v_2$ instead of $-v_2$, as done in Ref. \cite{FOPI:2011aa}. 
In Figs. \ref{fig:FOPI12v2y0} and \ref{fig:FOPI15v2y0} we show the elliptic flow  $(v_2)$ 
of protons, deuterons  and $^4$He  as a function of the scaled rapidity $y_0$ for  Au+Au collisions at $E_{kin}$=1.2 and 1.5 A GeV for $u_{t0} >0.4$ and the impact parameter range
$0.25<b_0<0.45$ for  three EoS.
One can see that in the considered $u_{t0}$ interval for the both energies  $v_2$ of protons is underestimated for all 3 EoS  as it was the case for the HADES data.  The SM EoS gives the largest $v_2$, followed by the hard EoS. The lowest elliptic flow comes from the soft EoS.  For deuterons and tritons the hard EoS gives slightly larger $v_2$ values than the SM EoS while the soft EoS gives again the lowest contribution.  The deuterons are best described by a H EoS whereas the triton data are very close to a SM EoS.

In  the lower right plot of Fig. \ref{fig:FOPI15v2y0} we compile $v_2(y_0)$ of the experimental results  and of the theoretical prediction with a SM EoS for protons, deuterons, $A=3$ clusters and $^4$He. This shows that the rapidity distributions of $v_2$ for the different cluster differ significantly. This is qualitatively reproduced by the PHQMD calculations, for the clusters even better than for free protons.    

\begin{figure*}
    \centering
\includegraphics[scale=0.4]{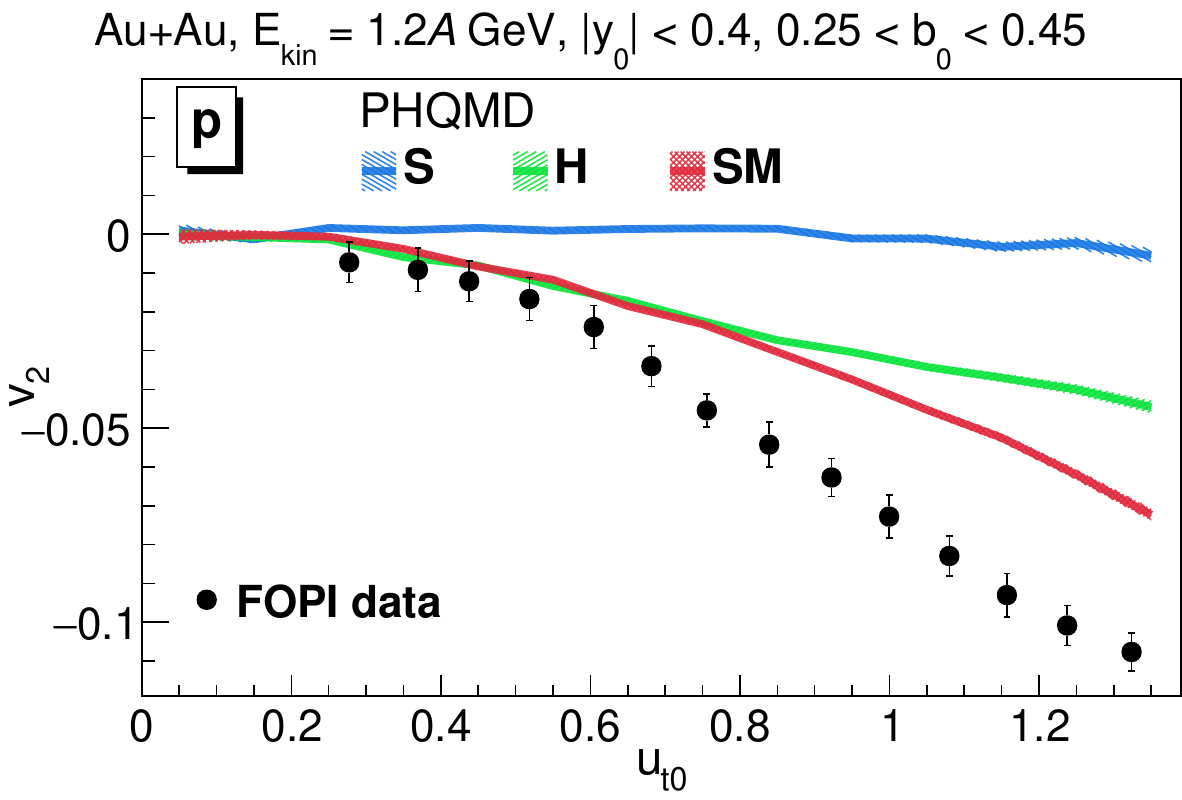}
\includegraphics[scale=0.4]{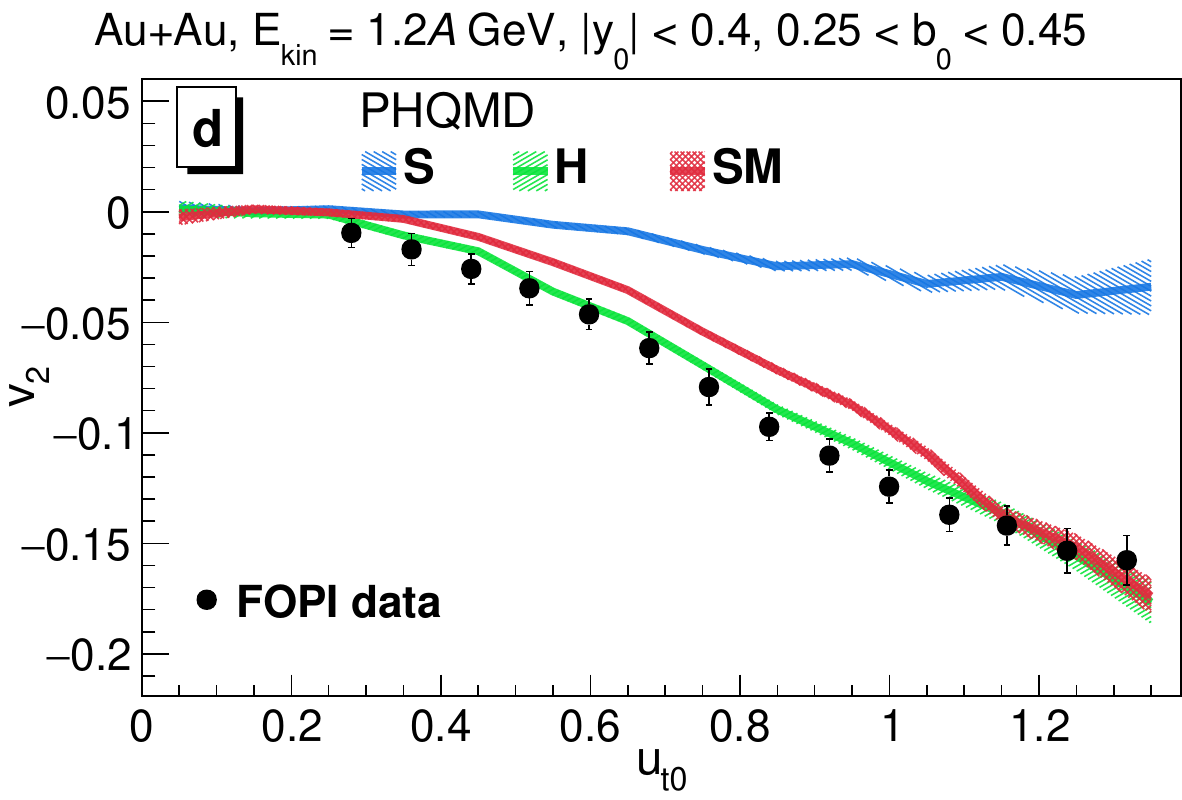}
\includegraphics[scale=0.4]{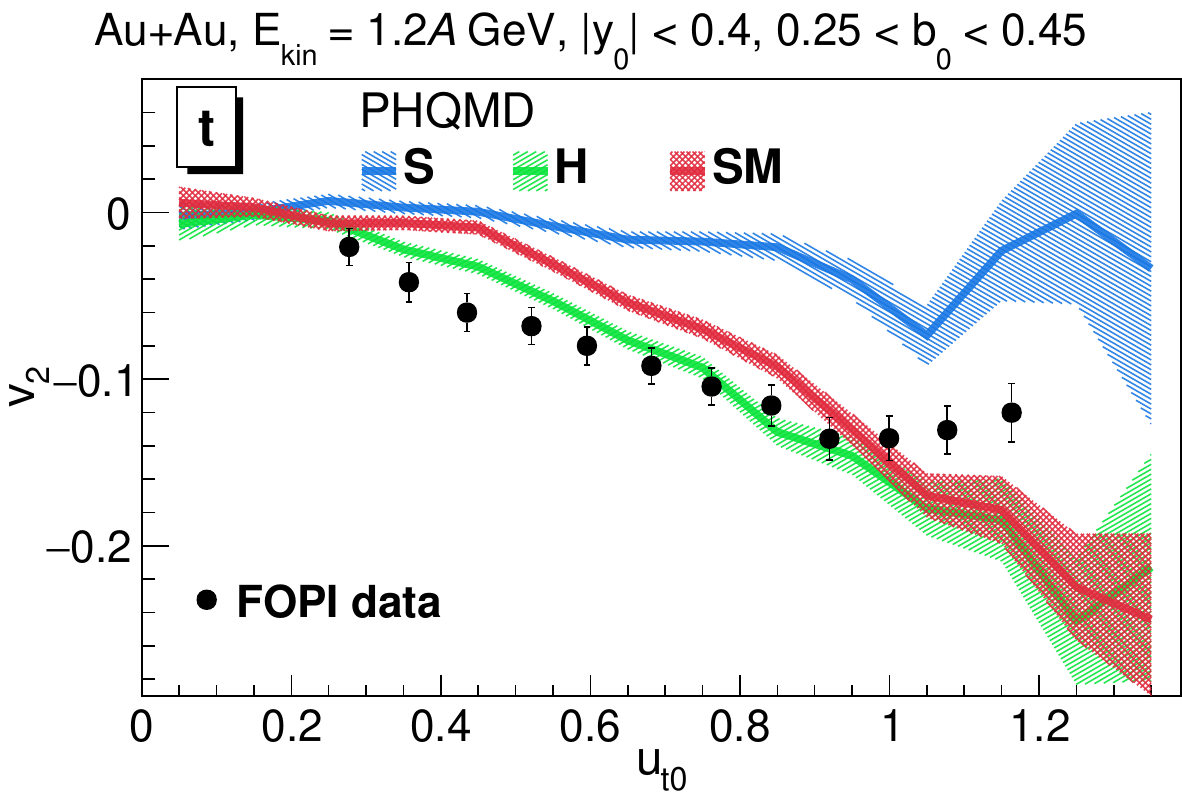}
\includegraphics[scale=0.4]{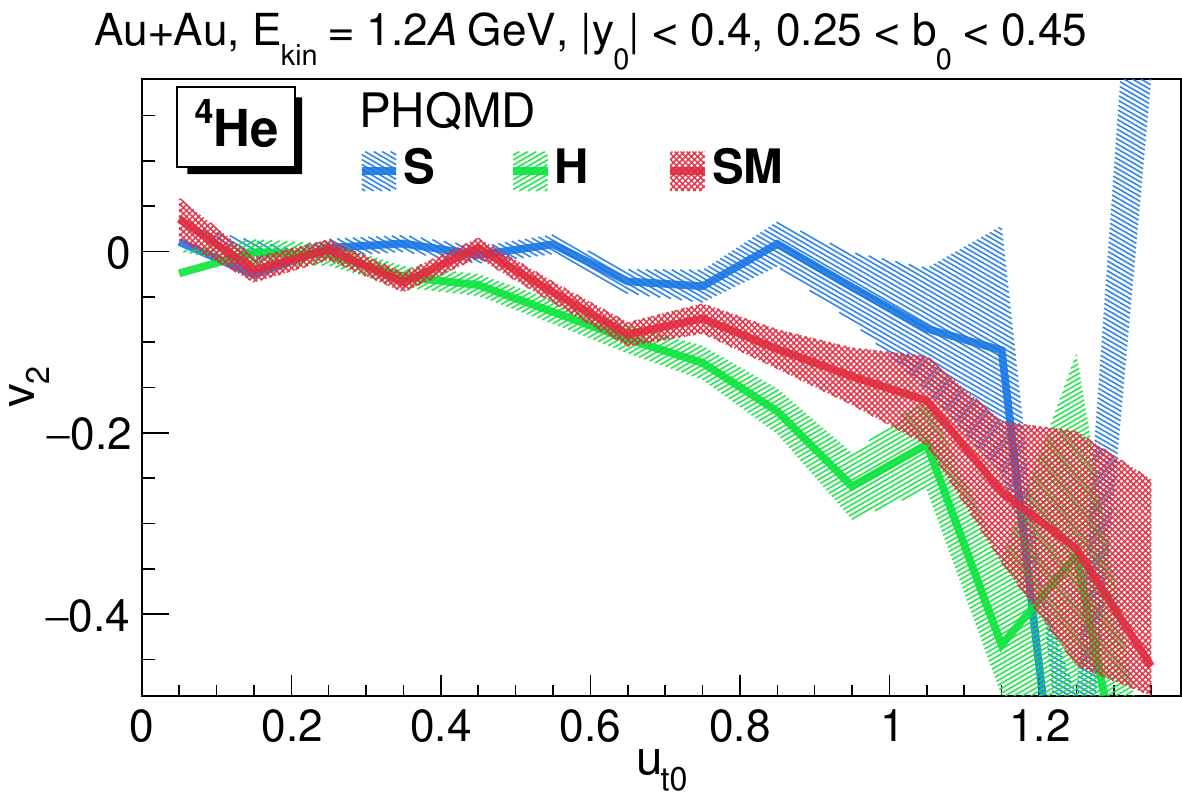}
\includegraphics[scale=0.4]{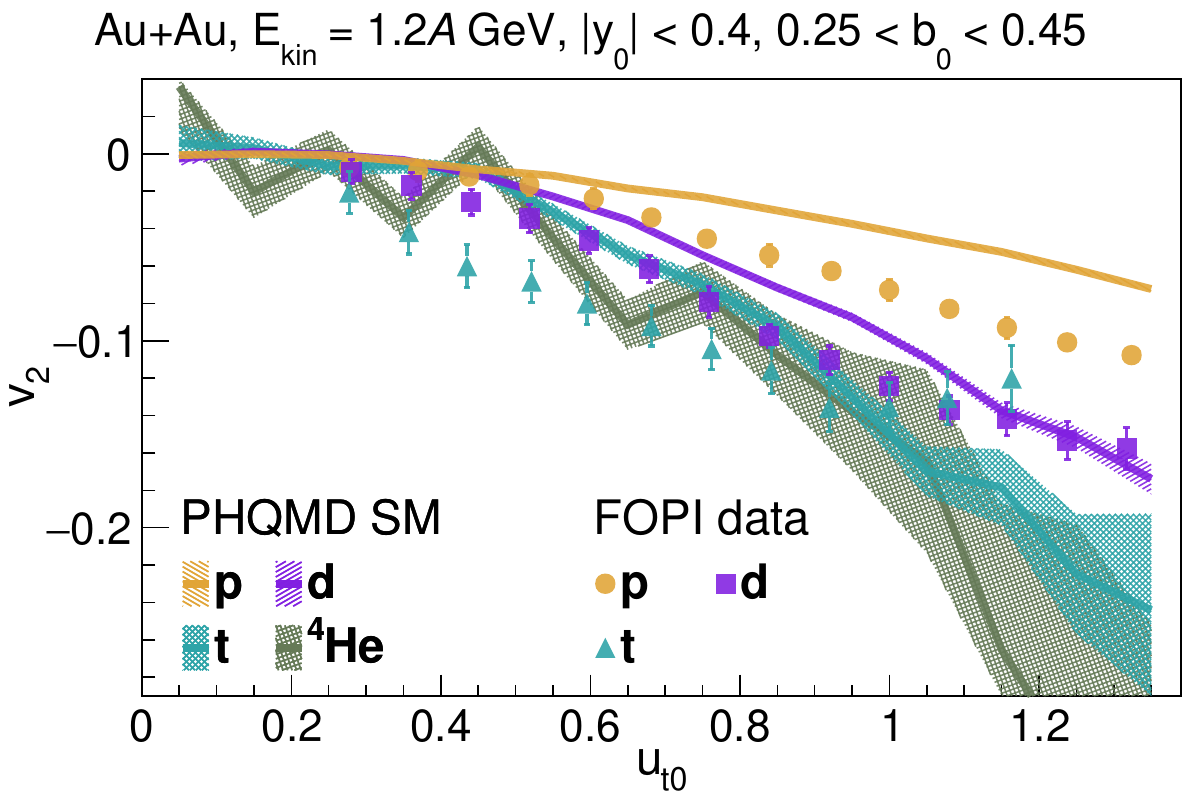}  
\includegraphics[scale=0.4]{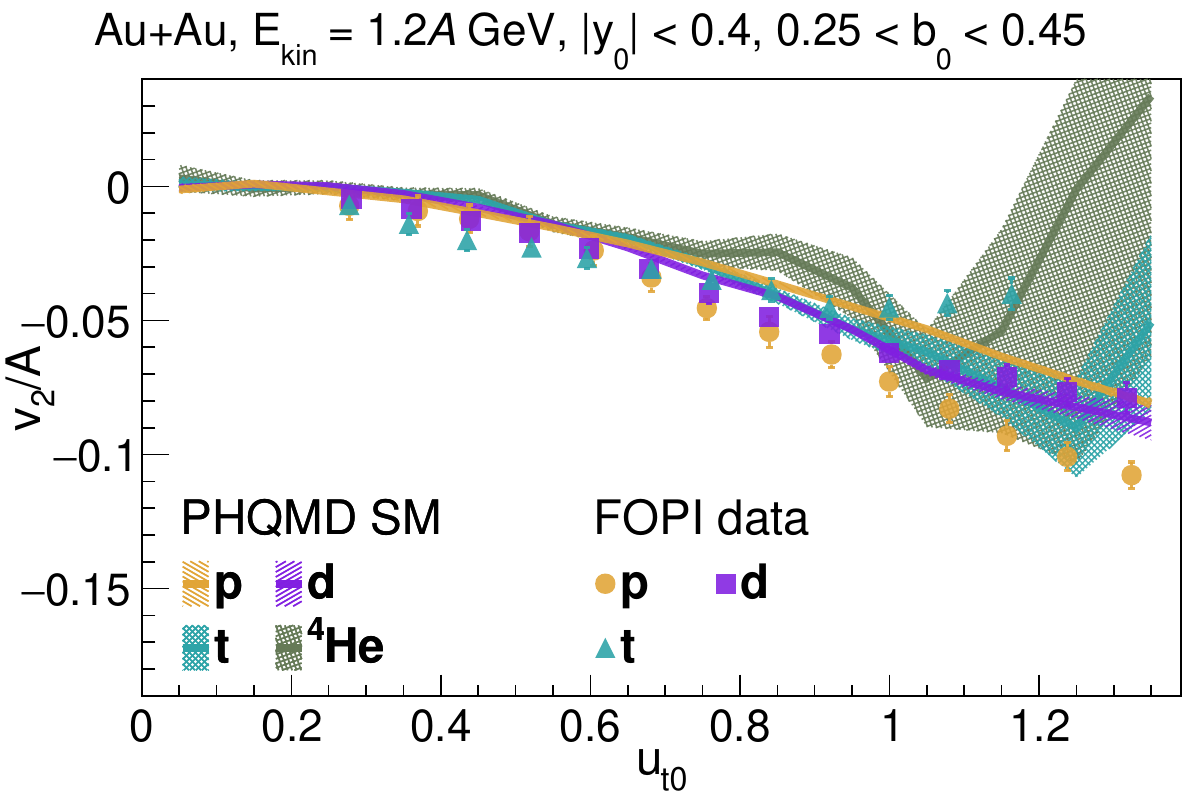}
    \caption{$v_2$ of protons (upper left), deuterons (upper right), tritons (middle left) and $^4$He (middle right) as a function of scaled transverse momentum $u_{t0}$ for  $|y_0|<0.4$ and the $0.25<b_0<0.45$ impact parameter range for Au+Au collisions at $E_{kin}$=1.2 A GeV. The left plot in the lower row shows  the compilation of $v_2(u_{t0})$ for protons, deuterons, tritons and $^4$He for the SM EoS;  the right plot shows the scaled $v_2/A (u_{t0})$. The colour code is the same as in Fig. \ref{Fig:FOPI12y0}. The FOPI experimental data are taken from Ref. \cite{FOPI:2011aa}.}
    \label{fig:Fopi_v2mid12}
\end{figure*}

\begin{figure*}
    \centering
\includegraphics[scale=0.4]{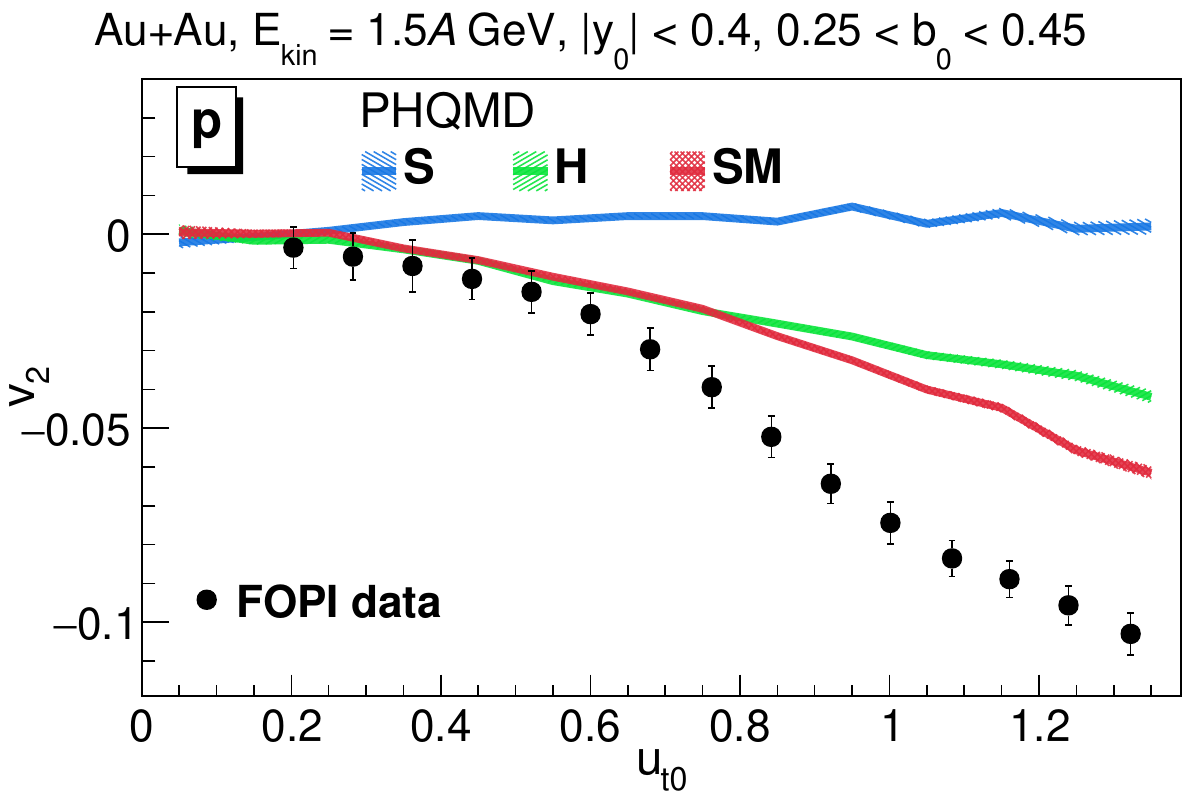}
\includegraphics[scale=0.4]{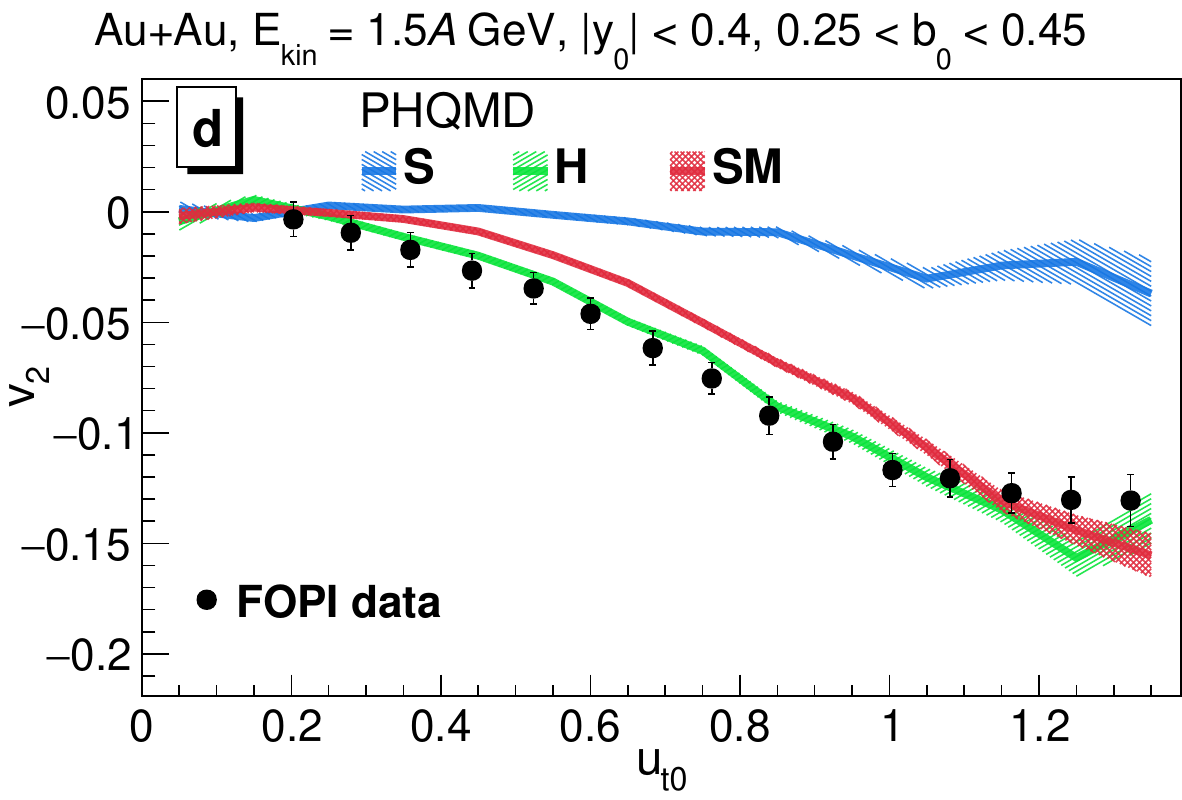}
\includegraphics[scale=0.4]{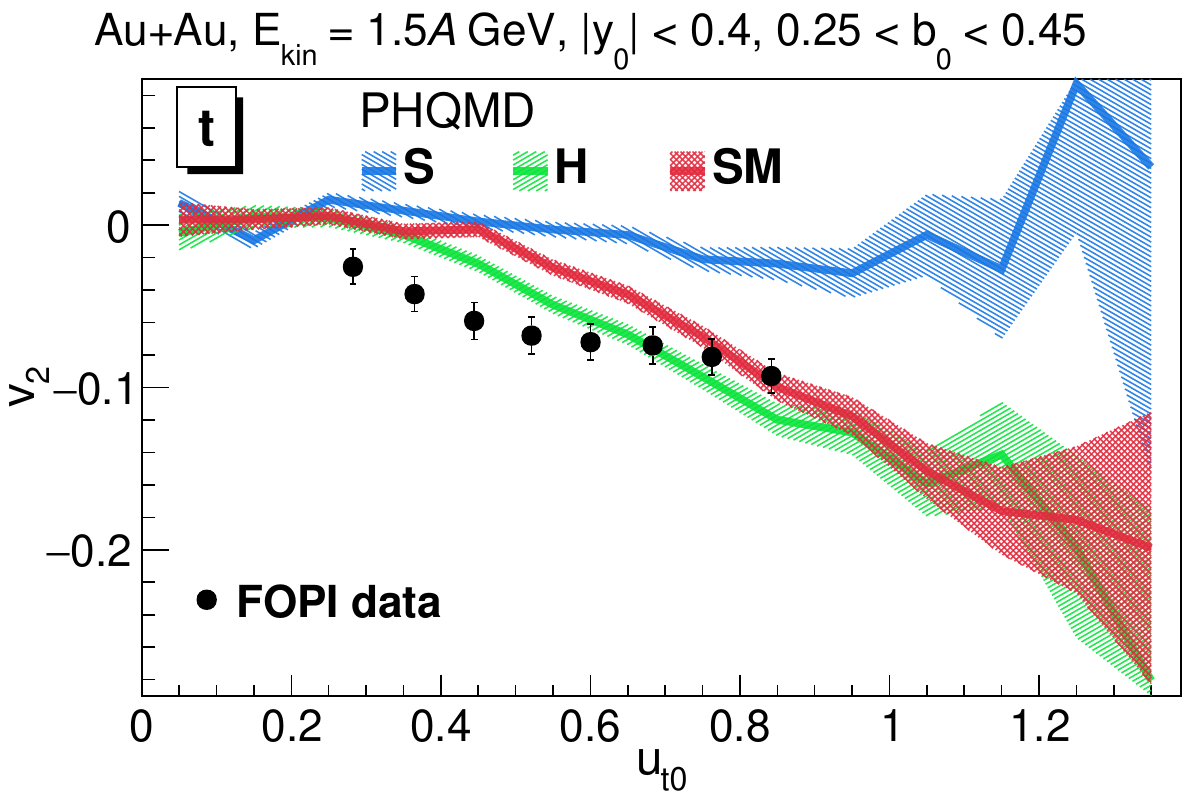}
\includegraphics[scale=0.4]{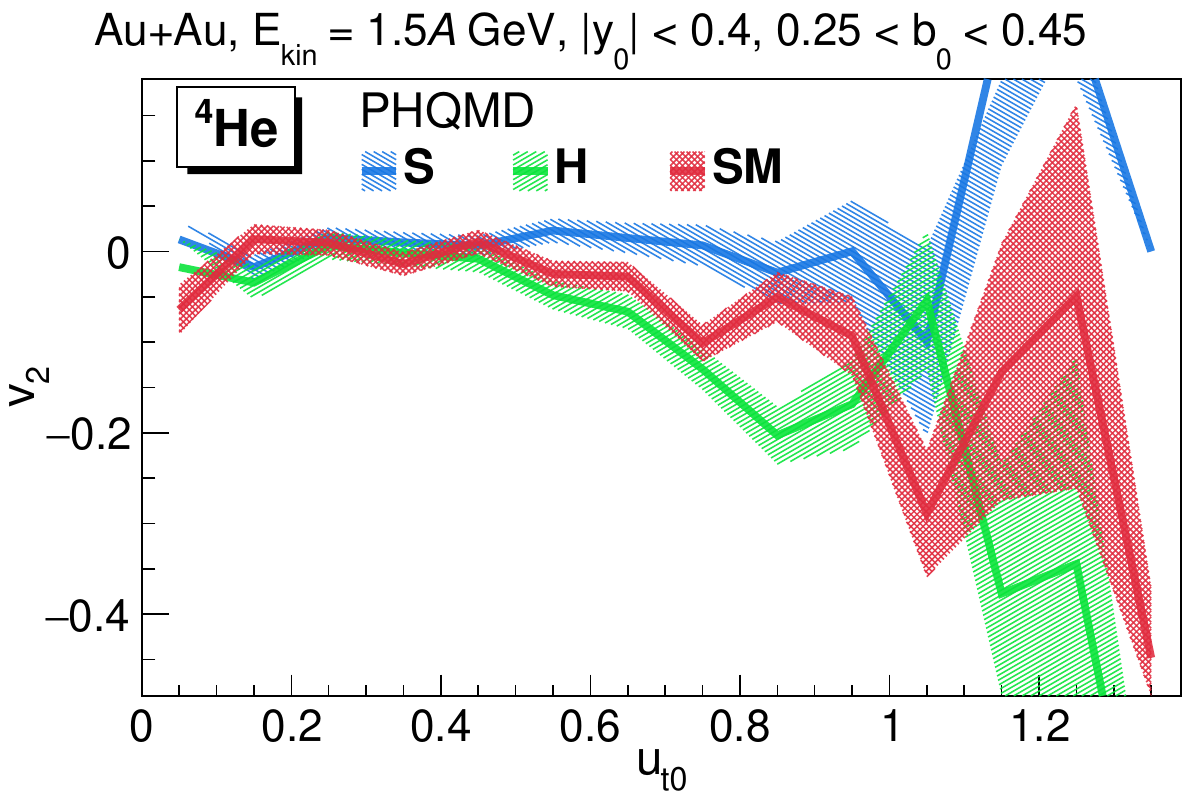}
\includegraphics[scale=0.4]{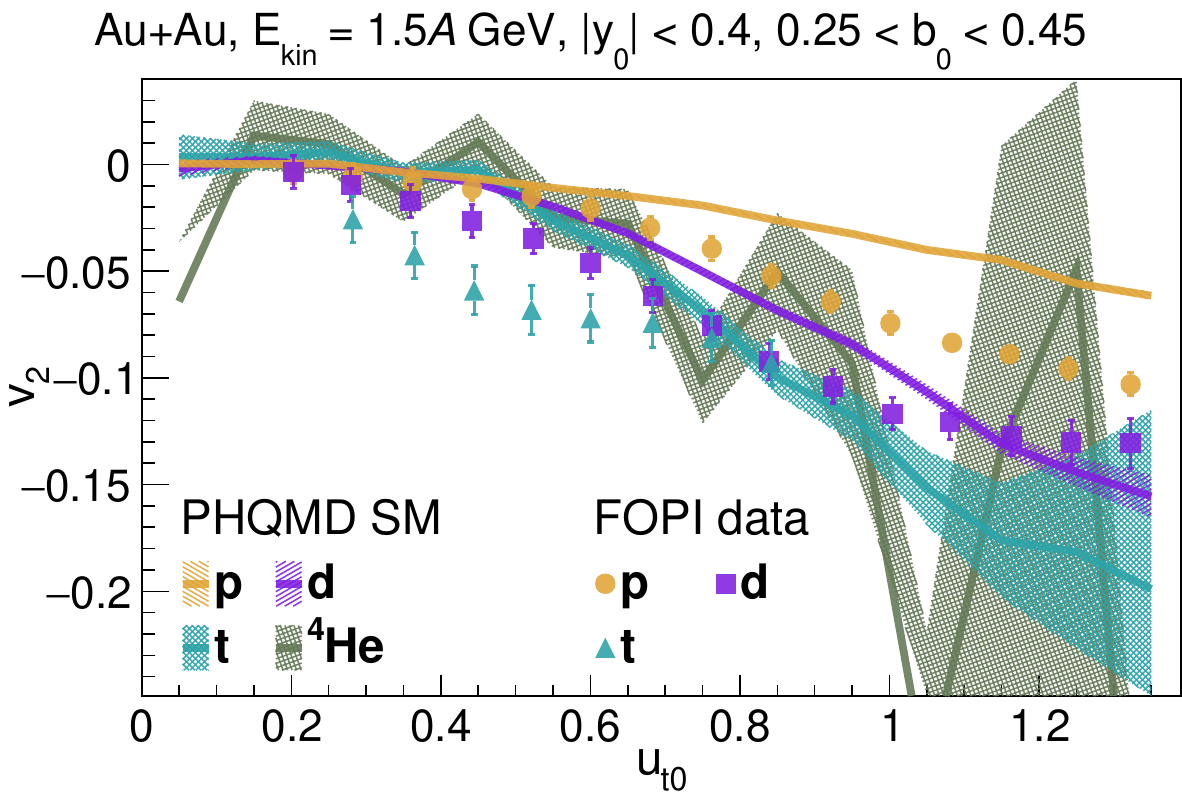}  
\includegraphics[scale=0.4]{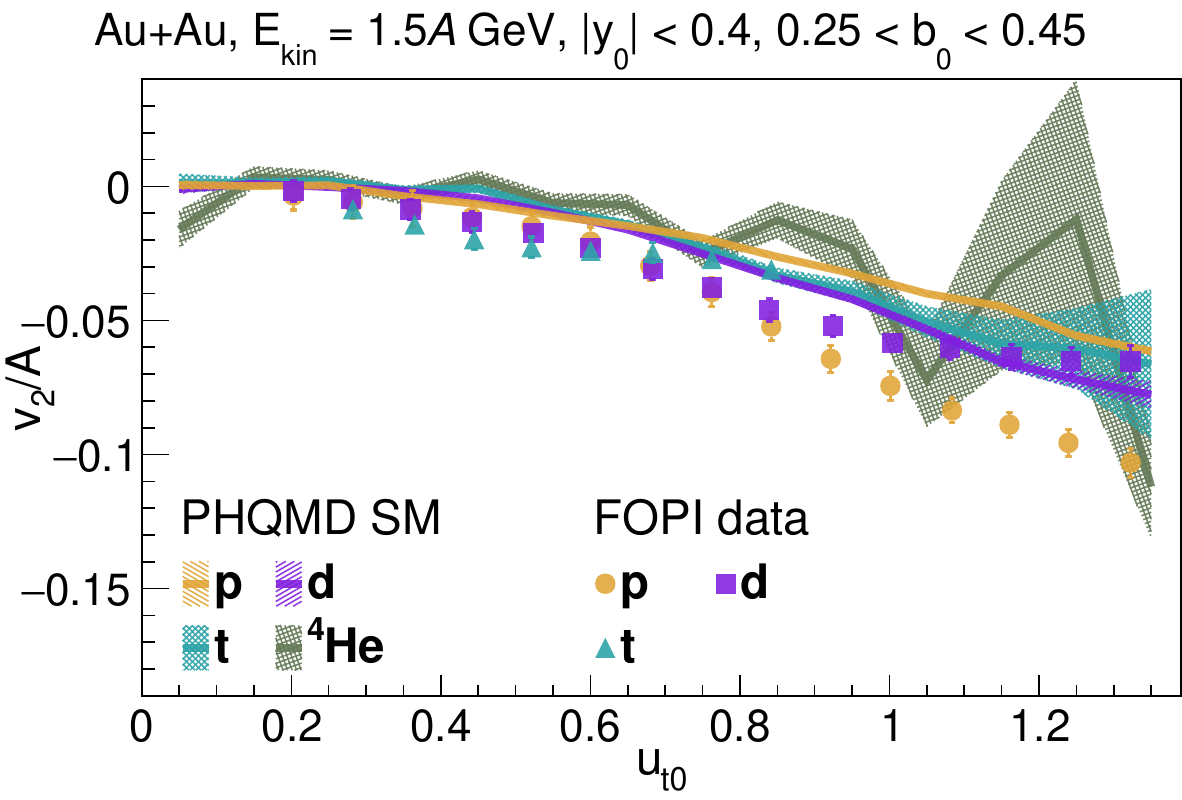}
 \caption{$v_2$ of protons (upper left), deuterons (upper right), tritons (middle left) and $^4$He (middle right) as a function of scaled transverse momentum $u_{t0}$ for  $|y_0|<0.4$ and the $0.25<b_0<0.45$ impact parameter range for Au+Au collisions at $E_{kin}$=1.5 A GeV. The plot on the lower row shows the compilation of $v_2(u_{t0})$ for protons, deuterons, tritons and $^4$He for the SM EoS; 
 the right plot shows the scaled $v_2/A (u_{t0})$.
 The colour code is the same as in Fig. \ref{Fig:FOPI12y0}. The FOPI experimental data are taken from Ref. \cite{FOPI:2011aa}.}    
\label{fig:Fopi_v2mid15}
\end{figure*}

As already stated, the $v_2(y_0)$ coefficients for the FOPI data in Figs. \ref{fig:FOPI12v2y0} and \ref{fig:FOPI15v2y0} are shown for the selected interval  of  $u_{t0}$. The results for $v_2(y_0)$ for $u_{t0} >0.4$  can be better understood if one studies the $u_{to}$ dependence of the flow explicitly.  

The experimental and the PHQMD results  for $v_2(u_{t0})$ at midrapidity, $|y_0|<0.4$,  are shown in Figs. \ref{fig:Fopi_v2mid12} and \ref{fig:Fopi_v2mid15}.
These figures show the  $v_2$ of protons, deuterons, tritons and $^4$He as a function of the scaled transverse momentum $u_{t0}$ for the $0.25<b_0<0.45$ impact parameter range  
for Au+Au collisions at $E_{kin}$=1.2 and 1.5 A GeV, respectively. 
One can see that for protons the $v_2$ dependence on $u_{t0}$, 
as predicted by PHQMD, is much stronger than that for the experimental data, an observation, which is a bit surprising in view of the fact that for the HADES data the $p_T$ dependence of $v_2$ is reasonably well described by PHQMD calculations with a SM EoS.  This explains the mismatch of $v_2(y_0)$ between theory and experiment  in Figs. \ref{fig:FOPI12v2y0} and \ref{fig:FOPI15v2y0}, where the lower $u_{t0}$ boundary is 0.4.
The deviation is stronger for large $u_{t0}$, i.e. for the region, which is selected for Figs. \ref{fig:FOPI12v2y0} and \ref{fig:FOPI15v2y0}.  
The $v_2$ dependence on $u_{to}$  of the experimental data for deuterons and tritons is well described by PHQMD employing a hard EoS, which results in slightly larger values of $v_2$ than the SM EoS, while the soft EoS underestimates clearly $v_2$  for protons and clusters.

The plots on the bottom rows of Fig. \ref{fig:Fopi_v2mid12}  and \ref{fig:Fopi_v2mid15} show the compilation of $v_2(u_{t0})$ for protons, deuterons, tritons and $^4$He for the SM EoS  at $E_{kin}$= 1.2 A GeV and 1.5 A GeV, respectively, in comparison with the FOPI experimental data. As seen in the left panels, also the FOPI data show a nontrivial dependence of the slope on the mass of the clusters. The difference between protons and clusters is substantially reduced if one displays, as discussed above for Fig. \ref{fig:v2pt_HADES_scaling},  scaled $v_2/A$ instead of $v_2$ as one can see from the bottom right panels of the figs. \ref{fig:Fopi_v2mid12}  and \ref{fig:Fopi_v2mid15}.  We see that (similar to HADES) the FOPI data for $u_{t0}<0.8$  as well as for the PHQMD calculations show close to midrapidity an almost perfect scaling  (which is also present if we replace in PHQMD $u_{t0}$ by $p_T/A$) at low $p_T$ ($u_T<0.8$).  For $u_{t0}>0.8$ neither the experimental results nor the calculations show scaling.

Summarizing the above results for $v_1$ and $v_2$, we conclude that comparing the PHQMD calculations with the data sets from HADES and FOPI one observes that overall the PHQMD calculations describe qualitatively the features of the data.
For both data sets we observe, using a SM EoS, a slight underestimation of the slope of $v_1(y)$ for protons as well as for clusters.  For the $v_1(p_T)$ the situation is more complex. For the HADES data the SM calculation underestimates slightly the $v_1(p_t)$ at high $p_T$ but reproduces the functional form whereas the PHQMD calculations with a SM EoS  agree with the FOPI $v_1$ data for $A=3$ and $^4He$ at large $u_{to}$ but miss the functional form of $v_1(u_{t0})$.

For $v_2(y)$ the functional form of the data is similar in all three data sets  and reproduced qualitatively as well by PHQMD calculations, best for a SM EoS, which underestimates nevertheless the $v_2$, especially at midrapidity.   The PHQMD calculations for the three data sets, i.e. for Au+Au for 1.23 A GeV from the HADES Collaboration and for 1.2 and 1.5 A GeV from the FOPI Collaboration, show a similar functional form. Please note, that the HADES and FOPI data presented here are for different centrality ranges.  While the HADES data is measured for 20-30\% most central events, the FOPI event selection is for the impact parameter range $0.25<b_0<0.45$, roughly corresponding to the 6-20\% most central events.  Differences in the quality of agreement between PHQMD calculations and data might therefore to some extent be attributed to the different methods of centrality definition, which should be modeled in a more detailed fashion in the future.

Without showing explicit plots, we note that - contrary to observed scaling of $v_2/A(p_T/A)$ at midrapidity - we did not observe any scaling for directed flow $v_1/A$ versus the scaled $p_T/A$.  This is also not seen in the experimental data which show a non-trivial $p_T$ dependence of $v_1$ for protons and clusters. Thus data on $v_1$ and $v_2$ of clusters can provide an extra information on the many-body interactions compared to proton data. 
This observation is consistent with recent PHQMD results at $\sqrt{s}_{NN}=3$  GeV reported in Ref. \cite{Zhou:2025zgn}.

\clearpage
\section{Sensitivity of \texorpdfstring{$v_1$}{v1} and \texorpdfstring{$v_2$}{v2} of deuterons  on the production mechanism: MST versus coalescence}
\label{mst_coal}

In this section we investigate the sensitivity of the flow observables $v_1$ and $v_2$ on the deuteron production mechanisms. For that we compare the PHQMD results for the "default" scenario, where deuterons are produced by the kinetic + MST mechanisms, with that produced by the coalescence mechanism. 
The coalescence mechanism is applied for a new set of events generated without kinetic deuterons such the all nucleons in the system are subject to the coalescence procedure 
as in other  models. We have checked that $v_1, v_2$ of deuterons from coalescence applied to the events with and without kinetic deuterons are very similar.

We stress that the PHQMD is a unique laboratory for such a comparison since all scenarios are integrated in the same code.

\begin{figure}[h]
    \centering
        \includegraphics[scale=0.4]{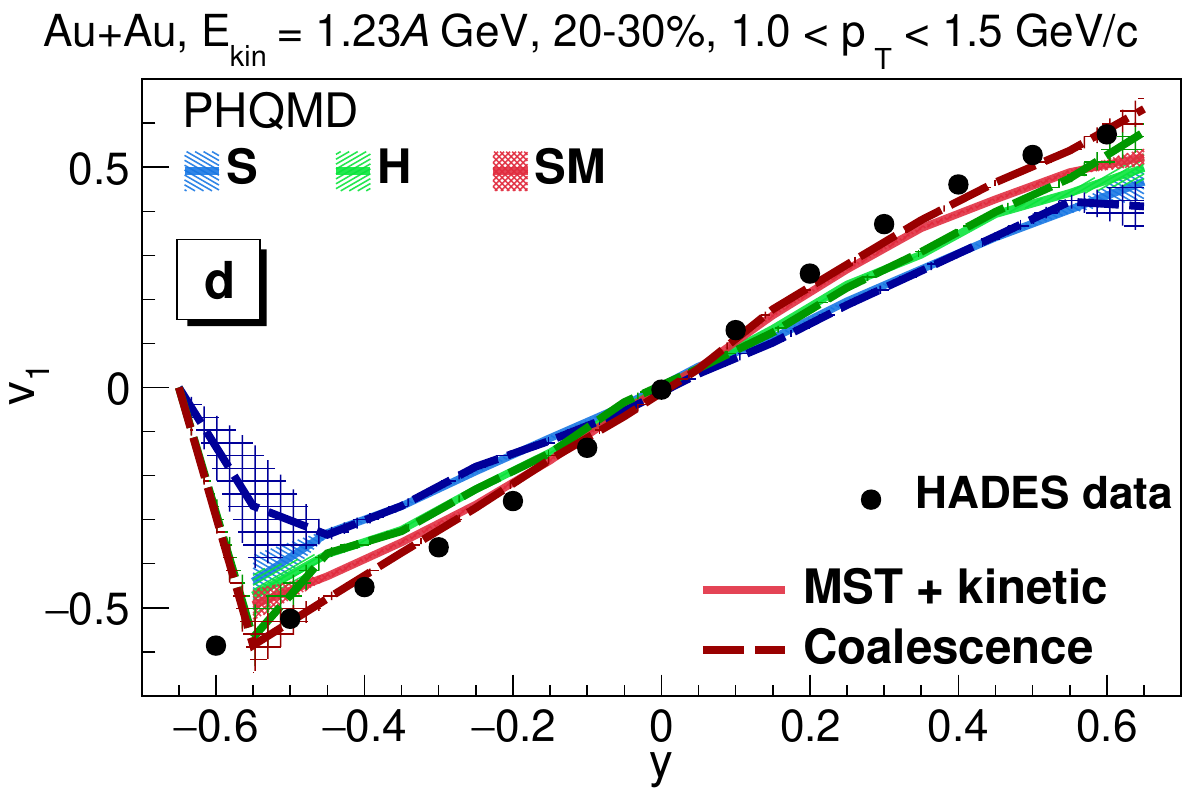}    
        \includegraphics[scale=0.4]{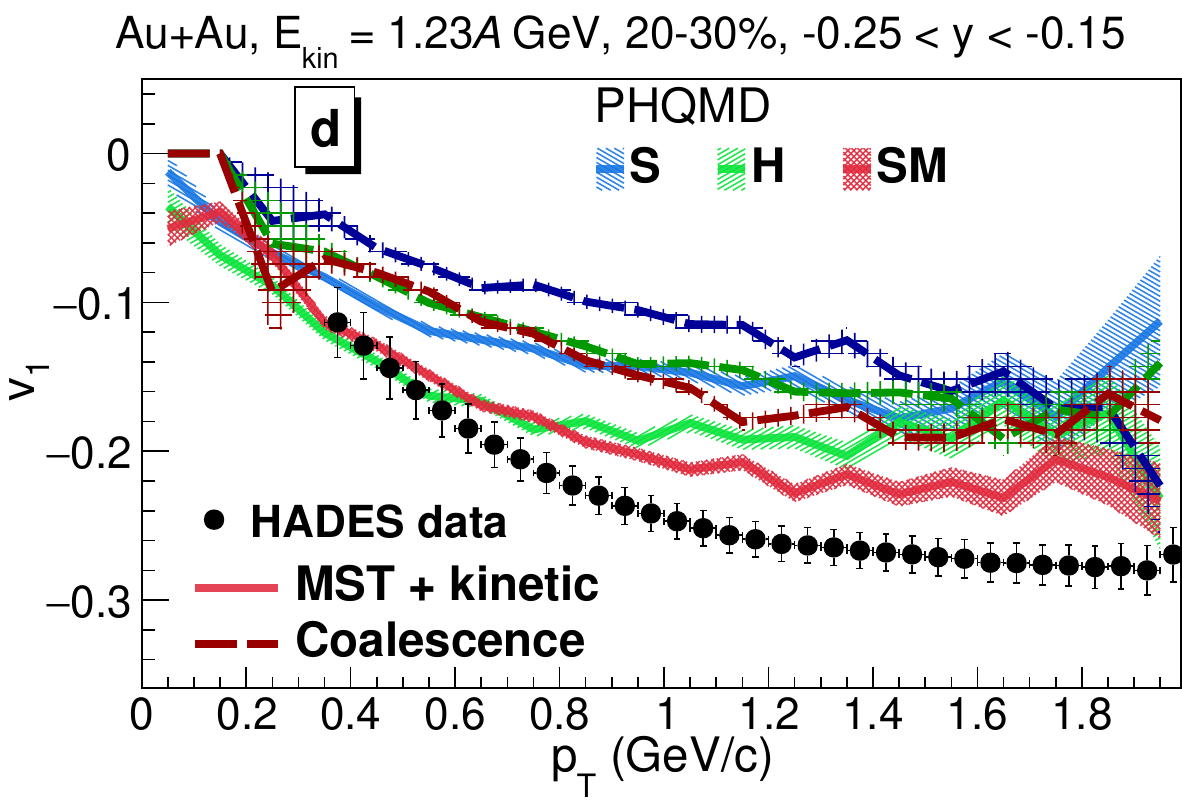}
    \caption{
    Upper plot: Comparison of the $v_1(y)$ of deuterons produced by kinetic + MST mechanisms (solid lines) with the coalescence mechanism (dashed lines)  for 20-30\% central Au+Au collisions at $E_{kin}=1.23$ A GeV for  $1.0 < p_T < 1.5$ GeV/c. 
    Lower plot: The comparison of $v_1(p_T)$ of deuterons produced by kinetic + MST mechanisms (solid lines) with the  coalescence mechanism (dashed lines) for 20-30\% central Au+Au collisions for $-0.25 < y < -0.15$.
   The blue lines "S" correspond to the PHQMD calculations with the "soft" EoS, the green lines "H" show the "hard" EoS, the red lines "SM" represent the momentum-dependent "soft" EoS. 
   The color coding for the coalescence results is the same but in dark colors.
    The HADES experimental data are taken from Ref. \cite{HADES:2020lob}.}
    \label{fig:coalescence_v1ypt}
\end{figure}

In the upper plot of  Fig. \ref{fig:coalescence_v1ypt} we present the comparison of $v_1(y)$ of  deuterons produced by the "default" scenario (PHQMD - kinetic + MST mechanism), (shown in Figs. \ref{fig:hadesv1y}  and \ref{pdtv1pt}, respectively)  to the $v_2$ obtained if one applies the  coalescence mechanism (dashed lines in dark colors)  for 20-30\% central Au+Au collisions at $E_{kin}=1.23$ A GeV for  $1.0 < p_T < 1.5$ GeV/c.  
The lower plot of Fig. \ref{fig:coalescence_v1ypt} shows the same comparison for $v_1(p_T)$  for the rapidity interval $-0.25 < y < -0.15$. Both deuteron production scenarios are confronted with the HADES data  \cite{HADES:2020lob}. 
One can see that for all EoS the $v_1$ of kinetic + MST deuterons is slightly larger than that of  coalescence,  what is in line with our findings in Ref. \cite{Kireyeu:2024woo}  that only a small part of  the nucleons, identified as being a part of deuterons by coalescence or by MST, are identical. 

\begin{figure}
    \centering
        \includegraphics[scale=0.4]{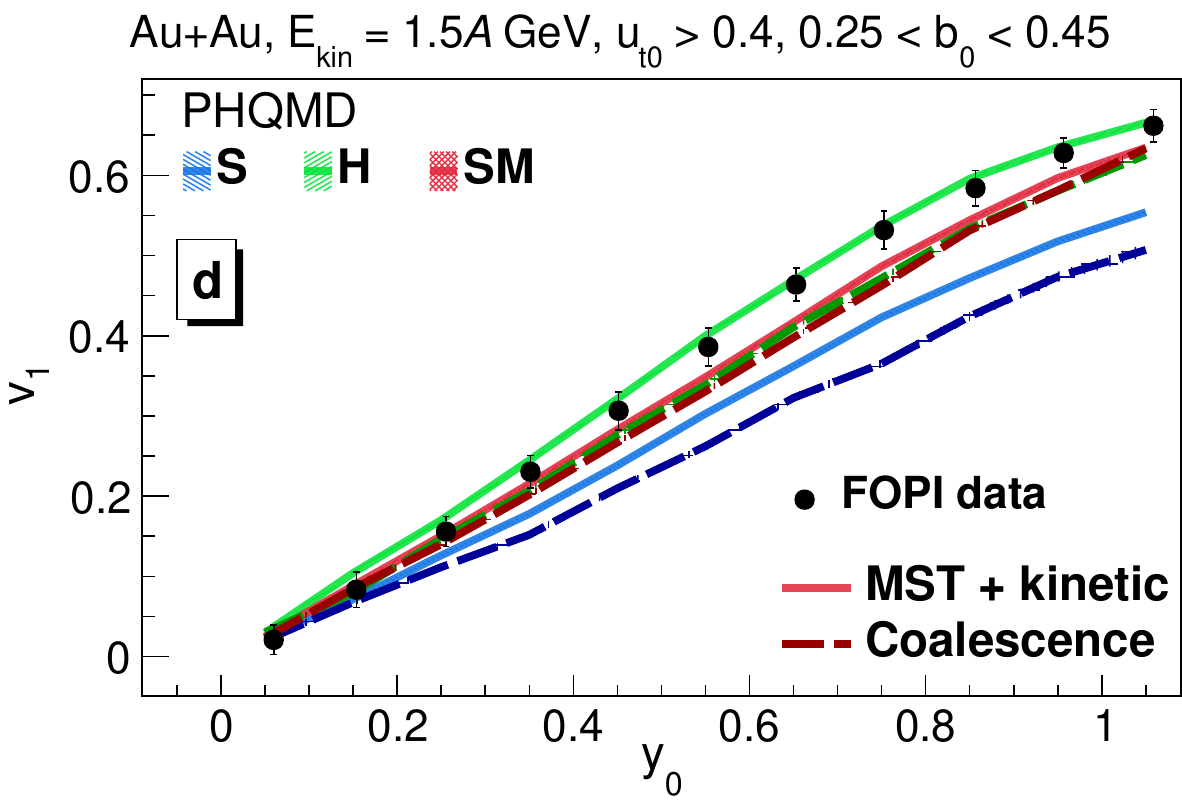}
        \includegraphics[scale=0.4]{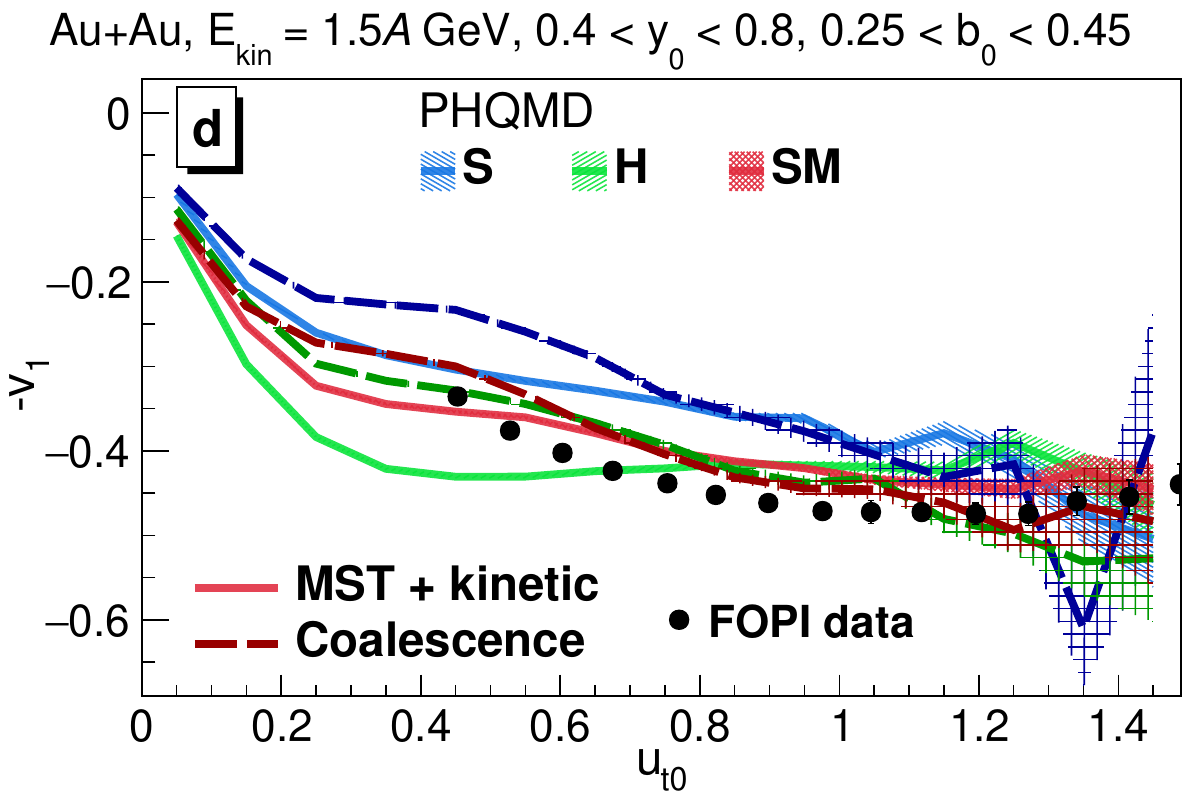}
    \caption{
    Upper plot: Comparison of the $v_1(y_0)$ of deuterons (upper right) produced by kinetic + MST mechanisms (solid lines)  with the coalescence mechanism (dashed lines)
    for Au+Au collisions at $E_{kin}=1.5$ A GeV for $u_{t0} >0.4$ and the impact parameter range
    $0.25<b_0<0.45$. 
    Lower plot: Comparison of $v_1(u_{t0})$ of deuterons produced by kinetic + MST mechanisms (solid lines) with the  coalescence mechanism (dashed lines) for Au+Au collisions at $E_{kin}=1.5$ A GeV  for $0.4 < y_0 < 0.8$ and centrality $0.25<b_0<0.45$
    The color coding is the same as in Fig. \ref{fig:coalescence_v1ypt}.
    The FOPI experimental data are taken from Ref. \cite{FOPI:2011aa}.}
    \label{fig:coalescence_v1y0ut0}
\end{figure}
Figure \ref{fig:coalescence_v1y0ut0} shows - similar to Fig. \ref{fig:coalescence_v1ypt} - a comparison of $v_1(y_0)$ and $v_1(u_{t0})$, but with respect to the FOPI data \cite{FOPI:2011aa}. The results for the kinetic + MST mechanisms are the same as in Figs. \ref{Fig:FOPI15y0} and \ref{Fig:FOP15u0}, respectively.

\begin{figure}
    \centering
        \includegraphics[scale=0.4]{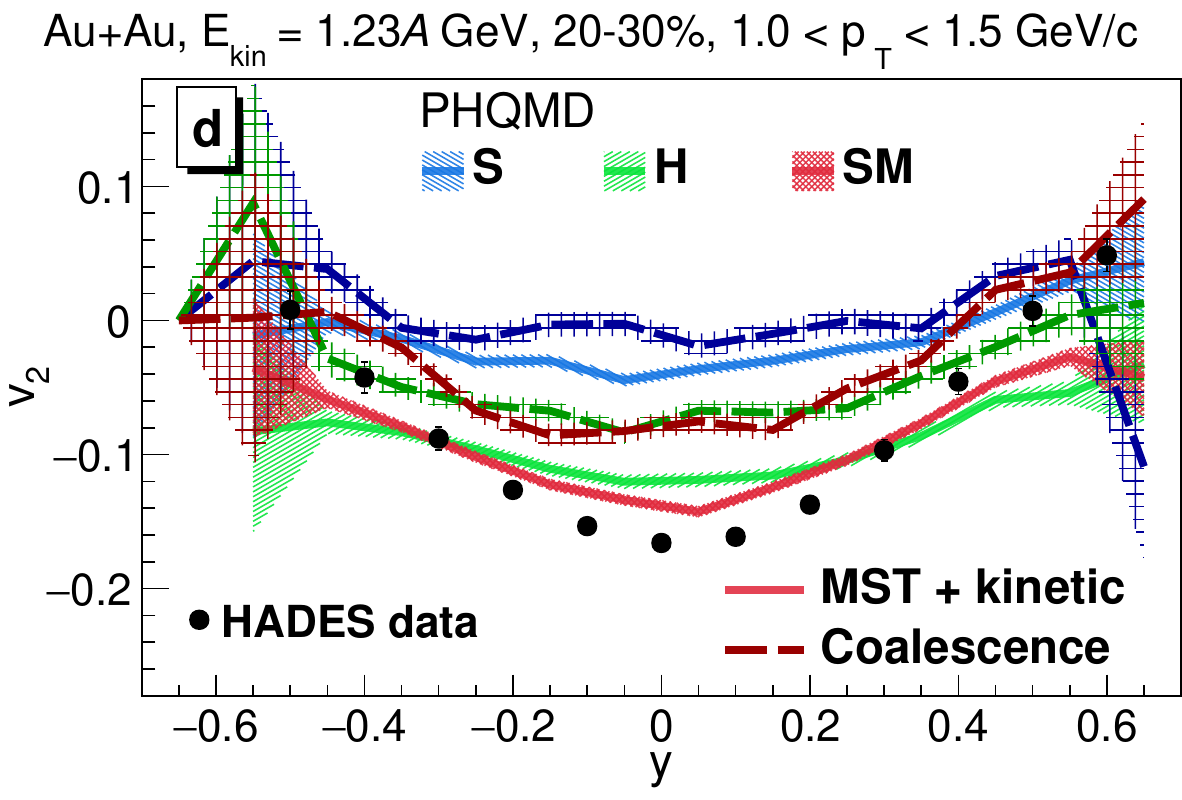}   
        \includegraphics[scale=0.4]{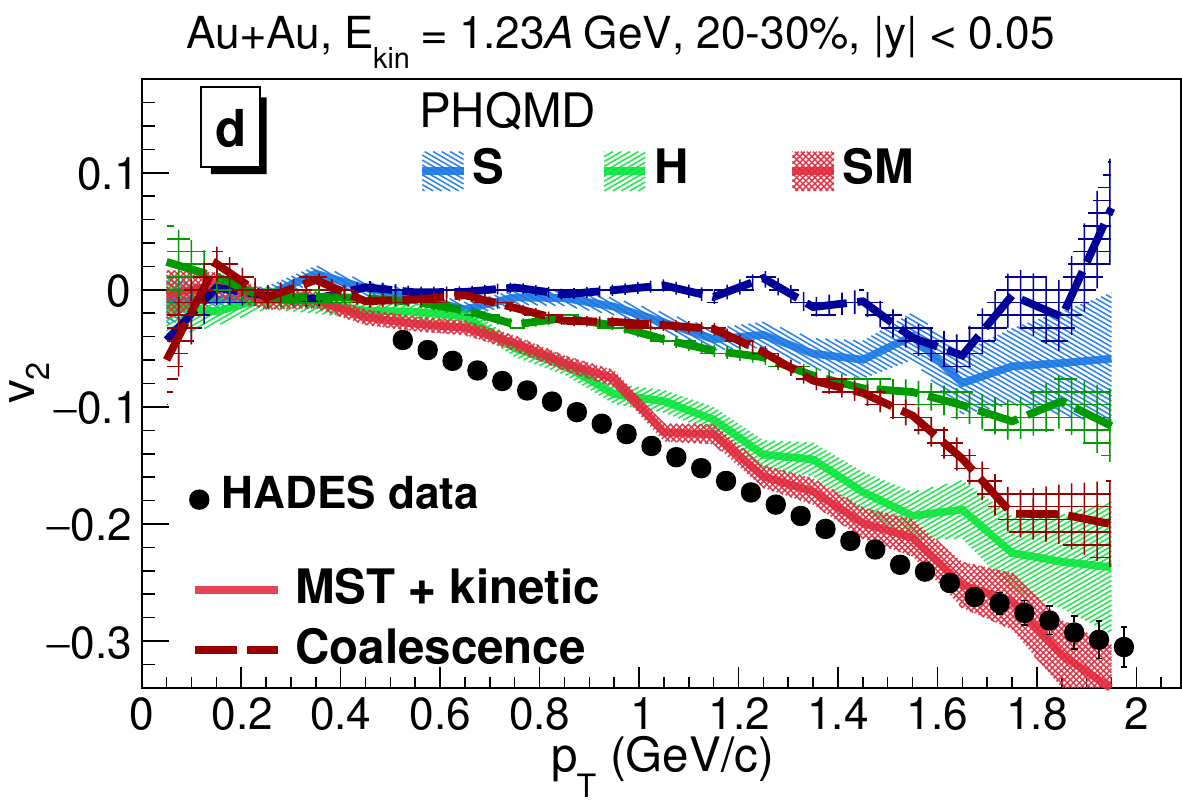}          
    \caption{
    Upper plot: Comparison of $v_2(y)$ of deuterons produced by kinetic + MST mechanisms (solid lines) with the coalescence mechanism (dashed lines)  for 20-30\% central Au+Au collisions at $E_{kin}=1.23$ A GeV for  $1.0 < p_T < 1.5$ GeV/c. 
    Lower plot: Comparison of $v_2(p_T)$ of deuterons produced by kinetic + MST mechanisms (solid lines) with the coalescence mechanism (dashed lines) for 20-30\% central Au+Au collisions for $|y| < 0.05$.
    The color coding is the same as in Fig. \ref{fig:coalescence_v1ypt}.
    The HADES experimental data are taken from Ref. \cite{HADES:2020lob}.}   
    \label{fig:coalescence_v2ypt}
\end{figure}
The upper plot of  Fig.  \ref{fig:coalescence_v2ypt}  displays the comparison of $v_2(y)$ of deuterons produced by the kinetic + MST mechanism (solid lines)  to the $v_2$ of deuterons produced by the coalescence mechanism (dashed lines)  for 20-30\% central Au+Au collisions at $E_{kin}=1.23$ A GeV for  $1.0 < p_T < 1.5$ GeV/c.  
The lower plot of Fig. \ref{fig:coalescence_v2ypt} shows the same comparison for $v_2(p_T)$ of nucleons measured in the rapidity interval $|y| < 0.05$. The both deuteron production scenarios are confronted with the HADES data  \cite{HADES:2020lob}.   
The results for the kinetic + MST mechanisms are the same as in Figs. \ref{fig:HADESv2yclast} and \ref{fig:v2pt_HADES}, respectively.

\begin{figure}
    \centering
        \includegraphics[scale=0.4]{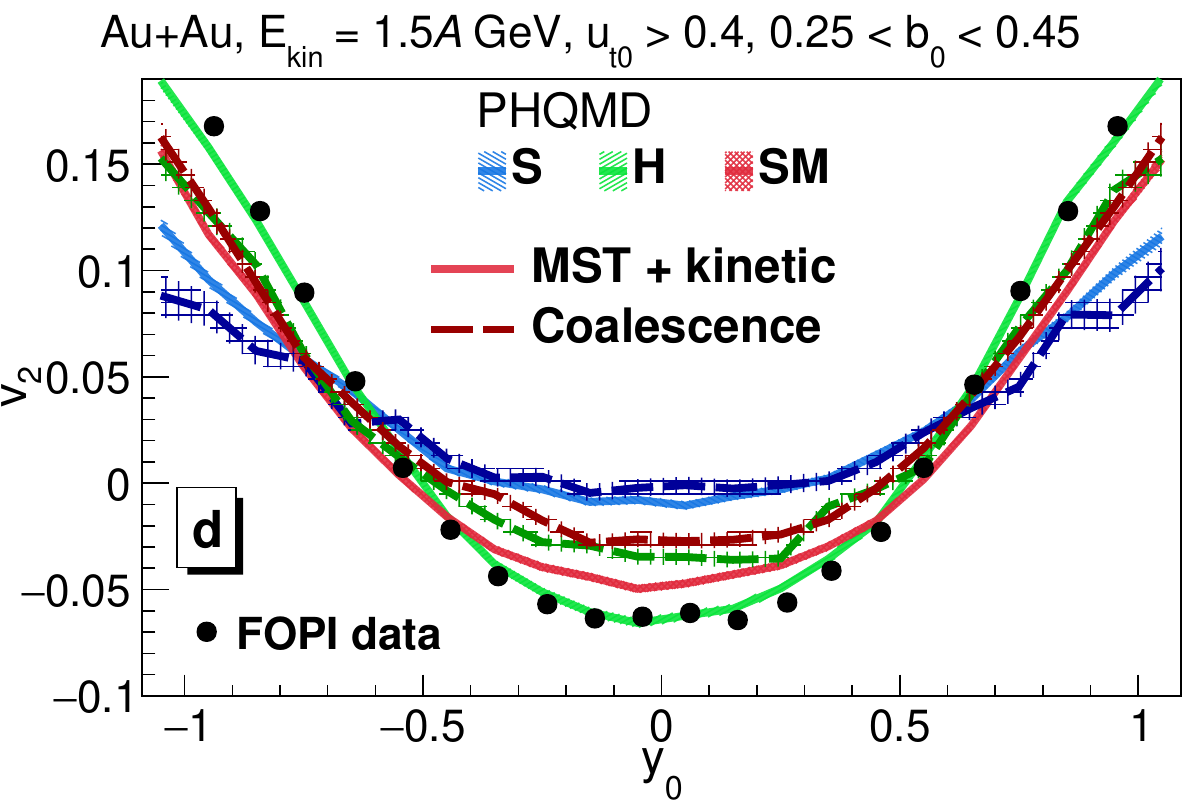}  
        \includegraphics[scale=0.4]{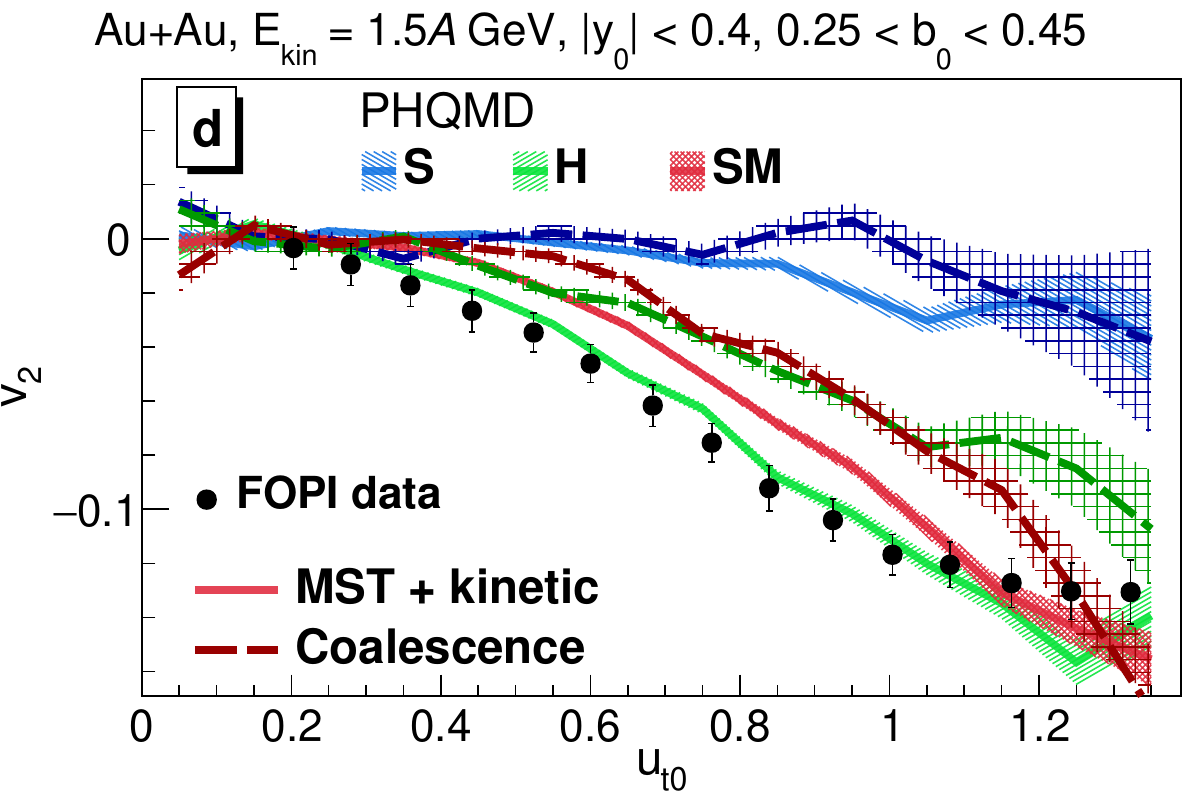}          
    \caption{
    Upper plot: Comparison of  $v_2(y_0)$ of deuterons (upper right) produced by kinetic + MST mechanisms (solid lines) with the coalescence mechanism (dashed lines)
    for Au+Au collisions at $E_{kin}=1.5$ A GeV for $u_{t0} >0.4$ and the impact parameter range
    $0.25<b_0<0.45$. 
    Lower plot: The comparison of $v_2(u_{t0})$ of deuterons produced by kinetic + MST mechanisms (solid lines) with the  coalescence mechanism (dashed lines) for Au+Au collisions at $E_{kin}=1.5$ A GeV  for $|y_0|<0.4$ and centrality $0.25<b_0<0.45$.
    The color coding is the same as in Fig. \ref{fig:coalescence_v1ypt}.
    The FOPI experimental data are taken from Ref. \cite{FOPI:2011aa}.}    
    \label{fig:coalescence_v2y0ut0}
\end{figure}

Figure \ref{fig:coalescence_v2y0ut0} shows - similar to Fig. \ref{fig:coalescence_v2ypt} - comparison of $v_2(y_0)$ and $v_2(u_{t0})$ but with respect to the FOPI data \cite{FOPI:2011aa} for Au+Au collisions at $E_{kin}=1.5$ A GeV for $u_{t0} >0.4$ and the impact parameter range  $0.25<b_0<0.45$.  The results for the kinetic + MST mechanisms are the same as in Figs. \ref{fig:FOPI15v2y0} and \ref{fig:Fopi_v2mid15}, respectively.

We find a substantial difference of the flow coefficients  $v_1$ and $v_2$ obtained with the standard PHQMD calculations, which include kinetic + MST deuterons, as compared to those obtained with the coalescence mechanism. The coalescence mechanism is used in the UrQMD and SMASH models, however in different ways \cite{Mohs:2024gyc}.  
This observation demonstrates that the flow coefficients $v_1$ and $v_2$ are very sensitive to the production mechanism (calculated within the same model) and can complement the information, which can be derived from $dN/dy$ and $p_T$ distributions \cite{Kireyeu:2024woo}, to identify experimentally the origin of the deuteron production in heavy-ion collisions.

\begin{figure}[h]
    \centering
        \includegraphics[scale=0.4]{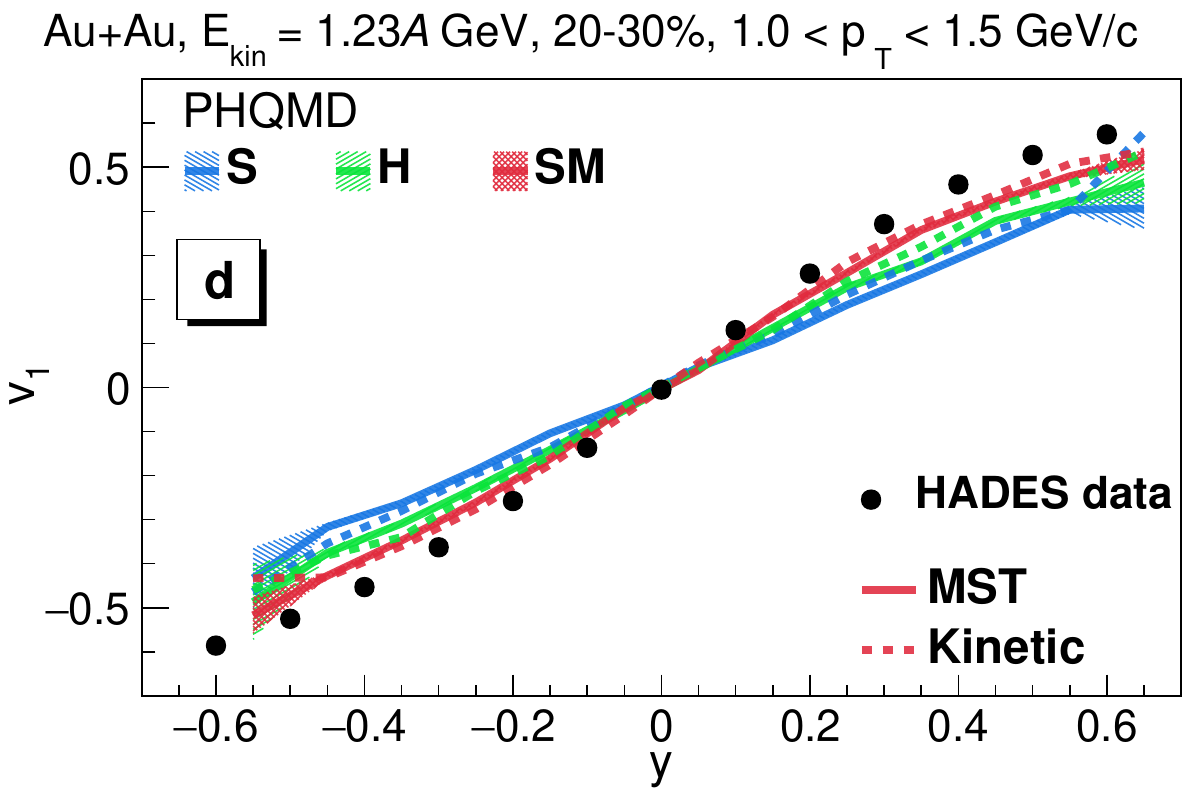}    
        \includegraphics[scale=0.4]{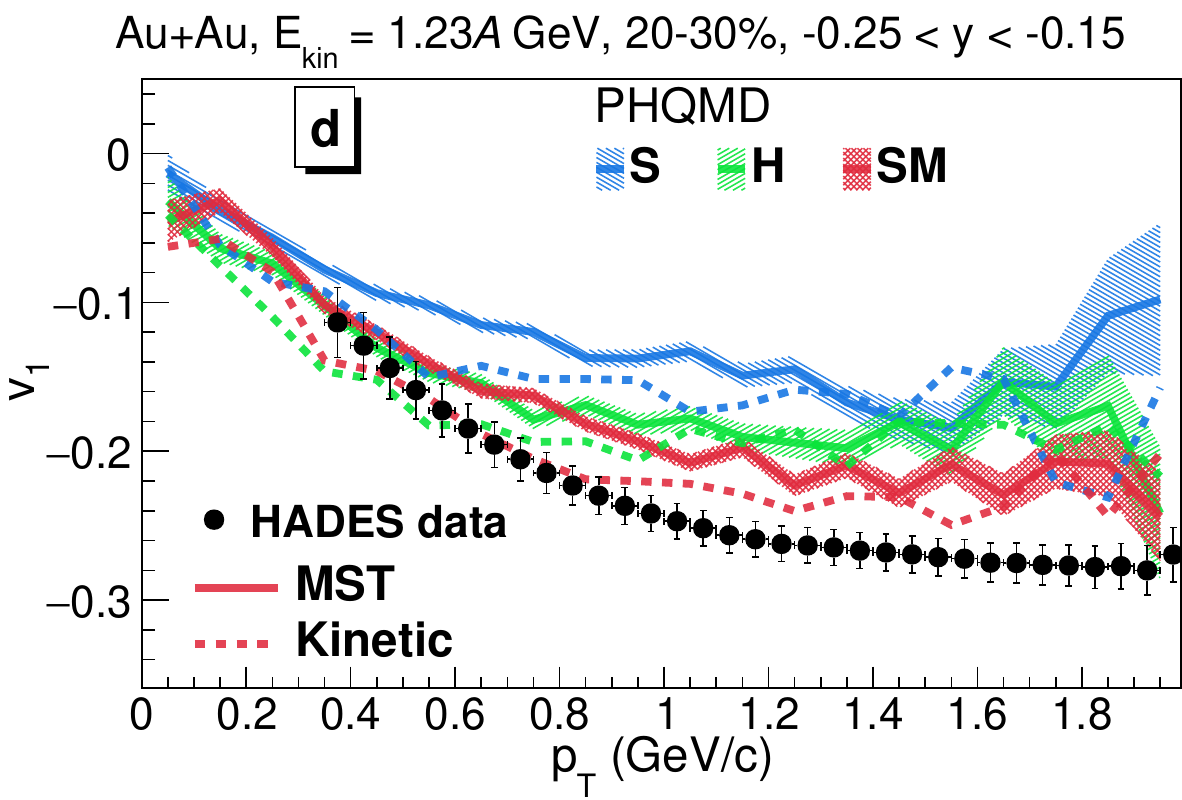}
    \caption{
Comparison of the $v_1(y)$ for  $1.0 < p_T < 1.5$ GeV/c (upper plot) and  $v_1(p_T)$ for $-0.25 < y <-0.15$ (lower plot) of kinetic (dashed lines) and MST deuterons (solid lines)  for 20-30\% central Au+Au collisions at $E_{kin}=1.23$ A GeV. 
   The blue lines "S" correspond to the PHQMD calculations with the "soft" EoS, the green lines "H" show the "hard" EoS, the red lines "SM" represent the momentum-dependent "soft" EoS. 
    The HADES experimental data are taken from Ref. \cite{HADES:2020lob}.}
    \label{fig:v1_mst_vs_kin}
\end{figure}

\begin{figure}[h]
    \centering
        \includegraphics[scale=0.4]{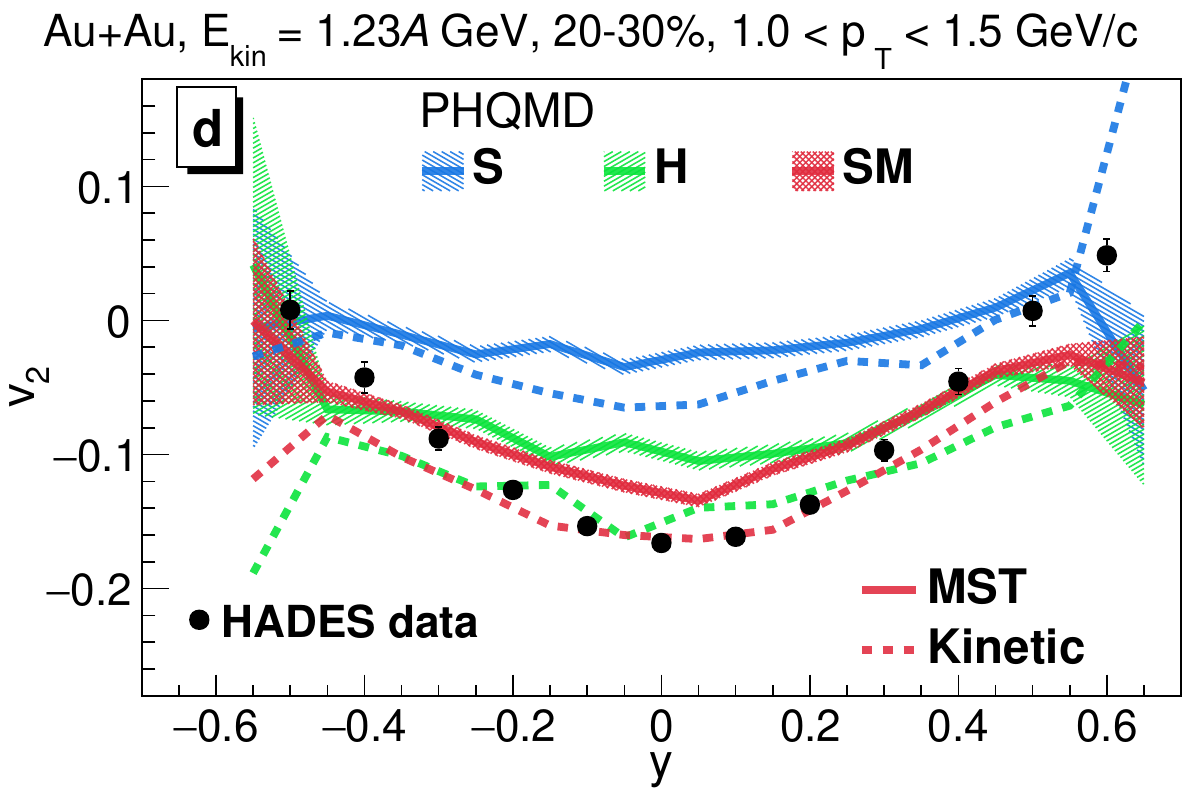}    
        \includegraphics[scale=0.4]{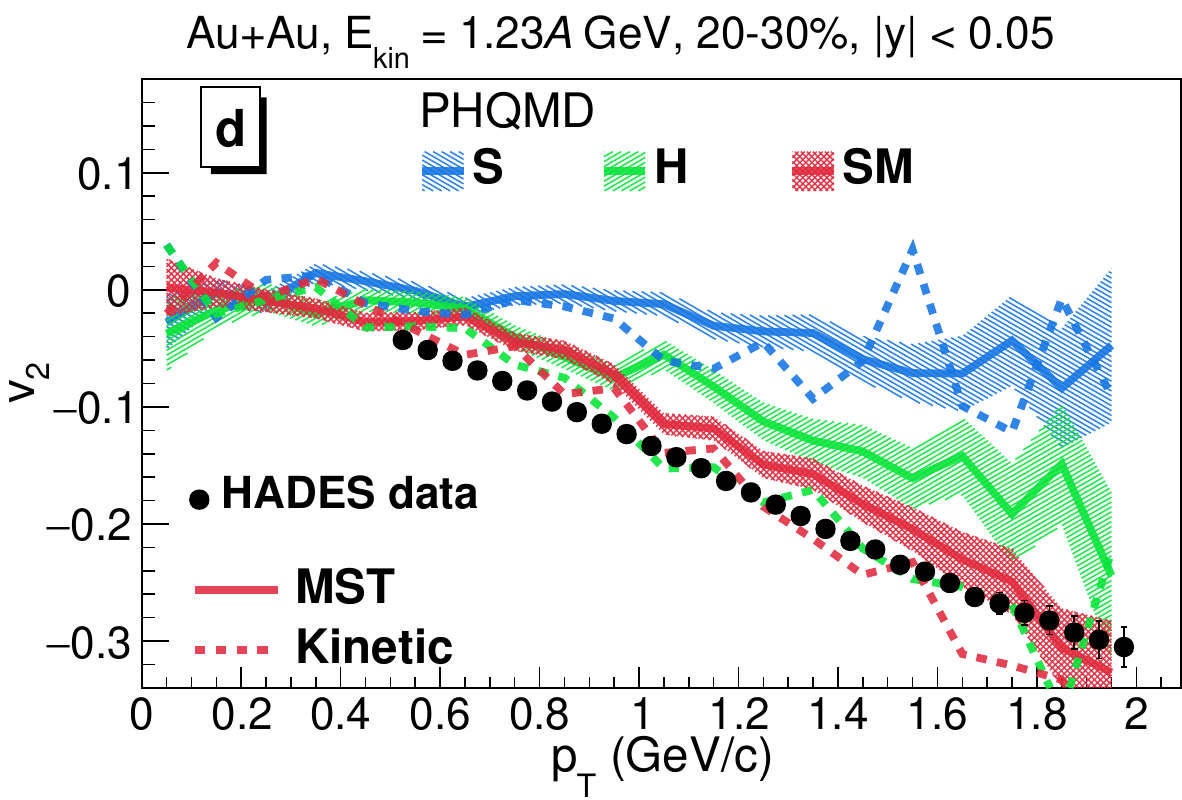}
    \caption{
Comparison of the $v_2(y)$ for  $1.0 < p_T < 1.5$ GeV/c (upper plot) and  $v_2(p_T)$ for $|y| < 0.05$ (lower plot) of kinetic (dashed lines) and MST deuterons (solid lines)  for 20-30\% central Au+Au collisions at $E_{kin}=1.23$ A GeV.
      Color coding as in Fig. \ref{fig:v1_mst_vs_kin}.
    The HADES experimental data are taken from Ref. \cite{HADES:2020lob}.}
    \label{fig:v2_mst_vs_kin}
\end{figure}

In Figs.   
\ref{fig:coalescence_v1ypt}, \ref{fig:coalescence_v1y0ut0}, \ref{fig:coalescence_v2ypt} and \ref{fig:coalescence_v2y0ut0} the coalescence mechanism is compared with the standard deuteron production in PHQMD, which combines the production by the kinetic mechanism and  the perturbative deuterons recognized by MST. It is relevant to separately show the sensitivity of the kinetic and MST deuterons to the EoS.   This is displayed in Figs. \ref{fig:v1_mst_vs_kin},  \ref{fig:v2_mst_vs_kin}, which are similar to Figs. \ref{fig:coalescence_v1ypt} and  \ref{fig:coalescence_v2ypt}, but
show the MST and kinetic contributions separately instead of the sum of both.
They present the comparison of the $v_1(y)$ and $v_2(y)$ (upper plots) and  $v_1(p_T)$ and $v_2(p_T)$  (lower plots) of kinetic (dashed lines) and MST deuterons (solid lines)  for 20-30\% central Au+Au collisions at $E_{kin}=1.23$ A GeV for S, H, SM EoS.
One can see that the kinetic and MST deuterons show a similar behavior with respect to EoS, reflected in the total $v_1$ and $v_2$.

\clearpage
\section{Summary}

In this study, we investigate the sensitivity of the directed flow $v_1$ and elliptic flow $v_2$ of protons and light clusters to the nuclear equation of state (EoS) at SIS energies using the PHQMD microscopic transport approach. For this purpose, the PHQMD model has been extended by incorporating a momentum-dependent potential in addition to the static Skyrme potential. We explore three types of EoS: soft (S), hard (H), and soft momentum-dependent (SM). The S and SM potentials share the same compressibility modulus of $K=200$ MeV (defined for infinite strongly interacting matter), while the hard EoS is characterized by a compressibility modulus of $K=380$ MeV. Furthermore, we examine the influence of the EoS on different deuteron production mechanisms implemented in PHQMD, including cluster recognition via the minimum spanning tree (MST) algorithm, hadronic kinetic reactions, and coalescence. To this end, we systematically compare the PHQMD predictions for flow observables and spectra with  available experimental data for protons and light clusters in the 1 A GeV energy regime.

Our findings are summarized as follows:
\begin{itemize}

\item
 We observed a pronounced sensitivity of proton and light-cluster rapidity and transverse momentum distributions to the EoS. Both the soft and soft momentum-dependent EoS produce similar results that differ significantly from those obtained with a hard EoS. Softening the EoS leads to a reduction of proton yield at midrapidity and an enhancement of light-cluster production.

\item 
The momentum dependence of the equation of state (EoS) has a significant impact on both the directed flow $v_1$ and the elliptic flow $v_2$—not only for protons, as previously established, but also for light clusters. The soft momentum-dependent (SM) interaction produces $v_1$ and $v_2$ values close to those obtained with a hard EoS and markedly different from those calculated with a static soft EoS, despite the identical compressibility modulus of $K=200$ MeV for the S and SM potentials (see Table 1). This indicates that the inclusion of momentum-dependent mean-field interactions modifies the collective dynamics beyond the effects associated solely with the static compressibility. Since elastic pA scattering experiments demonstrate that the nucleon–nucleon potential exhibits a pronounced momentum dependence, neglecting this feature in transport model analyses may lead to biased conclusions regarding the nuclear matter compressibility.

\item 
The comparison of PHQMD calculations employing the S, H, and SM EoS with experimental flow data for Au+Au collisions at $E_{\text{kin}}=1.23$ A GeV (HADES) and at $E_{\text{kin}}=1.2$ and $1.5$ A GeV (FOPI) shows that the SM interaction provides the best overall agreement with the measured proton $v_1$ and $v_2$. For light clusters, the calculations with SM and H EoS yield comparable results, whereas the static soft EoS ($K=200$ MeV) fails to reproduce the data. The PHQMD results obtained with the SM EoS slightly underestimate the experimental flow magnitudes, suggesting that the corresponding compressibility is marginally too low. A quantitative Bayesian analysis would allow a more precise determination of the compressibility parameter consistent with the available flow observables.

\item 
The rapidity and transverse momentum dependencies of $v_1(y)$, $v_2(y)$, $v_1(p_T)$, and $v_2(p_T)$ of light clusters differ significantly from those of protons, both in the experimental data and in the PHQMD calculations, which qualitatively reproduce the observed trends. At midrapidity, $v_2$ exhibits an approximate scaling with the cluster mass number $A$, consistent with the experimental observations. This scaling is nearly exact at low $p_T/A$ and becomes progressively violated at higher $p_T$. The flow observables thus provide a sensitive probe of the phase-space regions and dynamical conditions under which light clusters are formed during the heavy-ion collision.

\item 
The directed and elliptic flow coefficients, $v_1$ and $v_2$, of deuterons depend sensitively on the production mechanism implemented in the transport simulations. Two scenarios have been compared here:
deuterons produced by the combined kinetic + MST mechanism and deuterons produced by coalescence. The differences in the resulting flow coefficients are sufficiently large that, together with other observables, they may help to identify the cluster production mechanism realized in nature, which remains a subject of ongoing debate.
\end{itemize}

The PHQMD calculations, as well as recent results from SMASH and UrQMD (which employ different implementations of the equation of state), reproduce the functional dependence of the experimental flow data and achieve a quantitatively satisfactory agreement with the measurements. This represents a significant accomplishment of modern transport approaches, given that the elliptic flow $v_2$ is particularly sensitive to multiple model ingredients, including the mean-field potential, the treatment of nucleon–nucleon collisions, and the initialization of projectile and target nuclei in coordinate and momentum space. The similarity between the PHQMD and UrQMD results confirms earlier findings that a static hard EoS and a soft momentum-dependent EoS yield comparable flow magnitudes. The agreement between SMASH and PHQMD is more unexpected, as the respective implementations of the soft momentum-dependent interaction differ substantially. This aspect warrants further systematic investigation and coordinated efforts within the community.

To improve the precision of EoS constraints, a closer alignment between theoretical and experimental event selection procedures is essential. Specifically, the use of fixed impact parameter cuts in simulations should be replaced by centrality determinations consistent with experimental analyses. Moreover, the systematic uncertainties of $v_1$ and $v_2$ predicted by different transport models need to be quantified, including contributions from numerical configurations and from variations in the implementation of the momentum-dependent mean-field potentials in both mean-field and QMD frameworks.

However, when the analysis of subthreshold kaon production \cite{Hartnack:2005tr,Hartnack:2011cn} is taken into account, all observables that have so far been identified as sensitive to the nuclear equation of state point towards a soft momentum-dependent EoS, possibly with a compressibility modulus slightly higher than that currently used for a soft EoS. For neutron star physics, where the momentum dependence of the nucleon–nucleon potential is less relevant due to the low temperatures, the results from heavy-ion collisions indicate that the equation of state of cold symmetric nuclear matter remains soft up to densities of about $3\rho_0$, with a compressibility slightly larger than that adopted for a soft EoS. Given the compressibility, extracted from experiments
with giant monopole resonances, which are sensitive to densities around normal nulcear matter density, this implies that between normal nuclear matter density and three times that value, the compressibility does not change significantly. 
Future investigations may employ Bayesian inference techniques to extract, within a unified framework, the compressibility modulus, in-medium nucleon–nucleon cross sections, and effective masses constrained by a broad set of experimental observables. Beyond identifying parameter sets that best reproduce the data, such an approach would provide posterior probability distributions and parameter correlations, thereby quantifying uncertainties and degeneracies that cannot be accessed within the forward-modeling strategy, however is essential for a more robust determination of the nuclear equation of state.

\section*{Acknowledgements}
The authors acknowledge inspiring discussions with M. Bleicher, W. Cassing, H. Elfner, I. Grishmanovskii,  C.-M. Ko,  T. Reichert, N. Xu, O. Soloveva, T. Song,  Io. Vassiliev. 
We acknowledge fruitful discussions with  B. Kardan and his help for understanding of the HADES experimental data and acceptance.  
Furthermore, we acknowledge support by the Deutsche Forschungsgemeinschaft (DFG, German Research Foundation) grant BL982-3, by the Russian Science Foundation grant 19-42-04101 and by the GSI-IN2P3 agreement under contract number 13-70.
This study is part of a project that has received funding from the European Union’s Horizon 2020 research and innovation program under grant agreement STRONG – 2020 - No 824093.
The computational resources have been provided by the Center for Scientific Computing (CSC) of the Goethe University and the NICA LHEP offline cluster of the Joint Institute for Nuclear Research. V. Kireyeu acknowledge JINR for the Young Scientists and Specialists grant number 25-101-08. 


\bibliography{main.bib}

\end{document}